\documentclass{aa}

\usepackage{graphicx}
\usepackage{mathtools}
\usepackage{dsfont}
\usepackage{amssymb}
\usepackage{amsmath}
\usepackage{mathtools}
\usepackage{fixmath}
\usepackage[breaklinks=true]{hyperref} 
\usepackage{soul}
\usepackage{footmisc}
\usepackage{booktabs}
\usepackage{natbib}
\usepackage{url}
\usepackage{color}
\usepackage{soul}
\usepackage[varg]{txfonts}
\usepackage{array,multirow,graphicx}
\usepackage{twoopt}
\bibpunct{(}{)}{;}{a}{}{,} 
\makeatletter
\newcommandtwoopt{\citeads}[3][][]{\href{http://adsabs.harvard.edu/abs/#3}%
{\def\hyper@linkstart##1##2{}%
\let\hyper@linkend\@empty\citealp[#1][#2]{#3}}}
\newcommandtwoopt{\citepads}[3][][]{\href{http://adsabs.harvard.edu/abs/#3}%
{\def\hyper@linkstart##1##2{}%
\let\hyper@linkend\@empty\citep[#1][#2]{#3}}}
\newcommandtwoopt{\citetads}[3][][]{\href{http://adsabs.harvard.edu/abs/#3}%
{\def\hyper@linkstart##1##2{}%
\let\hyper@linkend\@empty\citet[#1][#2]{#3}}}
\newcommandtwoopt{\citeyearads}[3][][]%
{\href{http://adsabs.harvard.edu/abs/#3}
{\def\hyper@linkstart##1##2{}%
\let\hyper@linkend\@empty\citeyear[#1][#2]{#3}}}
\makeatother

\graphicspath{{./}{figures/}}

\date{Received 30 May 2023
    Accepted 25 July 2023}

\newcommand{\MTtwo}{\color{black}}

\titlerunning{Sun-like activity from a-coefficients}
\authorrunning{Benomar et al.}

\begin{document}

\title{Detecting active latitudes of Sun-like stars using asteroseismic a-coefficients}

\author{Othman Benomar\inst{\ref{inst1}}\inst{\ref{inst2}}\inst{\ref{inst3}} 
\and Masao Takata\inst{\ref{inst4}}
\and Michael Bazot\inst{\ref{inst5}}
\and Takashi Sekii\inst{\ref{inst1}}\inst{\ref{inst2}}
\and Laurent Gizon\inst{\ref{inst3}}\inst{\ref{inst6}}\inst{\ref{inst7}}
\and Yuting Lu\inst{\ref{inst8}}
}

\institute{Department of Astronomical Science, School of Physical Sciences, SOKENDAI, 2-21-1 Osawa, Mitaka, Tokyo 181-8588 \email{othman.benomar@nao.ac.jp, ob19@nyu.edu}\label{inst1}
\and Solar Science Observatory, National Astronomical Observatory of Japan, 2-21-1 Osawa, Mitaka, Tokyo 181-8588, Japan\label{inst2}
\and New York University Abu Dhabi, Center for Space Science, 
PO Box 129188, UAE\label{inst3}
\and Department of Astronomy, School of Science, The University of Tokyo, 7-3-1 Hongo, Bunkyo-ku, Tokyo, 113-0033, Japan \label{inst4}
\and Heidelberger Institut für Theoretische Studien, Schloss-Wolfsbrunnenweg 35, 69118 Heidelberg, Germany\label{inst5}
\and Max-Planck-Institut für Sonnensystemforschung, 37077, Göttingen, Germany \label{inst6}
\and Institut für Astrophysik, Georg-August-Universität Göttingen, 37077, Göttingen, Germany\label{inst7}
\and Department of Physics, The University of Tokyo,  Tokyo 113-0033, Japan\label{inst8}
}
%

\abstract {} 
{We introduce a framework to measure the asphericity of Sun-like stars using $a_1$, $a_2$ and $a_4$ coefficients, and constrain their latitudes of magnetic activity.}
{Systematic errors on the inferred coefficients are evaluated in function of key physical and seismic parameters (inclination of rotation axis, average rotation, height-to-noise ratio of peaks in power spectrum). The measured a-coefficients account for rotational oblateness and the effect of surface magnetic activity. We use a simple model that assumes a single latitudinal band of activity.}
{Using solar SOHO/VIRGO/SPM data, we demonstrate the capability of the method to detect the mean active latitude and its intensity changes between 1999-2002 (maximum of activity) and 2006-2009 (minimum of activity). We further apply the method to study the solar-analogue stars 16 Cyg A and B using Kepler observations. An equatorial band of activity, exhibiting intensity that could be comparable to that of the Sun, is detected in 16 Cyg A. However, 16 Cyg B exhibits a bi-modality in $a_4$ that is challenging to explain. We suggest that this could be a manifestation of the transition between a quiet and an active phase of activity. Validating or invalidating this hypothesis may require new observations.} 
{}


\maketitle
\section{Introduction} \label{sec:intro}

The observations from the space-borne instruments MOST \citep{Walker2003}, CoRoT \citep{Baglin2006a}, {\it Kepler} \citep{Borucki2010} and TESS \citep{Ricker2014TESS} have been important in advancing our knowledge of stellar interiors. This is particularly true for {\it Kepler}, which could observe continuously tens of thousand of stars for nearly four continuous years, enabling asteroseismic measurements that almost rival disk-integrated helioseismic measurements from a decade ago. 
Those precise measurements have for example enabled us to be better understand stellar rotation and its impact on stellar evolution. 
The asteroseismology of Sun-like stars is based on the study of the pressure modes that are excited by turbulent convection in the outer layers of these stars.
The acoustic modes may reach deep into the core or may {be localised} close to the stellar surface, giving access to the internal structure and dynamics of the stars.
Rotation has a critical impact on the stellar structure and evolution as it 
 induces a material mixing process \citep{Maeder2008}. Rotation is also an essential ingredient of the dynamo-effect \citep{Thompson2003} and can lead to a distortion of the star's shape due to the centrifugal force \citep{ Chandrasekhar1969}. 
For fast rotators, the rotational flattening must be taken into account. These stars show a variety of pulsation modes that are not present in slower rotators \citep[e.g.][]{Lignieres2006}. 

Despite its slow rotation rate, {the solar asphericity can be measured by helioseismology}. While the Sun is {seen as oblate by acoustic modes during the quiet phases of its activity cycle, the modes feel a more complicated shape during activity maxima}. During  solar maxima, 
the frequencies of acoustic modes increase slightly due to the presence of active regions.
The perturbation occurs near the surface and is stronger for modes that sense the active latitudes (below $\simeq 30^\circ$ for the Sun).
Physically the magnetic perturbations consist of several components that are not easy to disentangle (incl. stratification and wave speed perturbations \citep{Libbrecht1990,Antia2000,Dziembowski2000}. The magnetic perturbations affect the acoustic modes near the surface over only a few hundred kilometres, that is of order $10^{-5} R_{\odot}$, but is significant enough to be measured.

The origin of stellar activity is not well understood as it depends on the complex interplay between {rotation, convection and the magnetic field \citep{Brun2017b}}. 
Stellar magnetic cycles are observed in most cool stars \citep{Simon2002} 
over a large range of the electromagnetic spectrum and evidence of activity cycles is observed in X-ray \citep[eg.][]{Catura1975}, radio waves \citep{White1999,White2017}, chromospheric emission lines \citep{Vaughan1983,Olah2009,Olah2016}, and also through luminosity variation due to surface {magnetic activity} in the visible \citep{Hartmann1979,Silva-Valio2010,Ceillier2017}. On Sun-like stars, the level of activity is often observed to be cyclic, with activity periods ranging from a few years to decades. Although there are relationships between the stellar age, the rotation period, and the level of activity \citep{VanSaders2016},  the underlying mechanisms at play are not understood.

Since the advent of space-borne photometry and the observation of Sun-like pulsators by CoRoT, it became evident that helioseismic methods used for the Sun may be applied to asteroseismic observations. 
This has resulted in robust estimates of the average rotational splittings, which, in combination with the surface rotation rates inferred from photometric variability, indicate 
 that main sequence stars have nearly-uniform internal rotation rates, 
\citep{Gizon2013,Benomar2015,Nielsen2017}. 
All of the recent seismic studies of radial differential rotation agree that
angular momentum transport in the radiative zone is much more efficient than considered in theory, even in the case of stars more massive than Sun-like stars, such as $\gamma$-Doradus stars \citep[eg.][]{Mosser2012a,Gehan2018,Ouazzani2019}.
For the best Kepler observations,  asteroseismology showed evidence of
latitudinal differential rotation for main sequence stars \citep{Benomar2018Sci} and of the radial differential rotation in RGB \citep[e.g.][]{Deheuvels2012,Deheuvels2014}.

This paper aims at providing a framework to study stellar activity and its latitude by analysing its effect on pulsation frequencies. The proposed method involves the use of the $a$-coefficient decomposition \citep{Schou1994,Pijpers1997,Pijpers1998} on the stellar power spectrum, conveniently separating the perturbations caused  by rotation and asphericity.
The method is tested on  Sun-as-a-star data and on the solar-analogues 16 Cyg A and B, which are the two brightest stars in the initial Kepler observation field.

Only few successful measurements of the asphericity of other stars than the Sun have been made so far. Using ultra-precise measurements of the frequency splittings of time-harmonic (i.e. non-stochastic) low-degree p-modes, \cite{Gizon2016} inferred the oblateness of the  $\gamma$ Doradus–$\delta$ Scuti star
KIC 11145123 to be  $\Delta R =(1.8\pm0.6) \times 10^{-6} R \simeq 3\pm1$ km, i.e. smaller than expected from rotational oblateness alone, suggesting the presence of magnetic activity at low latitudes.  \cite{Bazot2019} measured the asphericity of the
solar-like pulsators 16 Cyg A and B and found that 16 Cyg A is likely prolate, implying that this star may have  low-latitude  magnetic activity on its surface. 

In the spirit of the study by \cite{Gizon2002AN}, we will perform monte-carlo simulations to demonstrate the possibility of inferring the even-$a$ coefficients from simulated oscillation power spectra to constrain the latitude of activity.
Unlike \cite{Gizon2002AN} who included only the $a_2$ coefficient in the  parametric model,  we will infer both the $a_2$ and $a_4$ coefficients.
We start in Section \ref{sec:splittings} by introducing the effects of rotation on pulsation frequencies and discuss the effect of the centrifugal force and of the activity on the mode cavities. Section \ref{sec:information_content} presents the assumptions required for the asteroseismic measurement of stellar activity. Section \ref{sec:bias} discusses the achievable accuracy of the seismic observables, while sections \ref{sec:results:Sun}  and \ref{sec:16CygAB} present the results on solar data and for 16 CygA and B. This is followed by a discussion and conclusion in section \ref{sec:discuss}.

\section{The effect of rotation and of magnetic activity} \label{sec:splittings}

This section presents the effect of rotation and of magnetic activity on pulsation modes, and introduces the frequency model used for asteroseismic data analysis.

\subsection{Frequency splittings}
Slowly rotating stars without significant magnetic activity are approximately spherical and it is common to describe the family of modes travelling inside it using spherical harmonics \cite[see e.g.][]{Unno1989}. If the departure from spherical shape remains small enough, it is convenient to keep the spherical representation for the equilibrium model and account for distortions through a perturbation analysis. All pulsations can then be described using a set of integers ($n$,$l$,$m$), namely the radial order, the mode degree and the azimuthal order, respectively. Acoustic pressure modes observed in Sun-like stars can be identified using their frequencies $\nu_{nlm}$.

\begin{figure}
\begin{center}
\includegraphics[angle=0, scale=0.29]{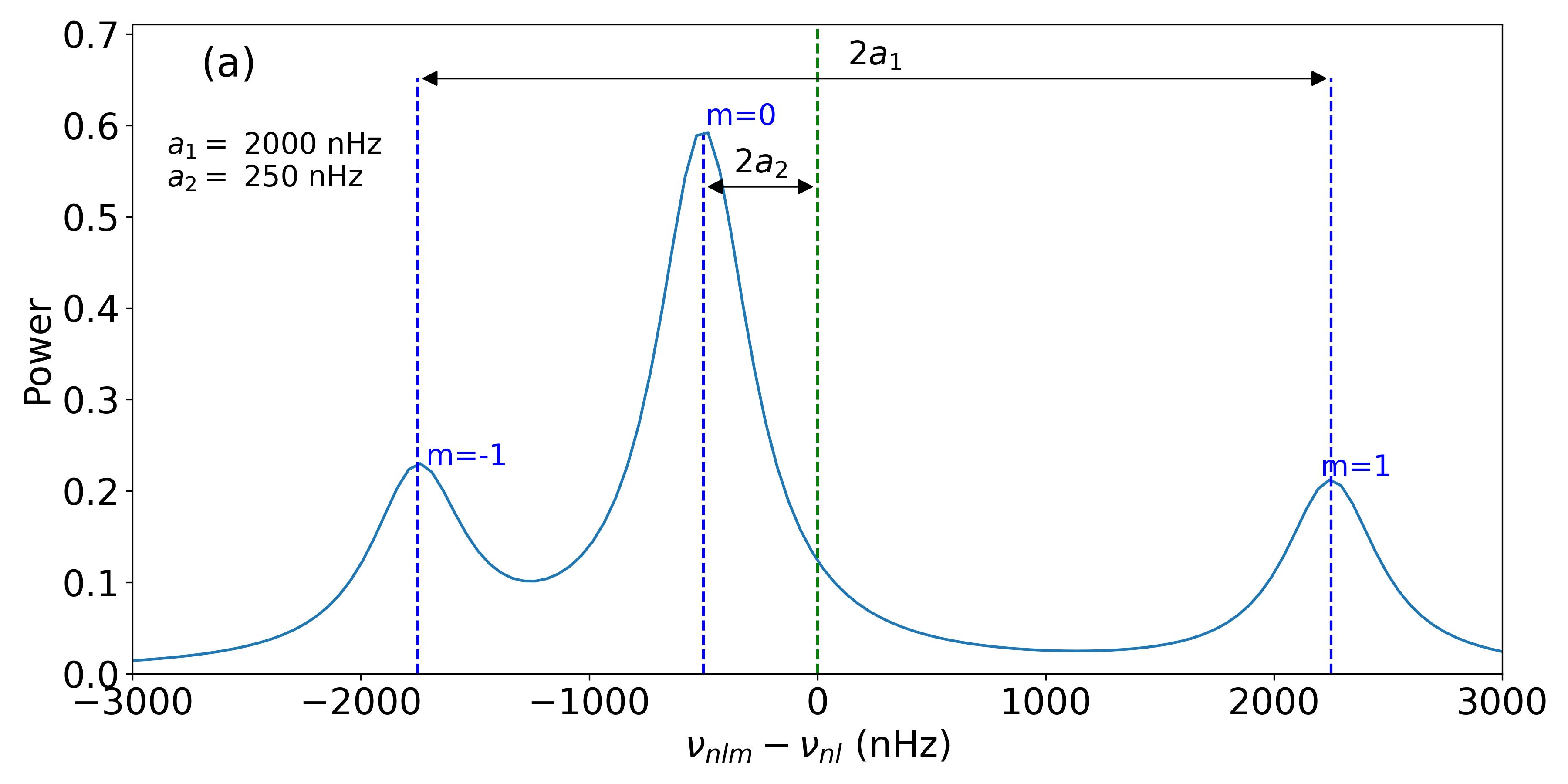} \\
\includegraphics[angle=0, scale=0.29]{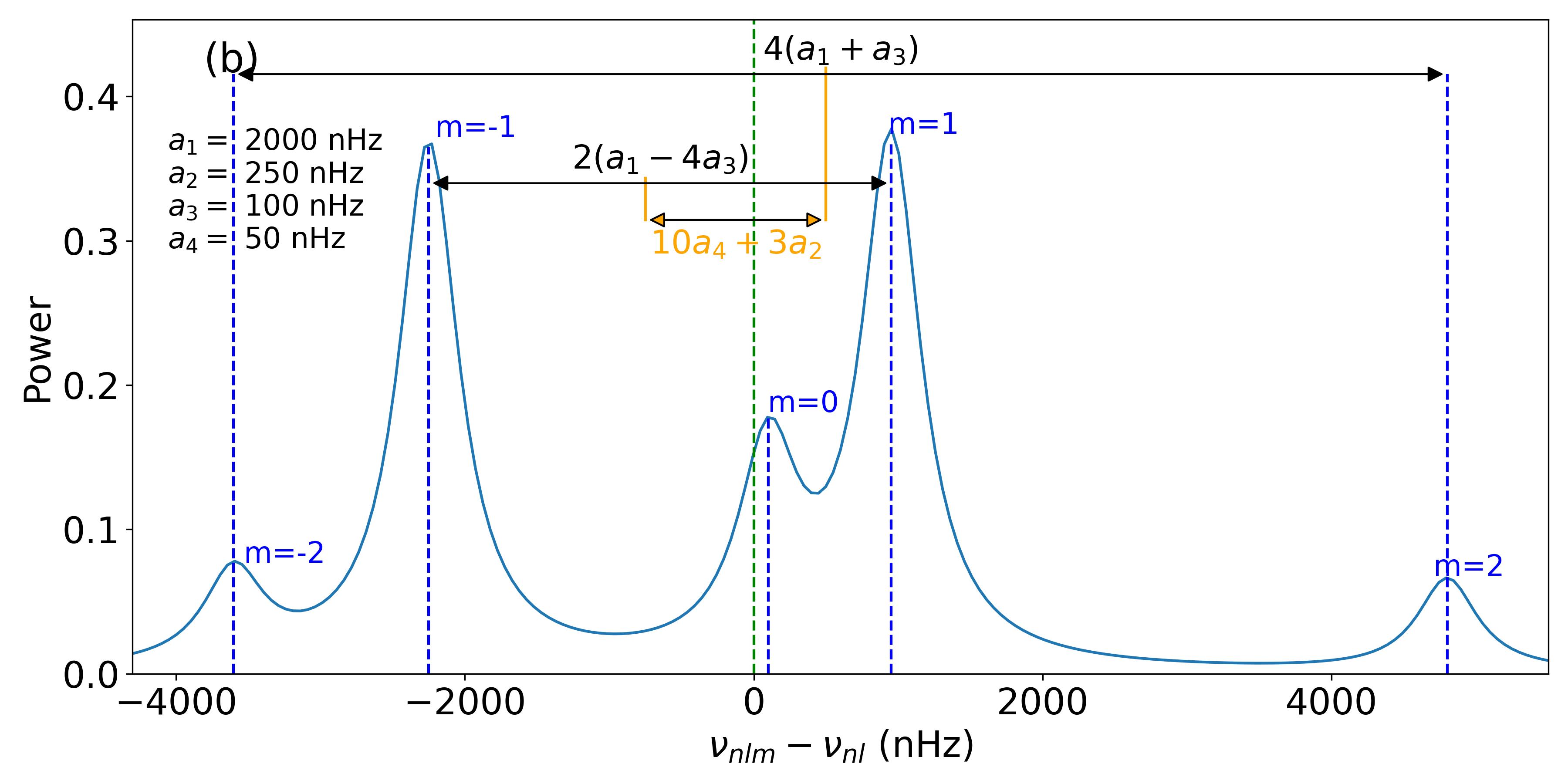}
\caption{Example of Lorentzian mode profiles showing the a-coefficients and their relationship with frequency spacings for $l=1$ (a) and $l=2$ modes (b). The orange spacing is $T_{n22} - T_{n21}$. $\nu_{nl}$ is the m-averaged frequency of each multiplet. The stellar inclination is $40^\circ$.} \label{fig:aj_visuals}
\end{center}
\end{figure}

In a non-rotating, non-active star, m-components are degenerate and cannot be resolved. When rotation or magnetic activity sets in, this degeneracy is lifted. The resulting frequency is treated as a perturbation to the degenerate frequency without rotation and activity, 
\begin{equation}
    \nu_{nlm} = \nu_{nl} + \delta\nu_{nlm},
    \label{Eq:1}
\end{equation}
with $\nu_{nl}$, the equilibrium eigenfrequency without rotation and activity and $\delta\nu_{nlm}$ the frequency splitting. These splitting may depend on  multiple physical effects perceived by the modes \citep{Libbrecht1990} within their cavity of propagation. These can be terms of order $\mathcal{O}(\Omega)$, with $\Omega$ the rotation rate estimated at the equator, $\Omega = \Omega(r, \theta = \pi/2)$. These depend directly on the stellar rotation profile $\Omega(r,\theta)$. Higher-order perturbations pertaining to the shape of the mode cavity can also exist.
Note that $\nu_{nl}$ differs from $\nu_{nl,m=0}$ as the $m=0$ components may have their frequency shifted by perturbations such as the magnetic activity (see e.g. Figure \ref{fig:aj_visuals}).

Splitting can be described using the Clebsch-Gordan a-coefficient decomposition \citep{Ritzwoller1991}, that corresponds to a representation of the splittings on a basis of polynomials $\mathcal{P}^{(l)}_j(m)$ {of degree $j$ in $m$},
\begin{equation}
    \delta\nu_{nlm} = \sum^{j_{max}}_{j=1} a_j(n,l) \mathcal{P}^{(l)}_j(m),
    \label{Eq:4}
\end{equation}
with the polynomials such that
\begin{equation}
    \sum_{m=-l}^l \mathcal{P}^{(l)}_i(m)\, \mathcal{P}^{(l)}_j(m) = 0  \,\,\, \mathrm{when }\,\,\, i \ne j.
    \label{Eq:4b}
\end{equation}

Here, $a_j(n,l)$ is the a-coefficient of order $j$ and $j_{max}=2l$ is the maximum order to which the decomposition must be carried for a given degree. The standard set of 
polynomials used in this expansion are those normalised as per described by \cite{Schou1994}.

This decomposition is extensively used in helioseismology and was used on Sun-like stars by \cite{Benomar2018Sci}. Asteroseismology has been so far unable to observe modes of degree higher than $l=3$ so that in the following the discussion is restricted to $l\leq3$ and $j_{max}=2l=6$, due to the selection rule of the $\mathcal{P}^{(l)}_{j}(m)$. This limitation is the consequence of full-disk integrated photometric observations. 
An example of splitting including odd and even $a_j$ coefficients is given in Figure \ref{fig:aj_visuals} for $l=1,2$.

This theoretical model leads to a natural interpretation of the observed splittings. One may decomposes these latter into their symmetrical, $S_{nlm}$, and anti-symmetrical, $T_{nlm}$, parts. These components can then be described as sums over, respectively, the odd and even a-coefficients,
\begin{align} \label{Eq:5:Snlm-Tnlm-1}
  S_{nlm} &= \frac{\nu_{n,l,m} - \nu_{n,l,-m}}{2m} = \frac{1}{m} \sum^{j_{max}/2}_{j=1} a_{2j-1}(n,l) \mathcal{P}^{(l)}_{2j-1}(m), \\ 
  \label{Eq:5:Snlm-Tnlm-2} T_{nlm} &= \frac{\nu_{n,l,m} + \nu_{n,l,-m}}{2} - \nu_{n,l,0} = \sum^{j_{max}/2}_{j=1} a_{2j}(n,l) (\mathcal{P}^{(l)}_{2j}(m) - \mathcal{P}^{(l)}_{2j}(0)).
\end{align}
This arises from the parity relation $\mathcal{P}_j^{(l)} (-m) = (-1)^j \mathcal{P}_j^{(l)} (m)$. These equations can be used to express the a-coefficients with $S_{nlm}$ and $T_{nlm}$. They also provide relations between the $\nu_{nlm}$ with the $a_j(n,l)$ (Appendix \ref{appendix:splittings}).

It can be seen from Appendix~\ref{appendix:Plm} that the sums in Eqs.~(\ref{Eq:5:Snlm-Tnlm-1}) and (\ref{Eq:5:Snlm-Tnlm-2}) involve, respectively, odd and even functions of $m$. Physically, this means that the symmetrical components of the splittings \citep{Gough1990b} result from large-scale perturbations sensitive to the prograde or retrograde nature or the waves, such as advection or the Coriolis force. On the other hand, the anti-symmetric splittings are caused by processes that are not affected by the propagation direction of waves. This may include the centrifugal force, that scales as $\mathcal{O}(\Omega^2)$, and whose effect on the oscillation frequencies varies as $m^2$. Magnetic fields or non-spherical deformations of the equilibrium structure will also contribute to the anti-symmetric splittings.

We further decompose the anti-symmetric splittings into a term depending on centrifugal-force-induced distortions and another one accounting for activity-related distortions,
\begin{equation}   
    T_{nlm} = \delta\nu^{(CF)}_{nlm} + \delta\nu^{(AR)}_{nlm}. 
\end{equation}
The symmetric part of the splitting corresponds to a term $\delta\nu^{(rot)}_{nlm}$ that stems from perturbations of order $\mathcal{O}(\Omega)$ and the total observed splitting is,
\begin{equation}  
    \delta\nu_{nlm} = \delta\nu^{(rot)}_{nlm} + \delta\nu^{(CF)}_{nlm} + \delta\nu^{(AR)}_{nlm}.
    \label{Eq:2}
\end{equation}
Using the Sun as an archetype of Sun-like star, it is possible to measure these contributions to $\delta\nu_{nlm}$ as shown in Sections \ref{sec:2.2} and \ref{sec:splittings:3}.

\subsection{Expressions for $\delta\nu^{(rot)}_{nlm}$, $\delta\nu^{(CF)}_{nlm}$ and $\delta\nu^{(AR)}_{nlm}$} \label{sec:2.2}
    
To the first order, the perturbation on the frequency due to rotation is,
\begin{equation}
     \delta\nu^{(rot)}_{nlm} = \frac{m}{2\pi} \int^{R}_0 \int^\pi_0  K_{nlm}(r, \theta) \Omega(r, \theta) r\,dr\,d\theta , 
     \label{Eq:6:Kernels}
    \end{equation}
where $R$ is the radius of the star and the kernel  $K_{nlm}(r, \theta)$ \citep{Hansen1977} expresses the
sensitivity of a mode to the rotation at the radial point $r$ and co-latitude $\theta$. $\Omega(r, \theta)$ is the rotation profile of the star.
It can be shown that $\delta\nu^{(rot)}_{nlm}$ actually only depends on symmetrical splittings 
\citep{Ritzwoller1991} which in turns depend only on odd coefficients. For example and for $l=3$, it is expressed as, 
\begin{equation}
    \delta\nu^{(rot)}_{nlm} \simeq \mathcal{P}^{(l)}_1(m)\,a_1 + \mathcal{P}^{(l)}_3(m)\,a_3 + \mathcal{P}^{(l)}_5(m)\,a_5. 
    \label{Eq:6:acoefs}
\end{equation}

Centrifugal forces typically distorts a spherical rotating sphere of gas into an oblate ellipsoid, elongated at the equator. Functional analysis shows that the contribution of centrifugal forces to the mode splitting scales with ${\Omega^2 R^3}/{\mathcal{G} M}$. Properly integrating higher-order terms of the perturbation expansion over the aspherical volume of the star and using asymptotic expressions for the equilibrium mode eigenfunctions (assuming $n$ is large enough), leads to the following expression for the centrifugal-force component of the frequency splitting \citep{Gough1984,Gough1990b},
\begin{equation}
    \delta\nu^{(CF)}_{nlm}= \frac{\Omega^2 R^3}{
    \mathcal{G} M} \nu_{nl} Q_{lm},
    \label{Eq:7}
\end{equation}
with $Q_{lm} \approx \frac{2}{3} \frac{l(l+1) - 3m^2}{(2l-1)(2l+3)}$ factor {depending on the density}. 
{\MTtwo
Equation (\ref{Eq:7}) means {that}
\begin{equation}
  \delta\nu^{(CF)}_{nlm} =
\mathcal{P}^{(l)}_2(m)\,a^{(CF)}_2(n,l)
\end{equation}
with
\begin{equation}
    a^{(CF)}_2(n,l)
    =
    -\frac{1}{2l+3}
    \frac{\Omega^2 R^3}{\mathcal{G} M} \nu_{nl}
\label{eq:a2CF}
\end{equation}
since
$-(2l+3)Q_{lm}=
\mathcal{P}^{(l)}_2(m)$.
}
 It should be noted that the contribution of the centrifugal-force-induced deformation to the frequency splittings can be described by a linear combination of the $\mathcal{P}_{2j}$ polynomials \citep{Gough1984, Gough1990b}. Since the stars are assumed to be slowly rotating, one may only retain terms of order $\mathcal{O}(\Omega^2)$, which correspond to the contribution of the $a_2$ coefficient alone.

One can further approximate 
{\MTtwo
equation (\ref{eq:a2CF})
}
in order to express $\delta\nu^{(CF)}_{nlm}$ as a function of quantities that can be obtained directly from the modelling of the acoustic power spectra of Sun-like stars \citep{Benomar2018Sci}. First, the stellar mean density of a Sun-like star scales to a good approximation with its large separation, that is the average distance in the frequency space between to modes of identical degree and consecutive orders \citep{Ulrich1986}. In solar units, this reads $\rho = (\rho_{\odot}/\Delta\nu_{\odot})\Delta\nu$, with the solar density $\rho_{\odot} = (1.4060 \pm 0.0005) \times 10^3$ kg\ m$^{-3}$ and $\Delta\nu_{\odot} = 135.20 \pm 0.25\ \mu$Hz \citep{Garcia2011b}. With an accuracy estimated to a few percents for Sun-like stars \citep{White2011}, the use of this scaling relation is thought to be a decent approximation.

The second simplification uses the fact that the Clebsch-Gordan coefficient decomposition of the frequency splitting imposes a one-to-one relationship between the a-coefficients and the coefficients of the decomposition of the velocity field into poloidal and toroidal components \citep{Ritzwoller1991}. Helioseismology suggests that the Sun rotates with a near constant angular velocity down to at least $r/R_{\odot} = 0.2$ \citep[e.g.][]{Thompson2003}, which is the maximum depth at which measurements from low-degree {p} modes are available. Furthermore, its outer-convective zone shows a differential rotation of $\simeq 30\%$ from the equator to the pole, which leads to an $a_{3,\odot}(n,l) \simeq 4$ nHz and to even smaller values for higher odds a-coefficients. This is significantly smaller than  $a_{1,\odot}(n,l) \simeq 420$ nHz. Therefore, the aforementioned one-to-one relation ensures that we can retain only the leading order in the expansion of the rotation rate and treat it as an average value, given in terms of seismic observables by 
{\MTtwo
$\Omega \simeq 2\pi a_1$}. 
This leads to
{\MTtwo
\begin{equation}
a^{(CF)}_2(n,l)
    \simeq
-\frac{\nu_{nl}}{2l+3}\, 
\frac{3\pi}{\mathcal{G} \rho_{\odot}}\frac{\Delta\nu^2_{\odot}}{\Delta\nu^2}a^2_1
.
    \label{Eq:8}
\end{equation}
}

Regarding $\delta\nu_{nlm}^{(AR)}$, there is no unambiguous theory to describe the effect of the near-surface magnetic activity on the shape of the cavity. Due to this and following \cite{Gizon2002AN}, 
a geometrical description is preferred to a physical model. {This description assumes that the corresponding wave-speed  perturbation separates in the latitudinal and radial coordinates.}
The proposed form of the perturbation in frequency is, 
    \begin{align}
        \delta\nu^{(AR)}_{nlm} &= \nu_{nl} \epsilon_{nl} \int_0^{2\pi} \int_0^\pi F(\theta|\boldsymbol{x}) |Y^m_{l}(\theta,\phi)|^2 \sin \theta\,d\theta\,d\phi \nonumber \\
        &= \nu_{nl}\epsilon_{nl} A_{lm}(\boldsymbol{x}) \nonumber \\
        &= \sum^{j_{\max}/2}_{j=0} \mathcal{P}^{(l)}_{2j}(m)\,a^{(AR)}_{2j}(n,l).
        \label{Eq:9}
    \end{align}
{The term $a^{(AR)}_{2j}(n,l)$ in equation (\ref{Eq:9}) refers to the combined contribution of a magnetic field and other perturbations in the stellar structure (e.g. stratification, temperature).}  The geometrical weight function $A_{lm}(\boldsymbol{x})$ describes the effect of an active zone at the  co-latitude $\theta$, on a mode of degree $l$ and azimuthal order $m$.  $A_{lm}(\boldsymbol{x})$ is the product of two contributions. First, the normalised spherical harmonics $Y^m_{l}(\theta,\phi)$ that decompose the magnetic activity effect over each modes. These are spherical-polar coordinates defined
in the inertial frame with a polar axis pointing in the direction
of the rotation axis.
Second, the weight distribution (or the shape of the active region) is defined by $F(\theta|\boldsymbol{x})$. Here, $\boldsymbol{x}$ refers to the parameters that are necessary to describe the function $F(\theta|\boldsymbol{x})$. 

{On the Sun, large active regions persist on the surface for 1-2 rotation periods and are randomly distributed in longitude over well defined latitudes. Here we assume that the corresponding perturbation can be approximated by  a function of latitude only. 
The general problem of distinct active regions on the differentially rotating surface would go beyond the present study \citep[see][for the case of a single long-lived active region]{Papini2019}.
In this paper, we only consider perturbations that are approximately steady in the inertial frame (i.e. latitudinal bands of activity)}.

In theory, a third integral over the radius is necessary to describe the dependence of the magnetic activity to the stellar depth. However, p modes are weakly sensitive to the deep structure inside stars. Here, the radial integral is replaced by $\epsilon_{nl}$, the overall activity intensity.
The frequency $\nu_{nl}$ allows a dimensionless $\epsilon_{nl}$, that can be compared between stars.
Section \ref{sec:information_content:Alm} further develops the required assumptions in order to obtain a reliable information content on the active region in asteroseismology.

    \subsection{Modelling the frequencies} \label{sec:splittings:3}

Combining equation (\ref{Eq:1}), (\ref{Eq:2}), (\ref{Eq:6:Kernels}), (\ref{Eq:8}) and (\ref{Eq:9}) leads to,

    \begin{align}
        \nu_{nlm} &= \nu_{nl} \left( 1 + \frac{3\pi}{(2l+3) \mathcal{G} \rho_{\odot}}\frac{\Delta\nu^2_{\odot}}{\Delta\nu^2}a^2_1 \right) \nonumber \\
        &\quad + \epsilon_{nl} A_{lm}(\boldsymbol{x}) + \frac{m}{2\pi} \int^{R}_0 \int^\pi_0 K_{nlm}(r, \theta) \Omega(r, \theta) r\,dr\,d\theta.
        \label{Eq:11}
    \end{align}

Using equation (\ref{Eq:1}) and (\ref{Eq:4}), $\nu_{nlm}$ can also be expressed using the a-coefficients  without loss of generality,
\begin{equation}
    \nu_{nlm}=  \nu_{nl} + \sum_{j=1}^{2l} \mathcal{P}^{(l)}_j(m)\,a_j(n,l).
    \label{Eq:12}
\end{equation}

It is possible to use equation (\ref{Eq:11}) in order to measure the activity parameters $\boldsymbol{x}$, by directly fitting the power spectrum. However, there are multiple benefits to use instead a two-step approach consisting in first using equation (\ref{Eq:12}) for the fitting of the power spectrum to evaluate the a-coefficients (method detailed in Appendix \ref{appendix:spectrum_analysis}). And then, fitting the coefficients obtained during the first step, using solely the equations (\ref{Eq:8}) and (\ref{Eq:9}) (detailed in Appendix \ref{appendix:aj_analysis}).
Firstly and in the general case, the evaluation of $A_{lm}(\boldsymbol{x})$ requires the precise computation of a double integral. This is a slow process, that increases the time necessary to fit the power spectrum\footnote{Using a MCMC method, from a day to a couple of weeks for a single star on a present-day CPU}. Postponing to an ulterior step the computation of $A_{lm}(\boldsymbol{x})$ reduces drastically the parameter space from a few tens of parameters to only a few\footnote{With our choice of $A_{lm}(\boldsymbol{x})$, three parameters are fitted. See section \ref{sec:information_content:Alm}.}, effectively making faster the convergence rate of the fitting algorithm and making it easier to explore the assumptions that have to be made on equation (\ref{Eq:11}) in order to have a functional approach in real cases (see the discussions in Section \ref{sec:information_content}).
Secondly, it allows us to decouple the observables from the physics, enabling to test various physical assumptions without having to re-perform the lengthy power spectrum analysis.
Finally, it eases the evaluation of the reliability of the activity determination, by enabling us to pinpoint the cause of biases (if any) on either the observables (eg. $a_2$, $a_4$, see Section \ref{sec:bias}) or on the interpretation of these observables in terms of physical parameters (see Section \ref{sec:3.3}).
The disadvantage of the two-step approach is that it requires more statistical assumptions, such as neglecting correlations and assuming Gaussian parameters. Although one can argue that it is possible to perform a hierarchical Bayesian analysis \citep[eg.][]{Hogg2010,Campante2016} to partly alleviate this issue, such an approach is generally slow and may cancel the benefits of the two-step approach.

\section{Information content in low order a-coefficients} \label{sec:information_content}

The general theoretical formulations of Section \ref{sec:splittings}, demonstrate how pulsation frequencies are expressed as function of the rotation, the centrifugal distortion and the activity of stars. However, observational limitations need to be accounted to enable a viable, robust model that extract as much possible the information content within currently existing data. This necessarily requires additional assumptions, for which the rational is detailed hereafter.

    \subsection{Factors contributing to the splitting accuracy and precision}
    
As explained in \cite{Kamiaka2018}, the pulsation height (H) and the noise background (N) are important factors that reduce the capabilities of asteroseismic analyses. Those are a complex function of the global stellar characteristics (mass, radius, age), of the convective motion at the surface of Sun-like stars, and also depend on the instrumental limitations. The mode height and the noise background are in fact difficult to evaluate {\it a priori}. However, the Height-to-Noise (HNR), defined as the ratio $H/N$ can be used to assess the quality of the spectrum of a pulsation mode. 
In fact, \cite{Kamiaka2018} show that the HNR at maximum of mode height $\widehat{HNR}$, can be used to study biases on the stellar inclination. This is because all main sequence Sun-like stars show a similar dependency of the height and of the width as a function of the frequency \citep[see e.g.][]{Appourchaux2014}. In Section \ref{sec:bias}, we propose the same, but on a-coefficients. 

The $\widehat{HNR}$ not only defines how many modes can be observed, but also the maximum degree of the modes that is observed. Considering specifically the Kepler observations, stars exhibit a $\widehat{HNR}$ for $l=0$ modes of up to 30, the highest HNR being observed for 16 Cyg A and 16 Cyg B. 
 \cite{Kamiaka2018} showed that the capability of distinguishing rotationally split components is of great importance if one wants to obtain a reliable asteroseismic inference of the rotation characteristics and of the stellar inclination. In particular, the ratio $a_1/\Gamma_{\nu_{max}}$ between the $a_1$ coefficient and the mode width determined at the maximum of mode amplitude  $\Gamma_{\nu_{max}}$ (see Figure \ref{fig:ref_star}), determines the expected bias on the stellar inclination. As shown in Section \ref{sec:bias}, $a_1/\Gamma_{\nu_{max}}$ also controls the importance of the bias for the other low order a-coefficients. 
Finally, another important factor is the spectrum resolution. The higher the resolution, the more resolved are the modes. Thus, observations $T_{obs}$ of several years are the most suitable in order to resolve and measure rotationally split components. Broadly speaking, observations exceeding a year and $a_1/\Gamma > 0.4$ are preferable to ensure a reliable measurement of $a_1$.

All of the limitations discussed above incite us to introduce assumptions to ensure robust measurements of a-coefficients (ie, mitigate biases).
 
 \subsection{Latitudinal profile of the activity} \label{sec:information_content:Alm}
 
Noting that it is challenging to measure low-degree a-coefficients for the Sun \citep[eg.][]{Toutain2001b}, we present here a minimal set of assumptions on even a-coefficients allowing us to constrain the asphericity of stellar cavities. 
One of the first aspects that has to be considered is the form of {$F(\theta | \boldsymbol{x})$, the function characterising the activity latitude $\theta$} (see equation (\ref{Eq:9})).

 \begin{figure*}
 \begin{center}
\includegraphics[angle=0, scale=0.75]{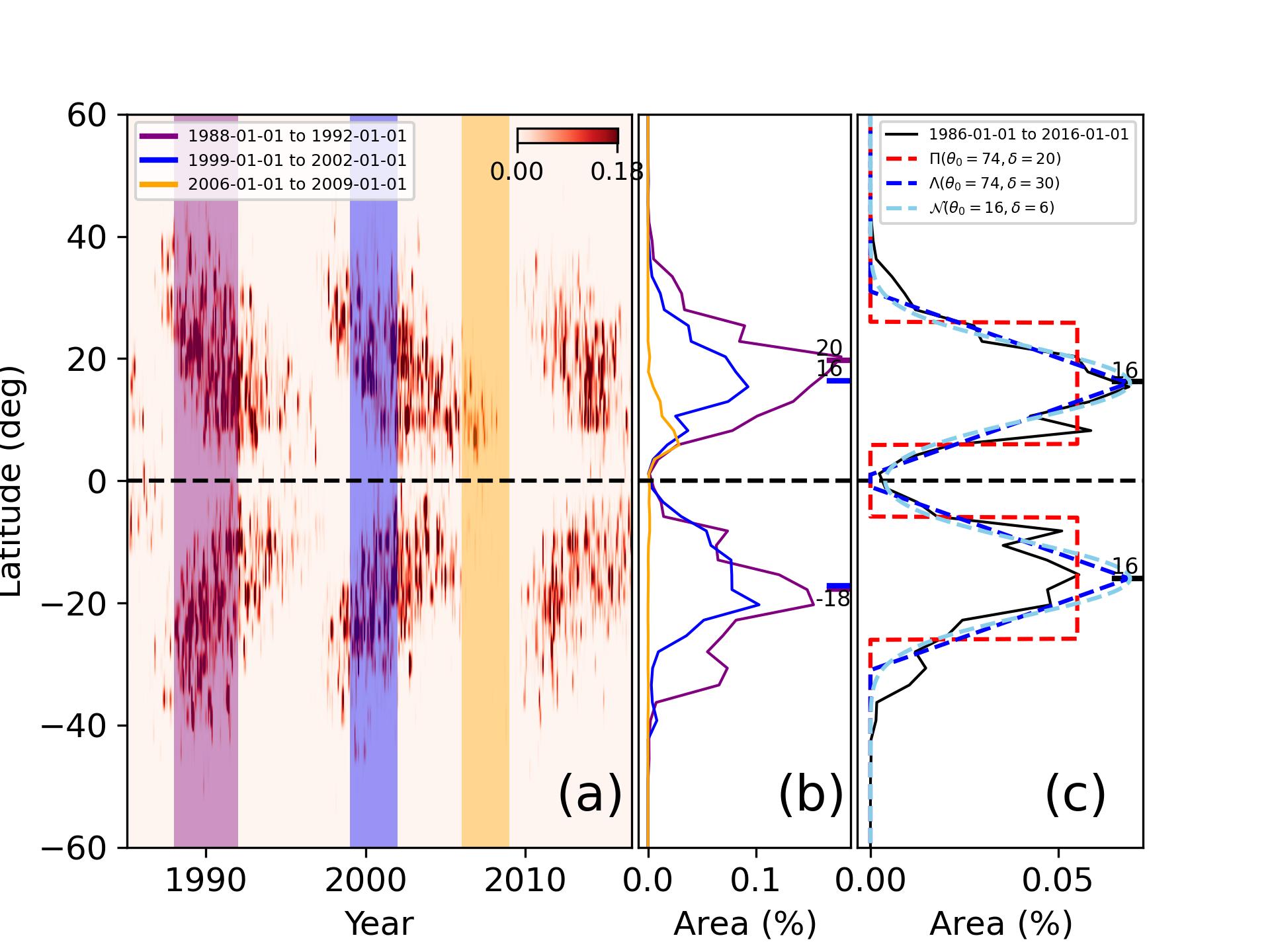}
\caption{Solar activity over time. (a) Butterfly diagram for the Sun. Vertical colour bands highlight periods of maximum of 1988-1992, 1999-2002 and minimum of activity of 2006-2009. (b) averaged spot area for the highlighted periods. Coloured ticks on (b) are the weighted mean for the activity latitude. {(c) Filters $F(\theta|\mathrm{X})$ used for this study superimposed to solar data (1986-2016). Parameters $\theta_0$ and $\delta$ are the latitude and the extension of the active region, respectively.}} \label{fig:Butterfly}
\end{center}
\end{figure*}

{The butterfly diagram of the Sun (Figure \ref{fig:Butterfly}) is used as a reference for this latitudinal dependence}. The data are from the Greenwich USAF/NOAA observatory \footnote{\url{https://solarscience.msfc.nasa.gov/greenwch.shtml}} and provide the daily area of the spots, counted manually over the period 1874-2016. 
The panel (a) on Figure \ref{fig:Butterfly} shows the butterfly diagram with colours representing the area covered by the spots in unit of percent of visible hemisphere. {It focuses on the observations after 1985 and covers two full solar cycles.} Vertical colour bands highlight three time intervals : 1985-1989 (Purple), 1999-2002 (Blue) and 2006-2009 (Yellow). The two first are during a maximum of solar activity while the last one is for a minimum of activity.
Due to the gradual migration of the spots over time, the longer observation period (1985-1989, 4 years) leads to a broader activity zone than the period 1999-2002. This indicates that the extension of the active region may not be trivial to measure in other stars because it will depend on the fraction of time the star is observed relative to the duration of its activity cycle. The activity cycles of Sun-like stars (if any) is a priori unknown, but Ca II H+K line emission and photometric studies \citep{Olah2009,Olah2016} suggest that it is of durations of roughly a few years to decades, as for the Sun. It indicates that over the course of several years, an active band as large as $\simeq 40^\circ$ may be expected.

Panel (b) of Figure \ref{fig:Butterfly} shows the cumulative area of spots as a function of the latitude and for the three considered periods. As noted earlier, the extension of the activity band is larger for the longest time-frame. The area of the spots during the active solar phase are symmetrical towards the equator. This suggests that when the activity is strong, $F(\theta|\boldsymbol{x})$ is almost north-south symmetric. This may be inaccurate for low activity phase, as shown for the period 2006-2009 {but because $|Y^m_{l}(\theta,\phi)|^2$ is also symmetrical towards the equator, }
this has no incidence on the $a^{(AR)}_{j}(n,l)$ coefficients.

During the minimum, the total average area of the spots is a few times lower than during the maximum. It is also narrower, such as the integral in equation (\ref{Eq:9}) is small, reducing $a^{(AR)}_{j}(n,l) \simeq 0$. 
During the phase of minimum of activity, the activity can effectively be considered as nonexistent (see Section \ref{sec:results:Sun} for the analysis on solar data), so that $a_2(n,l)$ is dominated  by the centrifugal term $a^{(CF)}_2(n,l)$ and the other even a-coefficients are null.

The overall latitudinal profile is seemingly following a bell-shape with sharp slopes {during periods of activity.
Figure \ref{fig:Butterfly}c compares data between 1986 and 2016 with three models for the active latitudes: a model using the gate function $F(\theta|\mathbf{x})=\Pi(\theta_0,\delta)$, a triangular function $\Lambda(\theta_0, \delta)$ and a Gaussian function $\mathcal{N}(\theta_0, \delta)$. The parameters of these were adjusted manually to approximately match the solar spot active latitudes profile. The triangular and the Gaussian functions describe equivalently the data, while the gate function initially proposed by \cite{Gizon2002AN} is roughly fitting the data. Because spots are strictly appearing at latitudes below $45^\circ$, the $\Lambda$ function may be the most suitable for the Sun. Nevertheless, these three functions are retained and compared here-further.}

    \subsection{Assumptions on the a-coefficients} \label{sec:3.3}
    
The most direct method for measuring the asphericity is to evaluate it directly for each mode ($n$,$l$), that is, measuring the terms $a_j(n,l)$. This being already difficult for low-order a-coefficients of the Sun {\citep{Chaplin2003}},  
it seems unreasonable to expect an accurate measurement of all individual $a_j(n,l)$ using asteroseismic data. These have a lower signal-to-noise ratio and severely reduced visibility at $l\ge3$ due to the integrated photometry. 
{After a trial and error process, jointly with power spectra simulations, we could identify a set of assumptions} ensuring reliable and precise measurement of a-coefficients.

We first consider a fictitious star rotating as a solid-body with a solar activity level ($\epsilon_{nl} \simeq 5 \times 10^{-4}$, \cite{Gizon2002AN}) with $a_1=1000$ nHz. The observed oscillation frequencies of 16 Cyg A are used here \citep{Davies2015, Kamiaka2018}. Split frequencies are derived using equations (\ref{Eq:1}), (\ref{Eq:2}), (\ref{Eq:8}), (\ref{Eq:9}), and converted into a-coefficients using equations (\ref{Eq:Appendix:a11:T}-\ref{Eq:Appendix:a63:T}).
The a-coefficients are linear functions of the frequency, which is expected {as $F(\theta|\theta_0, \delta)$ describes a single active region. Thus, }a reasonable assumption is to consider those as pure first order polynomial functions of frequencies. However, tests on artificial spectra  showed that it is often difficult to evaluate the slope of the a-coefficients. This is because the uncertainty on any $a_j$ is at least of the same order as its variations within the range of observed frequencies. This suggests that current asteroseismic data lack the resolution and the signal-to-noise to reliably measure the frequency dependence on the a-coefficients.

\begin{figure}
\begin{center}
\includegraphics[angle=0, scale=0.38]{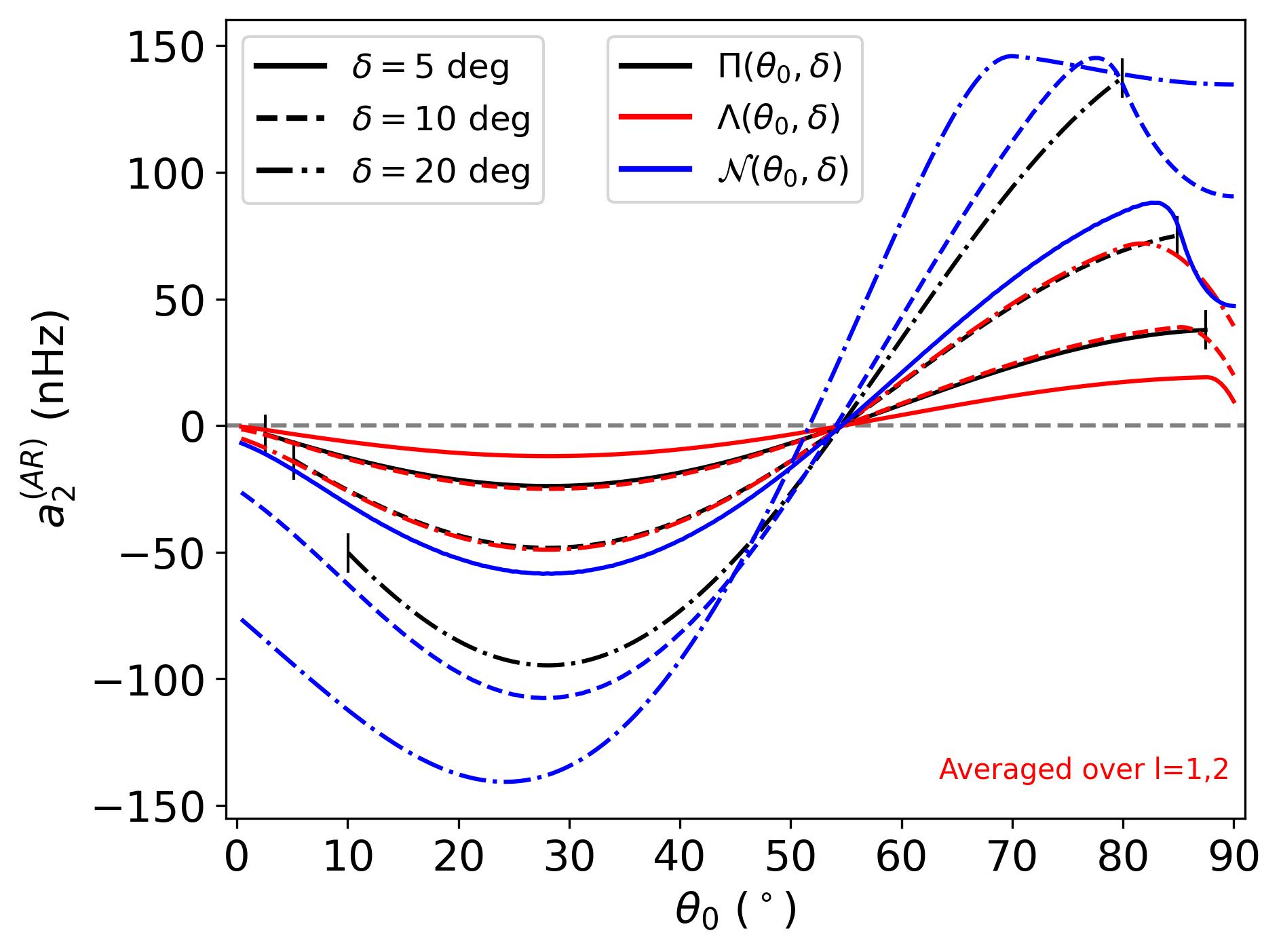}
\includegraphics[angle=0, scale=0.38]{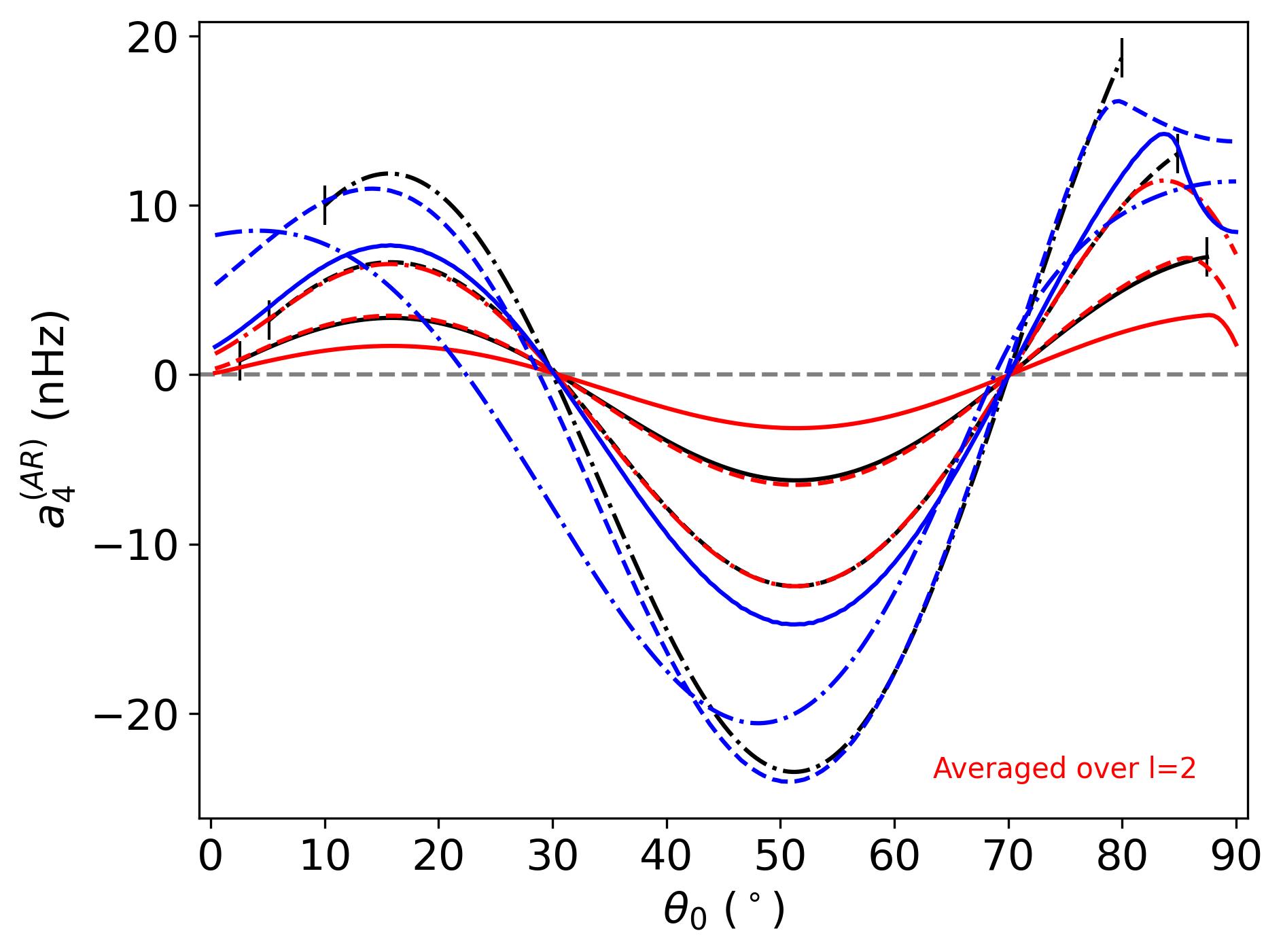}
\includegraphics[angle=0, scale=0.38]{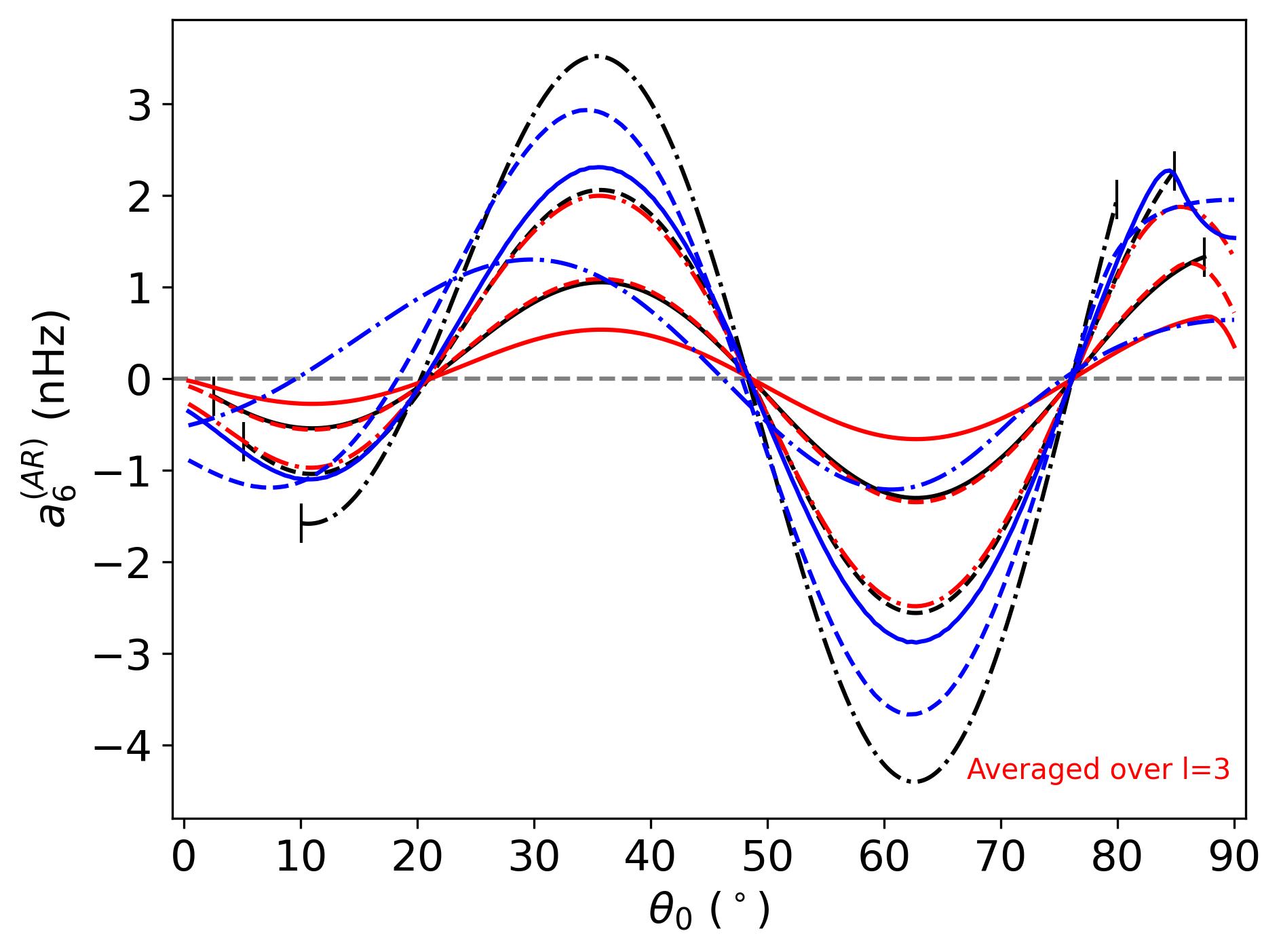}
\caption{Average $a^{(AR)}_j$ (no centrifugal effect accounted for) with $\epsilon_{nl}=5.10^{-4}$ in function of $\theta$ and $\delta$ {and for the gate ($\Pi$), triangle ($\Lambda$) and gaussian ($\mathcal{N}$) filter functions.} Uniqueness is guaranteed for $\theta$ only if at least two a-coefficients are measured. }
\label{fig:aj_mean}
\end{center}
\end{figure}

Figure \ref{fig:aj_mean} shows $a^{(AR)}_2$, $a^{(AR)}_4$, $a^{(AR)}_6$ {for activity described by $F = \Pi$ (black) or $\Lambda$ (red) or $\mathcal{N}$ (blue)}. These are the mean coefficients for the fictitious active star, as a function of $\theta_0$ and $\delta$. The figure indicates that there is a simple relationship between the a-coefficients, the co-latitude $\theta_0$ and the extension of the activity zone $\delta$, {independently of the shape of activity}. 
{In the case of $F= \Pi$, } the lines are cut near the pole and {the equator due to the condition $\theta_0 \ge \delta/2$ and $\theta_0 \leq \pi - \delta/2$}. {In the other profiles, edges effects (truncation) has noticeable effects near the equator. The definition of $\delta$ differs between the three profiles, which explains that a factor of a few in $a^{(AR)}_j$ is noticeable between $F= \Pi$, $\Lambda$ and $\mathcal{N}$ for a given $\delta$.} 
Note that adding the centrifugal effect reduces $a_2(n,l)=a_2^{(AR)}(n,l) + a_2^{(CF)}(n,l)$, because the coefficient $a_2^{(CF)}(n,l)$ is always negative. 

If an observation can constrain only a single a-coefficient (e.g. $a_2$), there often exists a degeneracy in $\theta_0$ as multiple value of $a_j$ can be obtained for a given $\theta_0$. As the figure shows, measuring two a-coefficients alleviates this degeneracy, provided that uncertainties are small enough. In other words, the accuracy on the inference of the active region using a-coefficients is ensured only if we can simultaneously constrain two a-coefficients (e.g. $a_2$ and $a_4$). This is essential to distinguish an activity near the pole ($\theta_0 \lessapprox 30^\circ$) from mid-latitude ($30^\circ \lessapprox \theta_0 \lessapprox 60^\circ$) or from near the equator ($\theta_0 \gtrapprox 60^\circ$).

It indicates that the loss of information resulting from the averaging does not affect the accuracy\footnote{It does however increases the uncertainty by a factor of a few.}.
Figure \ref{fig:aj_mean} also indicates that for a star with an activity intensity and an activity zone of extension commensurate with the one of the Sun ($\epsilon_{nl}
\simeq 5.10^{-4}$, $\delta \simeq 10^\circ$ {when $F=\Pi$}), the uncertainty on $a^{(AR)}_2$ must be approximately $\lessapprox 25$ nHz, the one for $a^{(AR)}_4 \lessapprox 10$ nHz and $a^{(AR)}_6\lessapprox 1.5$ nHz to be able to detect significant departures of the coefficients. This is to be compared with the spectral resolution of 7-14nHz for 2-4 years of observation, typical of the longest Kepler observations.
In the Figure \ref{fig:aj_mean}, $a^{(AR)}_2$ is obtained by averaging $l=1,2$, $a^{(AR)}_4$ by averaging $l=2$, while $a^{(AR)}_6$ is only constrained by $l=3$ modes. Averaging $a^{(AR)}_2$ and $a^{(AR)}_4$ over $l=1,2,3$ increases the maximum range of the a-coefficients by a factor $\approx 1.5$ for $a^{(AR)}_2$ and $\approx 2$ for $a^{(AR)}_4$, respectively, without changing the overall shape of the function. In Sun-like stars, the $l=3$ modes have a HNR at least ten times lower than $l=0$ modes because the height ratio between $l=3$ and $l=0$ modes is around 0.08 for the Sun \citep{Toutain1993,Toutain1998}. These modes are therefore difficult to observe.
Due to all of the above, it is extremely challenging to measure $a_6$ with current existing data. In the following, we will thus focus on assessing the relibability domains of $a_2$ and $a_4$ only.

\section{Bias analysis on $a_1$ $a_2$ and $a_4$} \label{sec:bias}

In order to understand the reliability of the inference on low-degree a-coefficient, it is necessary to perform a bias analysis. 
This requires to fit an ensemble of emulated spectra that are representative of Sun-like stars and to compare the results with the true inputs. 

\begin{figure*}
\begin{center}
\includegraphics[angle=0, scale=0.37]{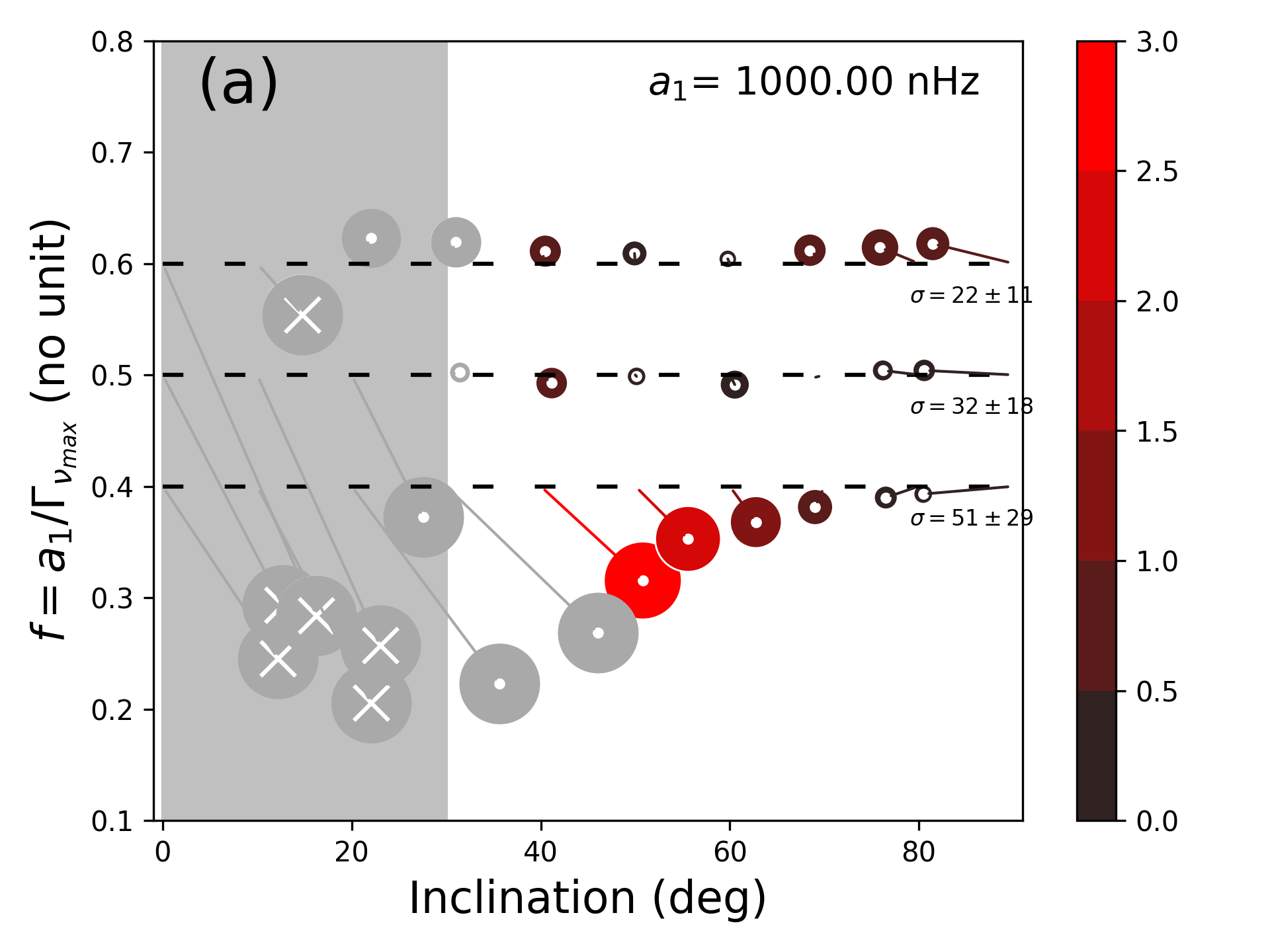}
\includegraphics[angle=0, scale=0.37]{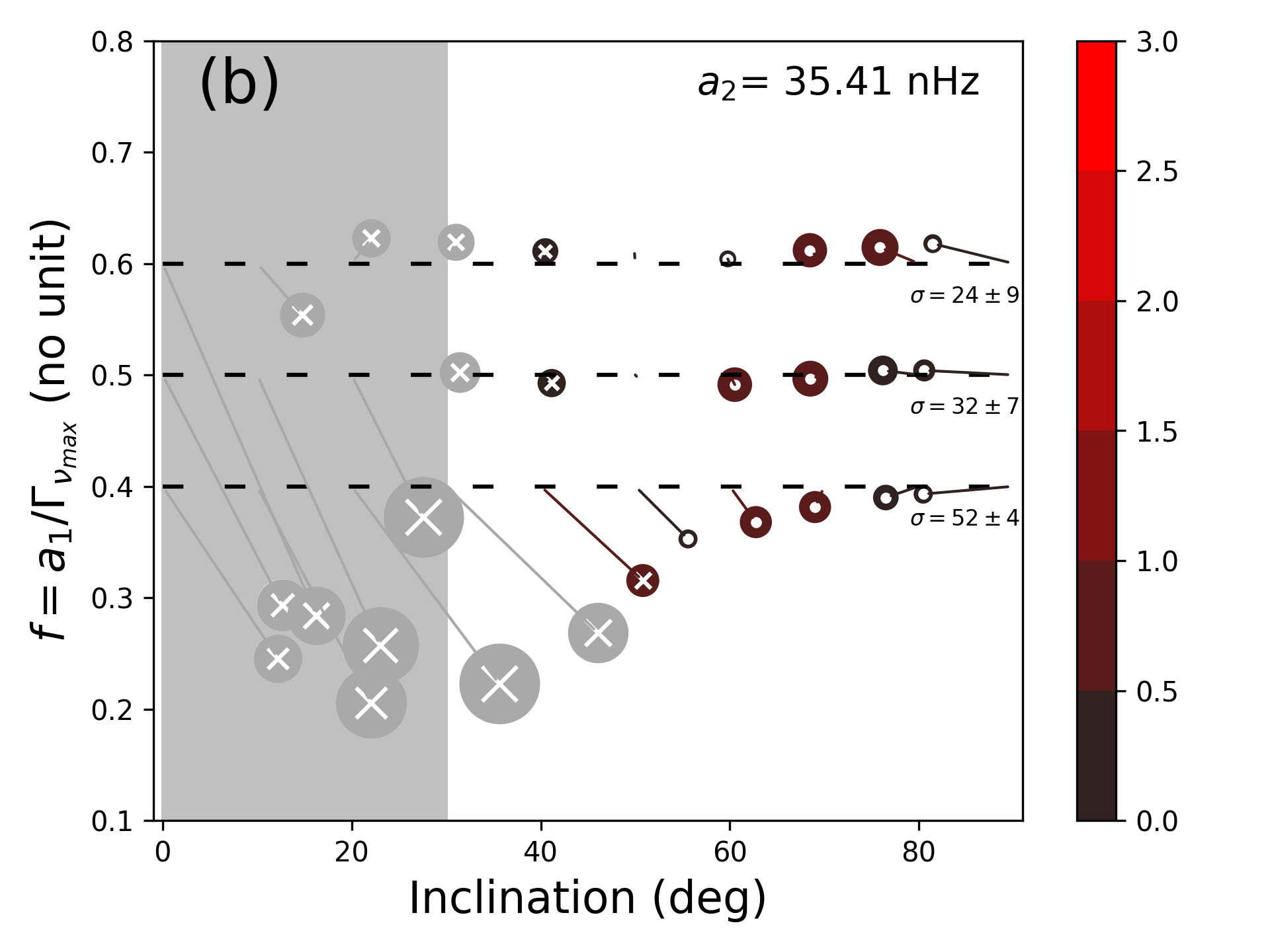} 
\includegraphics[angle=0, scale=0.37]{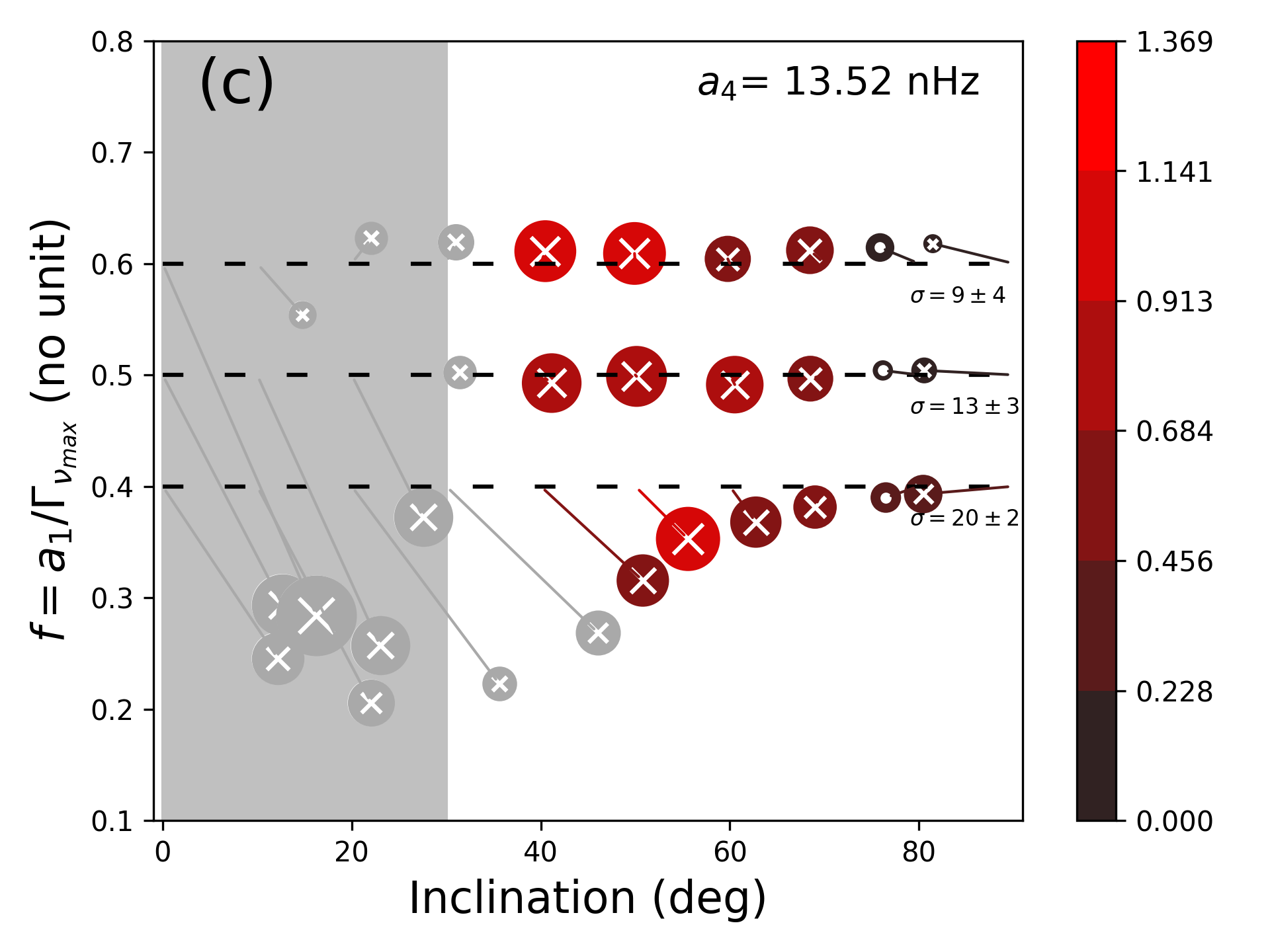} \\
\includegraphics[angle=0, scale=0.37]{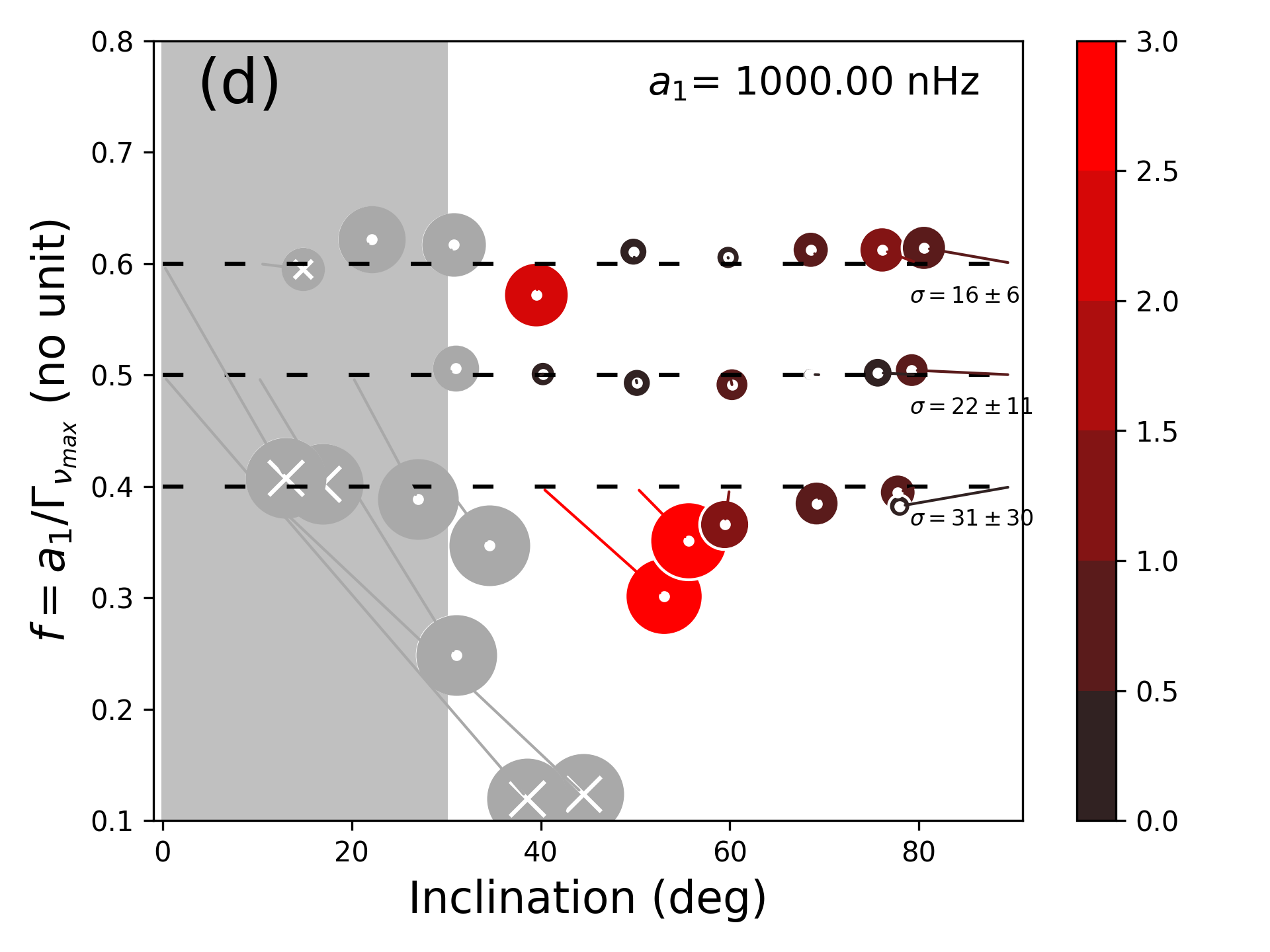}
\includegraphics[angle=0, scale=0.37]{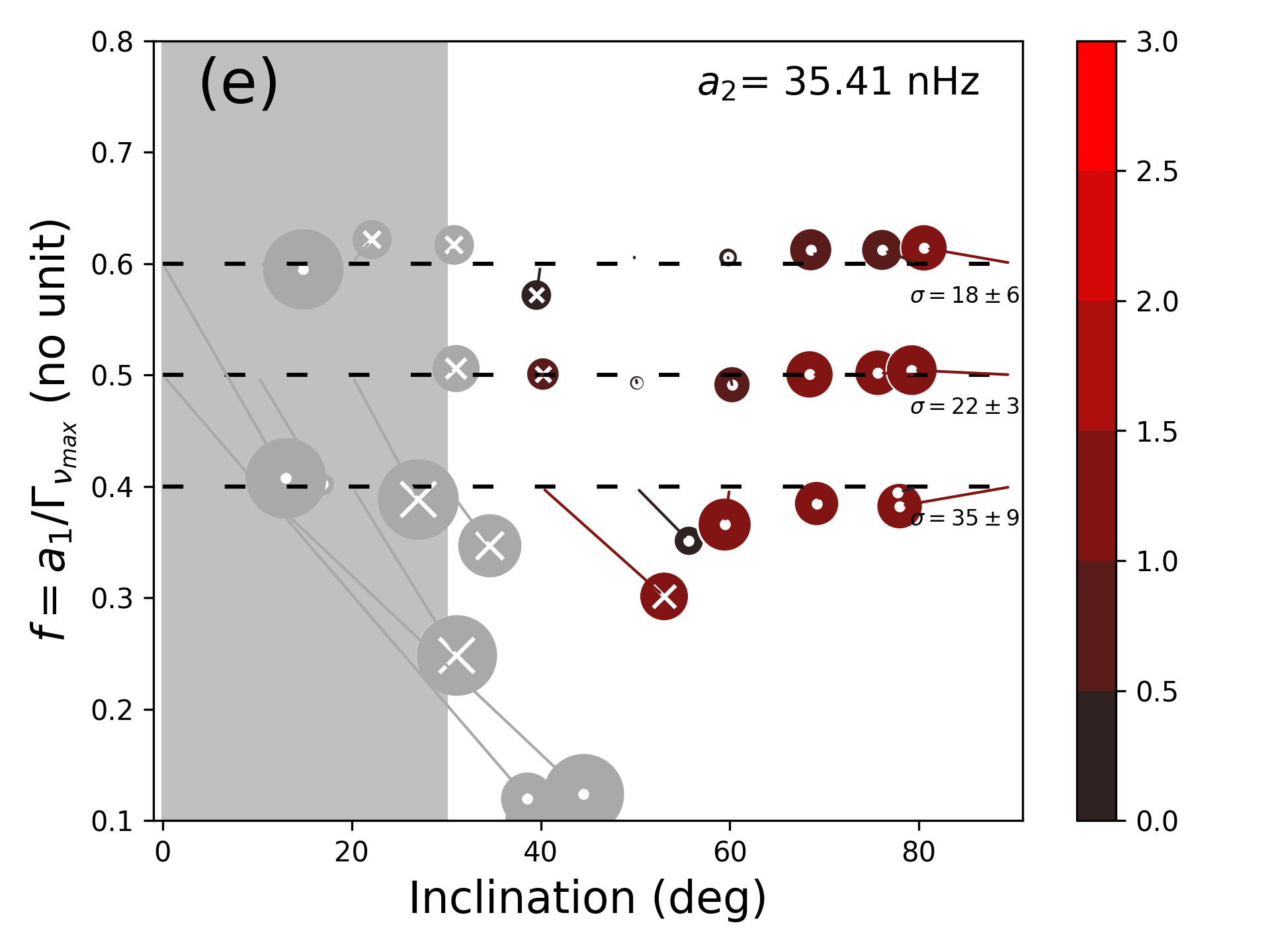}
\includegraphics[angle=0, scale=0.37]{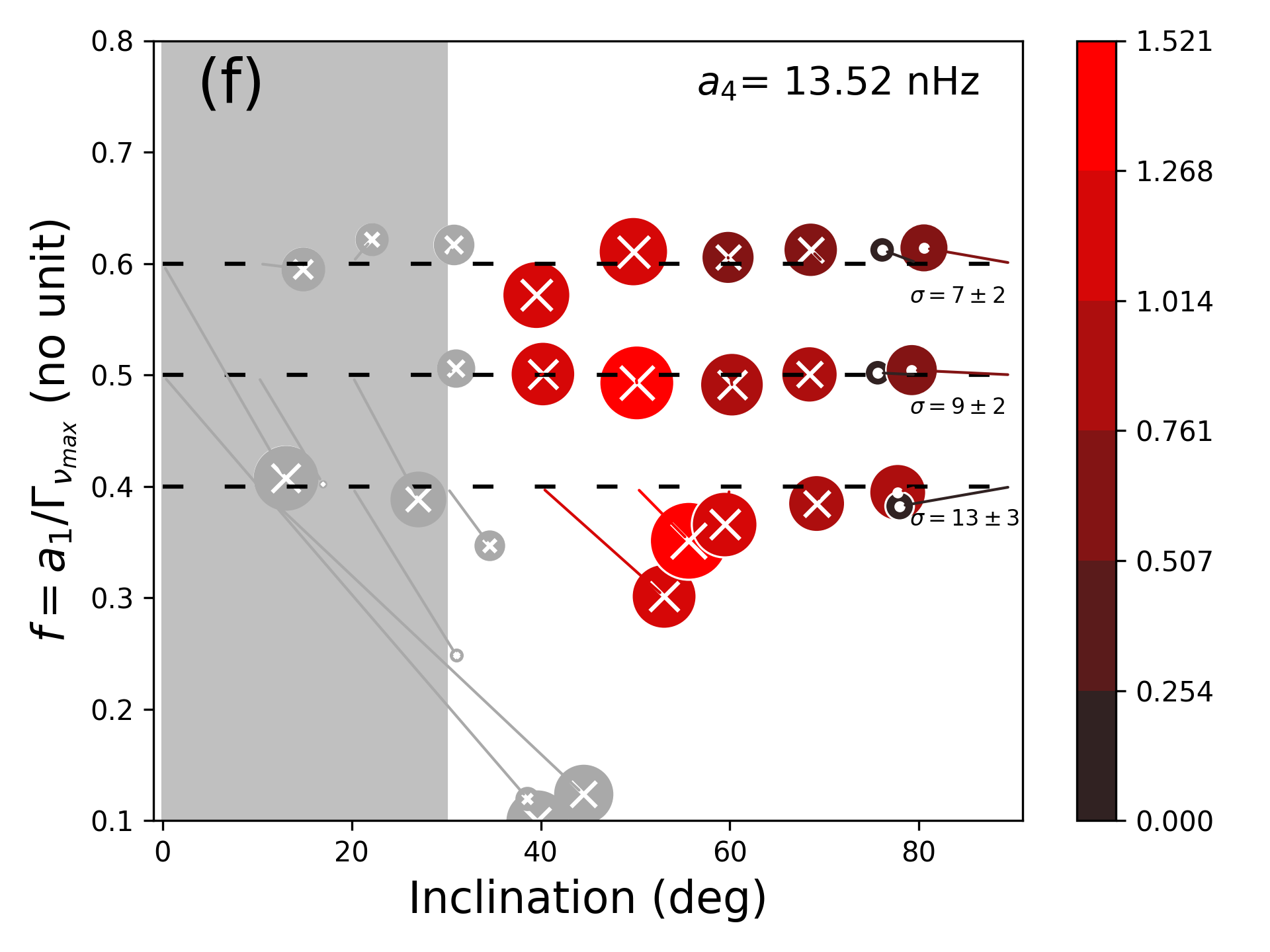}
\caption{Bias analysis for $\widehat{HNR}=30$ for an equatorial activity band ($\theta_0=85^\circ$,$\delta=10^\circ$) of similar intensity to the Sun ($\epsilon_{nl} = 5 \times 10^{-4}$), for $T_{obs} = 2$ years (top) and $T_{obs} = 4$ years (bottom). Colour and size of the circles indicate the modulus of the bias. The colour bar gives its scale normalised by the uncertainty, $b(a_j)/\sigma$. A white cross is for an underestimation. A white dot is for an overestimation. Below $30^\circ$ of inclination (gray area and symbols), the results are not reliable.}
\label{fig:bias:hnr30:Eq}
\end{center}
\end{figure*}

\begin{figure*}
\begin{center}
\includegraphics[angle=0, scale=0.37]{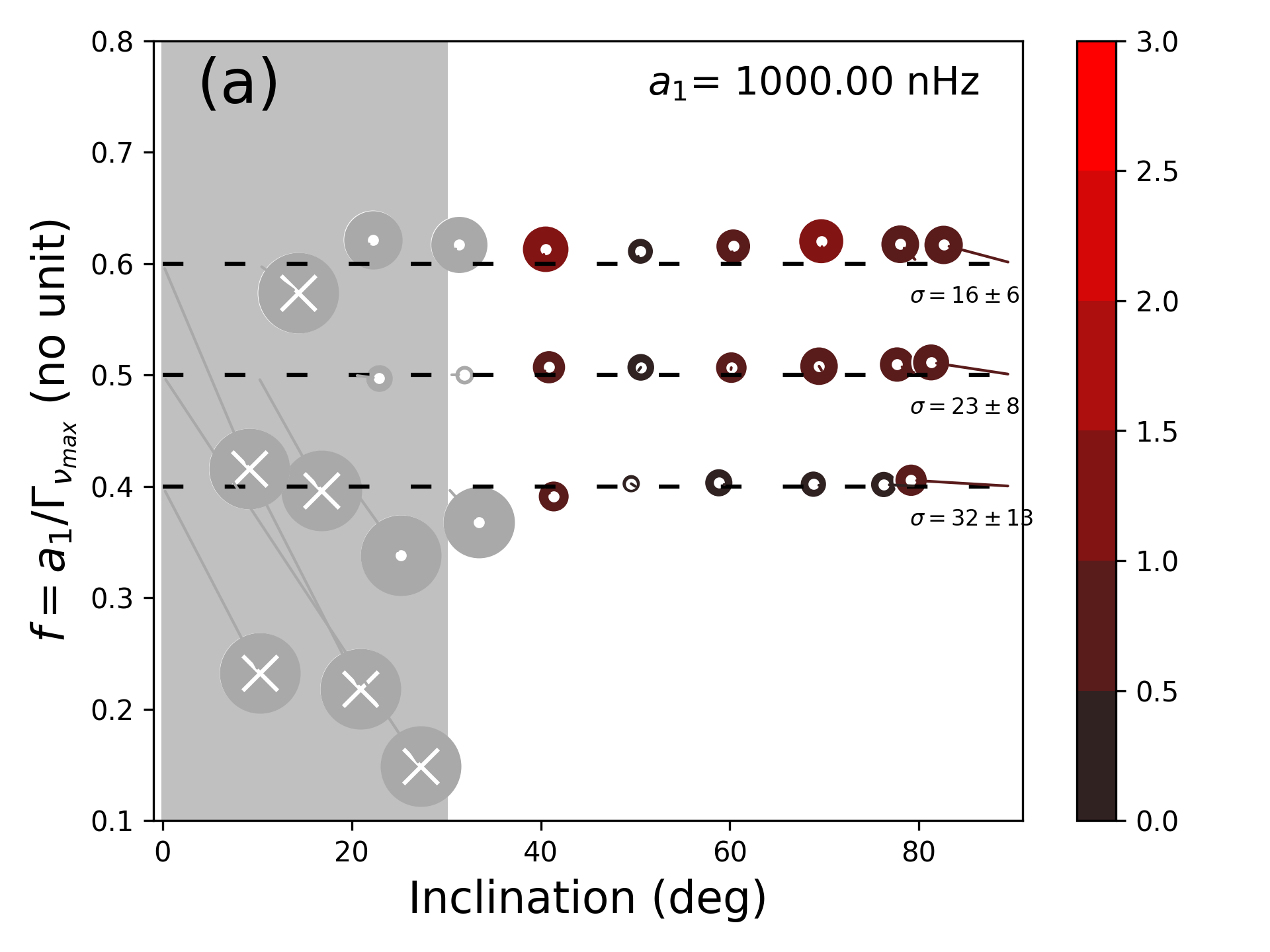}
\includegraphics[angle=0, scale=0.37]{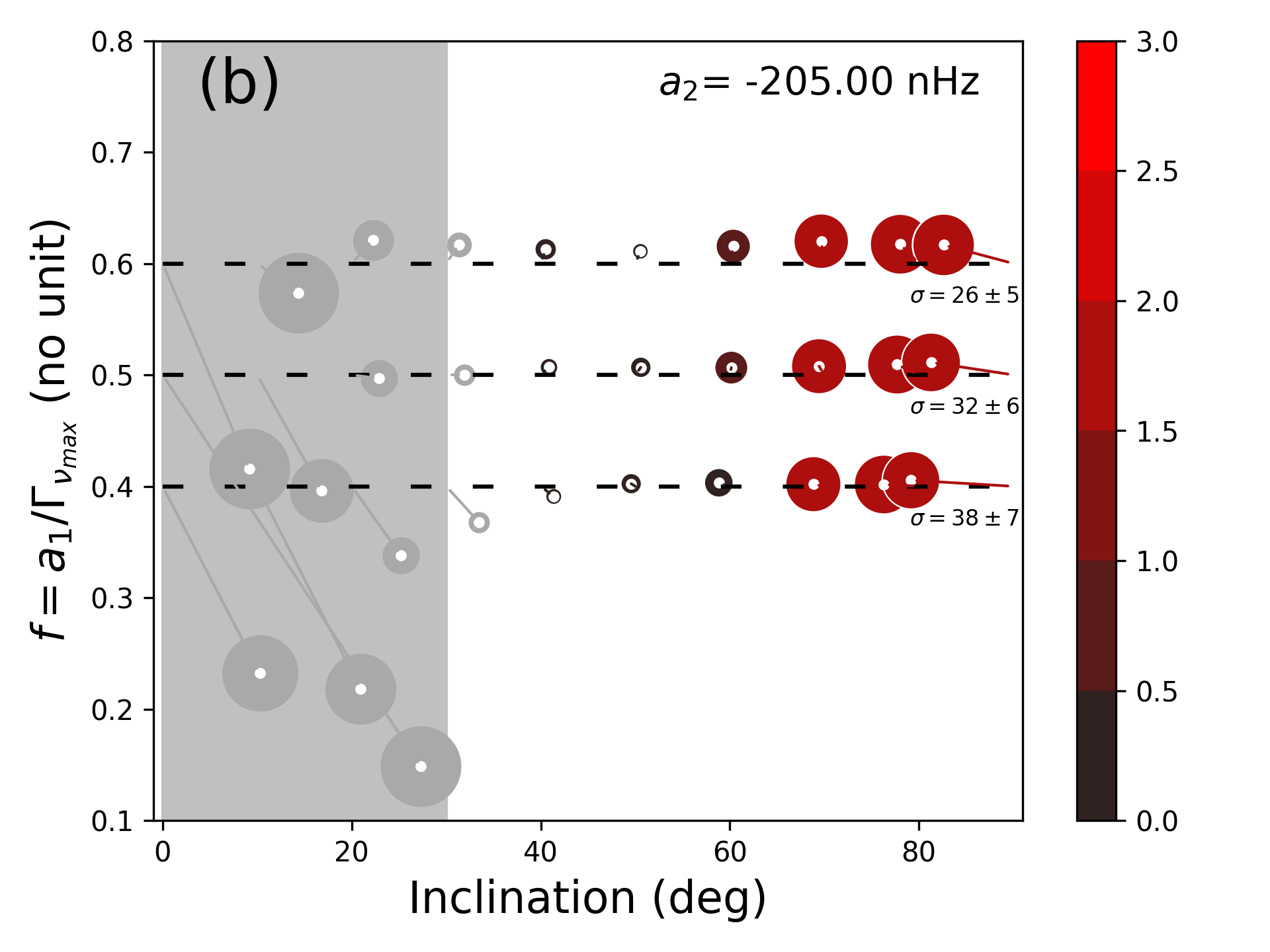}
\includegraphics[angle=0, scale=0.37]{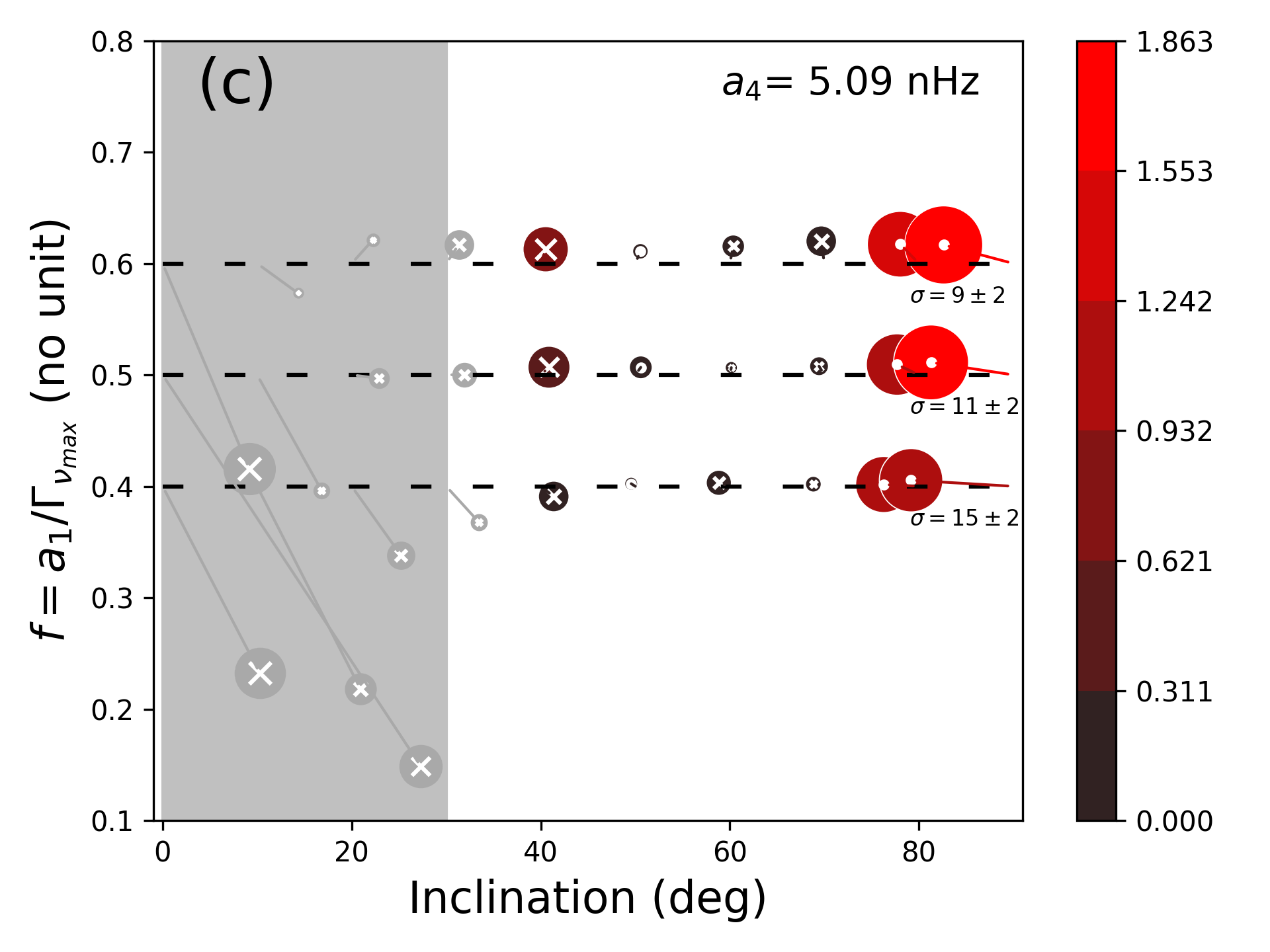} \\
\includegraphics[angle=0, scale=0.37]{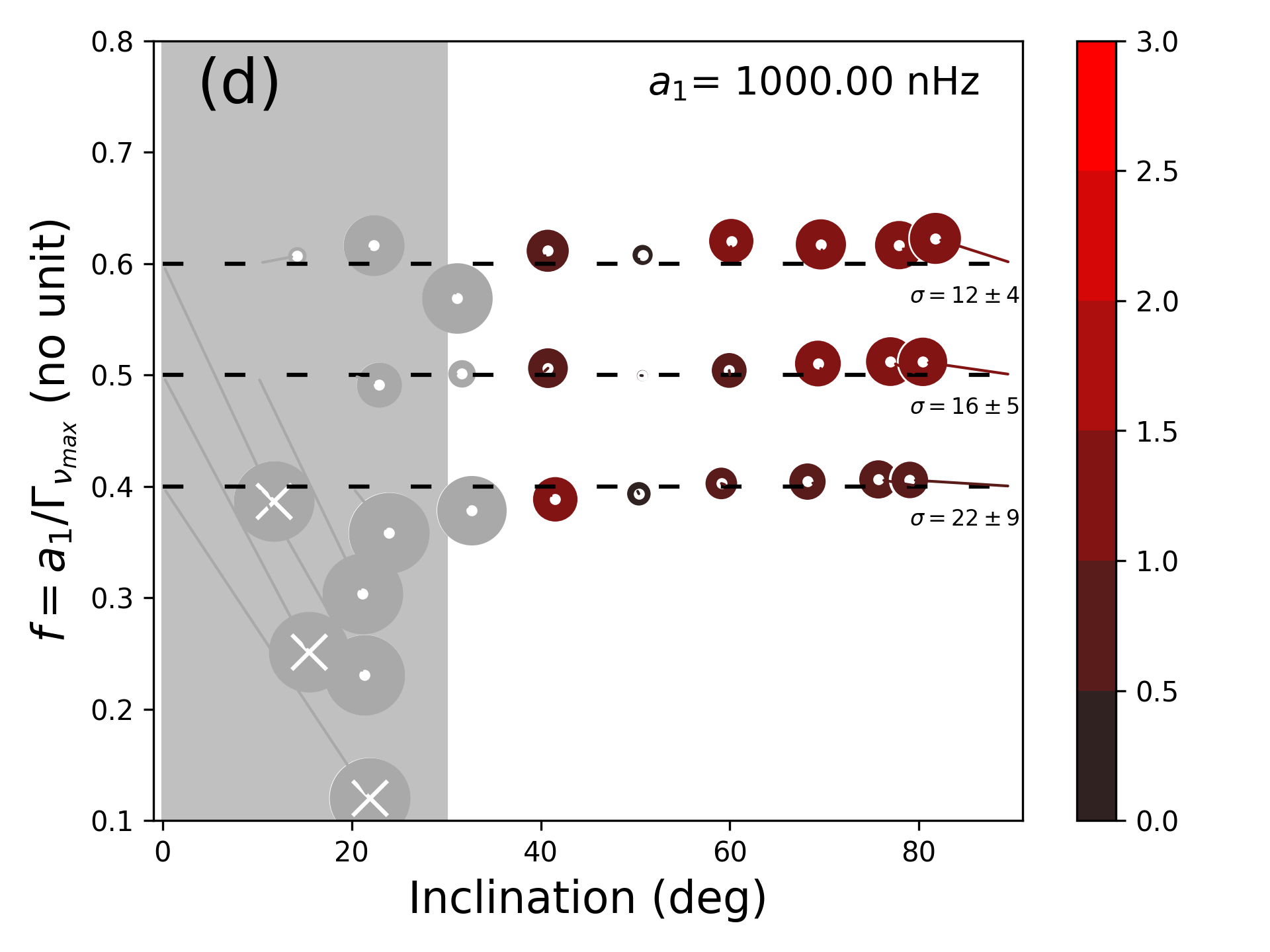}
\includegraphics[angle=0, scale=0.37]{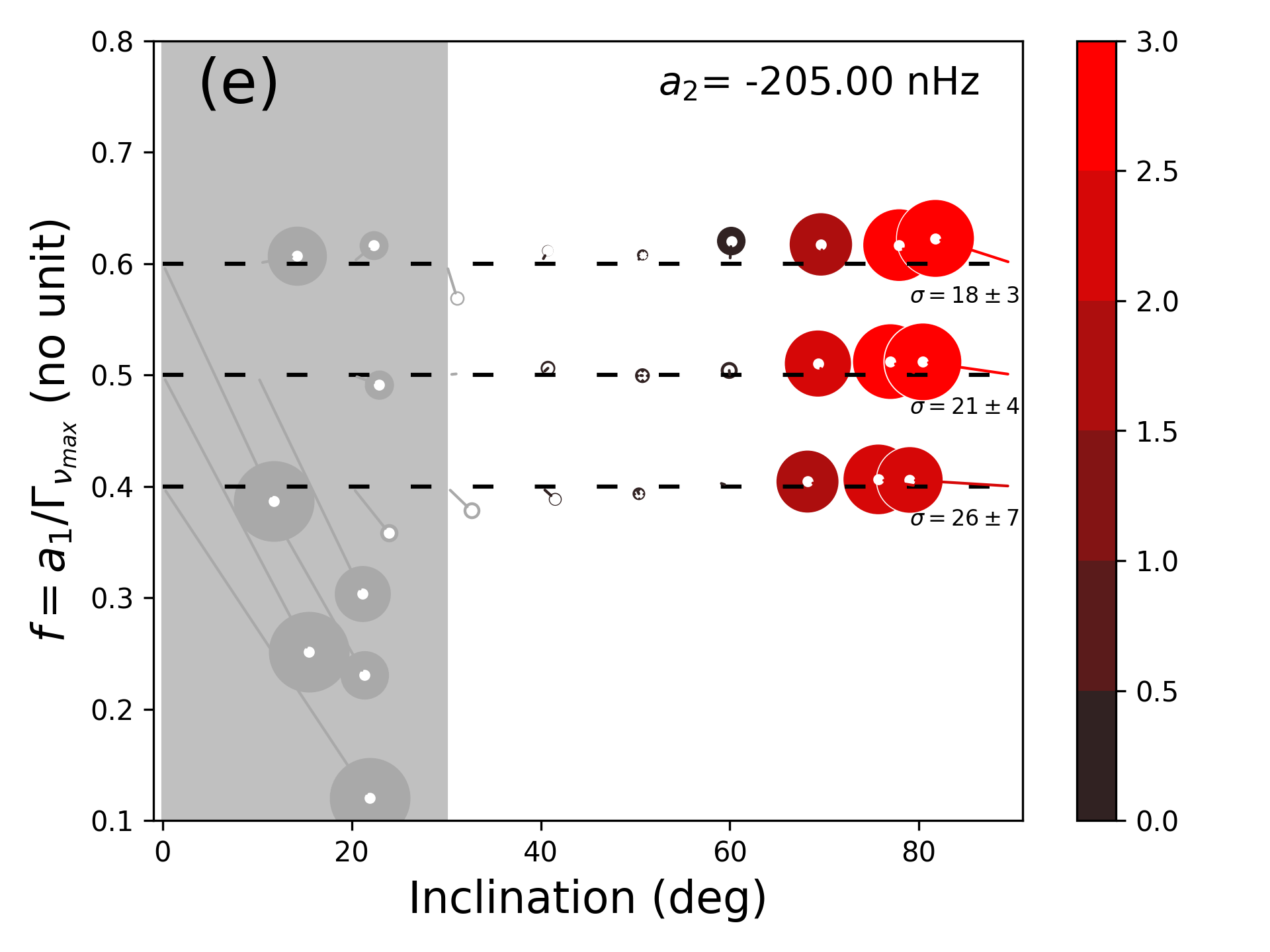}
\includegraphics[angle=0, scale=0.37]{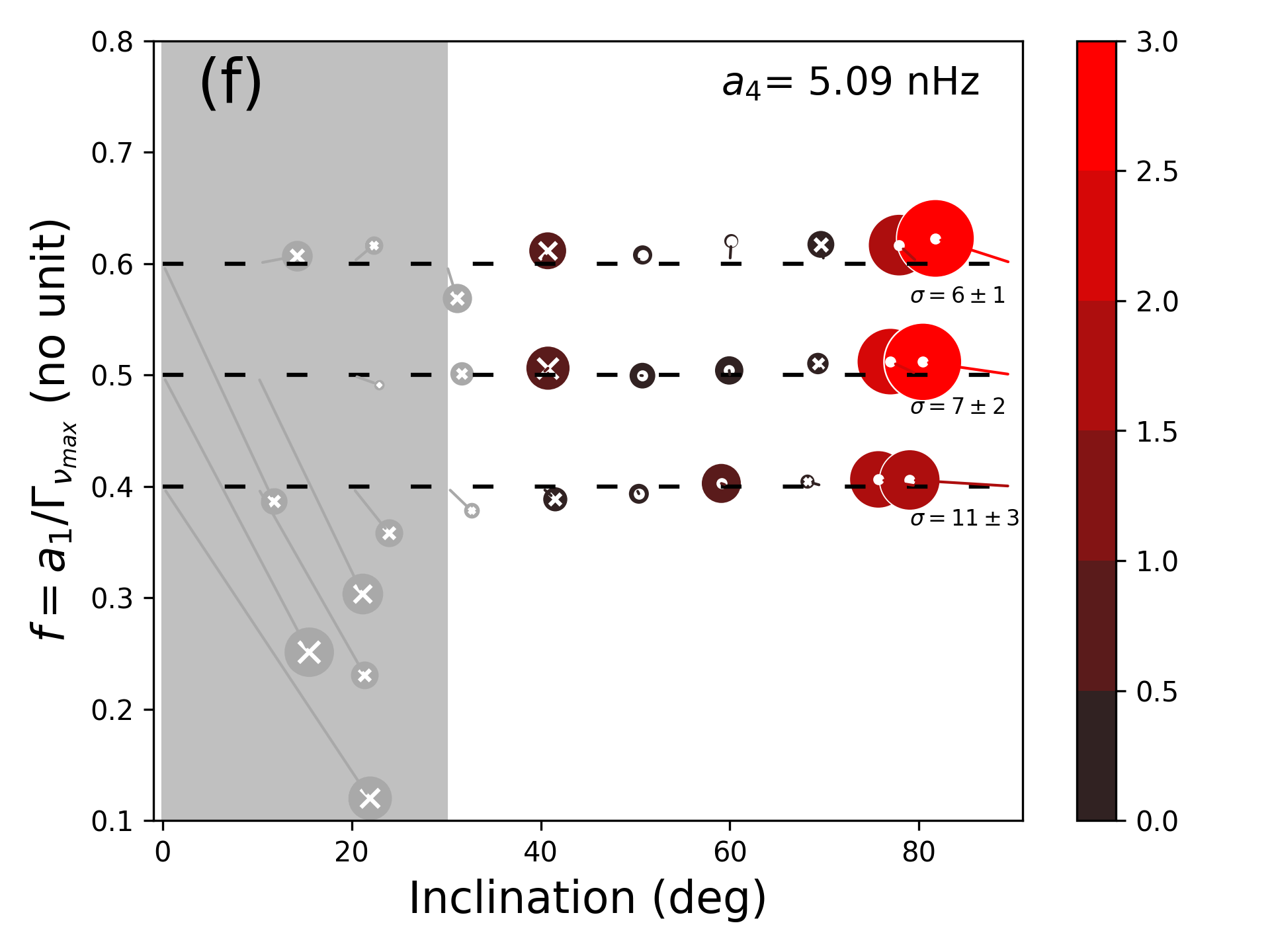}
\caption{Bias analysis for $\widehat{HNR}=30$ for a large polar activity cap ($\theta_0=22.5$,$\delta=45$) of similar intensity to the Sun ($\epsilon_{nl} = 5.10^{-4}$), for $T_{obs} = 2$ years (top) and $T_{obs} = 4$ years (bottom).}
\label{fig:bias:hnr30:Pol}
\end{center}
\end{figure*}

\subsection{Test setup}
The synthetic spectra use the frequencies, heights, widths and the noise background profile of 16 Cyg A as a template for the simulations. We want to specifically study the impact of the mode blending (effect of $a_1/\Gamma_{\nu_{max}}$), the stellar inclination ($i$), observation duration ($\mathrm{T_{obs}}$) and of the maximum Height-to-Noise background ($\widehat{HNR}$) on the accuracy of $a_1$, $a_2$ and $a_4$. Grids of spectra are built in the case of an equatorial band of activity and of a large polar activity. The grid parameters and their ranges are provided in Table \ref{tab:simu_vals} and discussed further below. 
The spectra are made using a spectrum simulator code\footnote{The tool suite used here along with the data inputs/outputs are available at  \url{https://github.com/OthmanB/Benomar2022/tree/version_2}.} that take the reference star (or template) and modify its properties to match the requirement of the user. 
The template is altered in terms of HNR following a similar approach as in \cite{Kamiaka2018}. The main difference is that we considered a frequency-dependent noise background. Heights are rescaled according to,
\begin{equation}
    H_{n,l=0} = \frac{\widehat{HNR}}{\widehat{HNR}_{\mathrm{ref}}} H_{\mathrm{ref}}(n,l=0) 
\end{equation}
where $\widehat{HNR}$ is the maximum HNR of the synthetic star, $\widehat{HNR}_{\mathrm{ref}}$ the maximum HNR of the reference star and $H_{\mathrm{ref}}(n,l=0)$ the $l=0$ heights of the reference star.

 \begin{figure}
 \begin{center}
    \includegraphics[angle=0, scale=0.50]{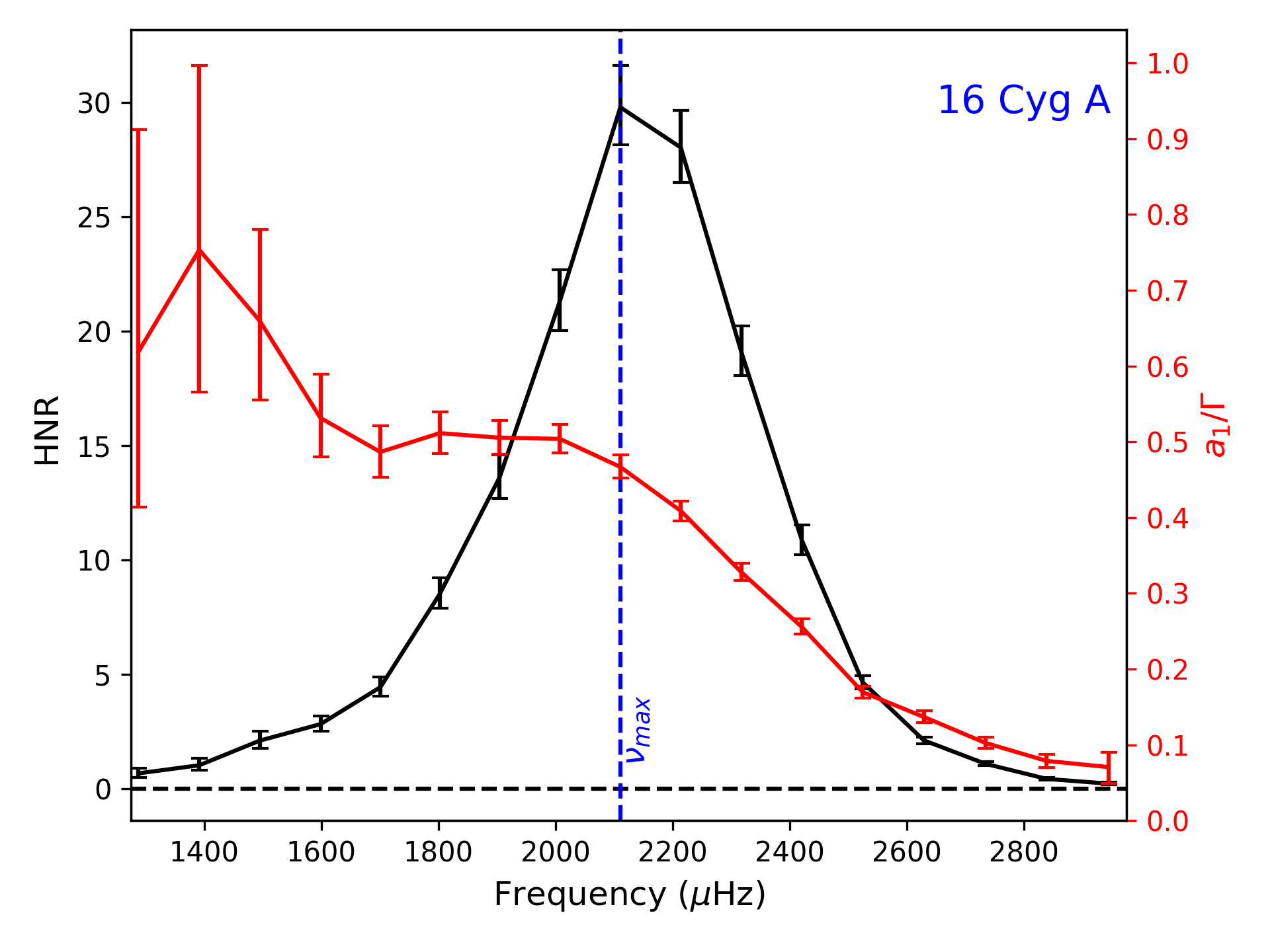}
\end{center}
\caption{HNR (black) and $a_1/\Gamma$ ratio (red) for 16 Cyg A, which is the reference star used to make simulations.} \label{fig:ref_star}
\end{figure}

As for the mode blending factor $f=a_1/\Gamma_{\nu_{max}}$, it is calculated fixing $a_1$ and altering $\Gamma_{\nu_{max}}$ such that,
\begin{equation}
    \Gamma_{\nu_{max}} = \frac{a_1}{f}.
\end{equation}
A fiducial value of $a_1=1000$ nHz ($\simeq 2.4$ the solar rotation) is used in the simulation. Note that this differs from \cite{Kamiaka2018} where the splitting was modified in order to obtain the desired mode blending factor.

As shown in Table \ref{tab:simu_vals}, three HNR cases are investigated, ranging from 10 to 30. Two observation durations are considered: 2 years and 4 years. This is representative of the best Kepler observations \citep{Davies2015} and of future observations from PLATO \citep{PLATO2014}. Similarly to the test cases of \cite{Gizon2002AN}, $a_2$ and $a_4$ are determined for an equatorial activity of extension of the same order as in the Solar case ($\delta\simeq10^\circ$), and in the case of a large polar cap ($\delta=45^\circ)$. Both situations assume an activity of the same order as for the Sun.
The priors are set in a similar manner as they would if the power spectrum was from a real star. The evaluation of the mode parameters is performed using a MCMC method, like for actual stars (see Appendix \ref{appendix:spectrum_analysis} for further details), but on the limit-spectrum (no noise). Fitting the limit spectrum allows to assess the systematic errors by calculating the expectation value of the probability density function. The expected uncertainty can also be known by computing the standard deviation.

\begin{table}
\begin{center} 
 \caption{Parameters set to construct artificial power spectra. $a_2$ and $a_4$ are determined assuming active regions located either in an equatorial band or in a polar cap.}
 \begin{tabular}{c|c|c}  
    Variable     &  \multicolumn{2}{c}{Values}     \\ \hline
    HNR        & \multicolumn{2}{c}{10, 20, 30}    \\ 
    $\mathrm{T_{obs}}$ (years) &   \multicolumn{2}{c}{2, 4}  \\
    $a_1/\Gamma_{\nu_{max}}$ & \multicolumn{2}{c}{0.4, 0.5, 0.6} \\ 
    $\epsilon_{nl}$     & \multicolumn{2}{c}{$5.10^{-4}$}   \\
    $a_1$ (nHz)      & \multicolumn{2}{c}{1000 }         \\ \hline
    Act. Region            &  $\theta_0=85^\circ$, $\delta=10^\circ$ (Eq.)  & $\theta_0=22.5^\circ$, $\delta=45^\circ$ (Pol.)\\ \hline
    $a_2$ (nHz)      & 35.41                    &    -205.06  \\
    $a_4$ (nHz)      & 13.52                    &    5.09  \\ \hline
 \end{tabular}
\end{center} \label{tab:simu_vals}
\end{table}

\begin{table*} 
\begin{center}
 \caption{Summary of the main parameters for the Sun and 16 Cyg A and B used to infer the activity. fiducial biases b($a_2$) and b($a_4$) for 16 Cyg A and B are set after inspection of Figure \ref{fig:bias:hnr30:Eq}. The lowest (highest) solution of $a_4$ for 16 Cyg B is has a probability of $30\%$ ($70\%$). }  \label{tab:real_vals}
 \begin{tabular}{c| c|c|c|c}
                            & Active Sun  & Quiet Sun   & 16 Cyg A  & 16 Cyg B \\ \hline
  $\widehat{HNR}$           & 100                     &     70                &  30             &  29     \\ 
  $\Delta\nu$ ($\mu$Hz)     & 135.1                   &     135.1              &  103.35         & 116.92 \\ 
  $a_1$  (nHz)              & $421\pm 10$             &     $410\pm9$          &  $614\pm37$     & $607\pm78$  \\ 
  $a_1/\Gamma$              &  $0.45\pm0.03$          &     $0.40\pm0.02$      & $0.45\pm0.04$       &  $0.58\pm 0.1$\\
  inclination ($^\circ$)    & $90$ (fixed)               &     $90$ (fixed)          & $45\pm4$        & $35\pm3$\\
  $a_2$ (nHz)               & $80\pm19$               &     $11\pm21$          &  $19.6\pm8.6$   & $18.5\pm23.5$ \\
  $a_4$ (nHz)               & $5.0\pm10.5$            &     $2.1\pm9.8$        &  $2.9\pm8.9$    & $-27.9\pm6.5$  or $-1.0\pm6.7$ \\
  $a^\mathrm{(CF)}_2$ (nHz) &   $-6.5\pm0.3$          &     $-6.1\pm0.3$       &  $-17.0\pm2.2$   & $-15.5\pm4.0$           \\
  $a^\mathrm{(AR)}_2$ (nHz) &  $86.2\pm18.7$          &     $16.7\pm20.4$      &  $39.0\pm20.6$   & $34.0\pm21.8$       \\  
  b($a_2$) (nHz)         &      /                  &         /               &     -10         & -10   \\
  b($a_4$) (nHz)         &      /                  &         /               &     -10         & -10    \\ \hline
\end{tabular}
\end{center} 
\end{table*}

 \subsection{Results} \label{sec:4:2}
 
 To appreciate the level of inaccuracy achieved when measuring $a_j, j\in[1,2,4]$ coefficients, the bias is calculated for $a_j$, $f=a_1/\Gamma_{\nu_{max}}$ and $i$,
 \begin{eqnarray} \label{eq:bias_triptic}
     b(a_j) = \mathrm{E}[a^{\mathrm{meas}}_j] - a^{\mathrm{true}}_j, \nonumber \\
     b(F) = \mathrm{E}[F^{\mathrm{meas}}] - F^{\mathrm{true}},  \\
     b(i) = \mathrm{E}[i^{\mathrm{meas}}] - i^{\mathrm{true}}. \nonumber
 \end{eqnarray}  
 
 Here, the letter E refers to the expectation value (median) from the probability density function.
 
 Figure \ref{fig:bias:hnr30:Eq} shows the resulting bias maps for a-coefficients as a function of the blending factor $F$ and of the stellar inclination for a-coefficients corresponding to an equatorial activity band. Figure \ref{fig:bias:hnr30:Pol} is the same but for a large polar activity cap. The bias is represented as a projected three dimensional vector using the three quantities defined by equation (\ref{eq:bias_triptic}) and for $j\in[1,2,4]$. To evaluate the importance of the bias on $a_j$, the plot shows $b(a_j)/\langle\sigma_{a_j}(i,F)\rangle_{i}$, with $\langle\sigma_{a_j}(i,F)\rangle_{i}$ the average standard deviation of the probability density functions obtained by the MCMC sampling over $i \in [30^\circ, 90^\circ]$. The value of $\langle\sigma_{a_j}(i,F)\rangle_{i}$ (noted $\sigma$ for convenience hereafter) and its variance along the inclination axis is also shown on the plots. A value of $b(a_j)/\sigma$ greater than one indicates that the inaccuracy exceeds the typical expected uncertainty at $1\sigma$ and may lead to significantly biased results during the subsequent analyses of the a-coefficients. 
 A negative bias (underestimation) for $a_j$ is indicated by a cross within the circle while a positive bias (overestimation) by a dot. The darker the colour, the smaller the norm of the normalised bias. The size of the circle symbols is also proportional to the bias, but capped to 3, to avoid excessively large symbols when $i<30^\circ$ (see next paragraph for explanations). 
 The case of observations of duration of 2 years are in the top figures (a, b, c) while the 4 years observation case is shown in bottom figures (d, e, f). 
 
 We first note that the biases on inclination and on $F$ are consistent with results from \cite{Kamiaka2018}. Regions of stellar inclination below $30^\circ$ are found to provide very unreliable results and the median of the inclination is biased toward $\simeq 80^\circ$ when the true input is $i=90^\circ$. 
 Meanwhile, biases in $a_1$ are found to often exceed $50\%$ in the gray area\footnote{This is likely due to the fit mis-identifying the $l=2, m=\pm1$ with $l=2, m=\pm2$, effectively doubling or halving the $a_1$ coefficient.}, indicating that for inclination below $\simeq 30^\circ$, the median estimator of the probability density function might not be trusted. This justifies also the range for the calculation of $\sigma$. In the following we therefore focus the discussion of the figures for the region above $30^\circ$. 
 
  In the case of an equatorial zone of activity, we note that measurements of $a_1$ may have negligible inaccuracies when the blending factor $F$ is above $0.4$, but starts to be significantly deteriorated at (and probably below) $f=0.4$ because $\vert b(a_1)\vert /\sigma$ exceeds 1.5 when $i<50^\circ$. Split components are then overlapping significantly leading to important degeneracies between a-coefficients, mode widths and the stellar inclination. 
  Meanwhile, we note that the biases on $a_2$ and $a_4$ remain mild compared to the uncertainty ($\vert b(a_2)\vert/\sigma < 1$ and $\vert b(a_4)\vert /\sigma < 1$), even for $f=0.4$. 
  
  In the case of a polar active region, $\vert a_2 \vert$ represents $\simeq 20\%$ of the $a_1$ coefficient. In these conditions, we note that $b(a_1)/\sigma$ does not exceeds the unity (see (a) and (d)), indicating a good accuracy. The terms $\vert b(a_2)\vert /\sigma$ and  $\vert b(a_4)\vert /\sigma$  are also lower than 0.5, provided that the stellar inclination is between $30^\circ$ and $70^\circ$.
  This indicates that a large $a_2$ coefficient is generally associated with  a higher accuracy for all a-coefficients. The plot also demonstrates that a large zone in the parameter space has moderate to small bias. 
  
  Interestingly, figures for $\widehat{HNR}$ of 10 and 20 (Figures \ref{fig:bias:hnr10:Eq}-\ref{fig:bias:hnr20:Pol} in Appendix \ref{appendix:extra_bias_maps}) lead to similar conclusions. Therefore, even in less favourable $\widehat{HNR}$ conditions, the expected uncertainty on the measurement generally encompass the bias provided that the stellar inclination exceeds $\simeq 30^\circ$ (ensuring that $m \neq 0$ have significant amplitudes) and that the mode blending factor is above $0.4$. This indicates that for a majority of stars observed by Kepler, the accuracy may not be a major issue when measuring $a_1$, $a_2$ and $a_4$ and provided that a careful assessment of F and $i$ is performed.
  Note however that the average uncertainty $\sigma$ does usually increase when the $\widehat{HNR}$ and/or the observation duration decrease, which will have an impact on the precision of the determination of the active region.
  
 \begin{table*} 
\begin{center}
 \caption{Statistical summary for the activity parameters and their model log-marginal {likelihood $\mathrm{ln(P(\boldsymbol{O} \vert M_{AR})}$ when $F(\theta|\mathbf{x}) = \Pi(\theta_0, \delta)$, $\Lambda(\theta_0, \delta)$ or $\mathcal{N}(\theta_0, \delta)$. $\mathrm{ln(P(\boldsymbol{O} \vert M_{CF})}$ is the log-marginal likelihood for a pure centrifugal effect.} The Activity significance is the probability that the activity is necessary to explain the data and derived from the log-marginal likelihoods. Uncertainty on the activity significance is less than $0.25\%$.} \label{tab:real_vals:inference}
 \begin{tabular}{cc||c||c||c|c||c|c|c|c} 
            &                & \multirow{3}{*}{Active Sun}         & \multirow{3}{*}{Quiet Sun}    & \multicolumn{2}{c||}{16 Cyg A}  & \multicolumn{4}{c}{16 Cyg B} \\ 
            &                &                         &                  & \multirow{2}{*}{No bias}  & \multirow{2}{*}{bias corr.} & \multicolumn{2}{c|}{No bias} & \multicolumn{2}{c}{bias corr.} \\ 
             &               &.                        &                  &          &            & $a_4\simeq -27.9$ & $a_4\simeq-1.0$ & $a_4\simeq -27.9$ & $a_4\simeq-1.0$ \\ \hline
  &   $\ln(P(\boldsymbol{O} \vert \mathrm{M}_{CF}$)  &   $-10.809$        &  $-0.367$  &   $-1.856$   & $-3.889$   & $-10.596$       & $-1.228$ & $-5.910$  & $-2.951$      \\  \hline\hline
\multirow{6}{*}{\rotatebox[origin=c]{90}{\parbox[c]{1cm}{\centering $\Pi(\theta_0, \delta)$}}}
  &   $\epsilon_{nl}$ ($10^{-4}$) & $7.6^{+25.5}_{-5.3}$   &  $3.7^{+19.6}_{-3.1}$   &   $5.3^{+19.6}_{-4.1}$   & $6.4^{+24.8}_{-4.8}$  &  $12.5^{+31.9}_{-7.5}$   & $4.6^{+20.8}_{-3.7}$  & $11.5^{+39.4}_{-7.9}$ & $5.1^{+23.8}_{-3.9}$  \\
  &  $\theta_0$ (deg)          &  $76^{+8}_{-7}$    &  $60^{+19}_{-41}$ &   $71^{+13}_{-14}$ & $79^{+8}_{-10}$ &  $58^{+3}_{-3}$    & $67^{+14}_{-25}$ &  $61^{+3}_{-4}$ & $78^{+8}_{-12}$ \\
  &   $\delta$ (deg)            &    $7^{+16}_{-5}$  &  $2^{+11}_{-2}$   &   $4^{+16}_{-3}$   & $5^{+15}_{-4}$  &  $8^{+17}_{-6}$    & $3^{+13}_{-2}$ & $6^{+14.8}_{-4.6}$ & $4^{+15}_{-3}$        \\ \cmidrule(rl){3-10} 
  &   $\ln(P(\boldsymbol{O} \vert \mathrm{M}_{AR}$)  &   $-0.685$        &  $-0.347$  &   $-0.482$   & $-0.656$   & $-0.878$        & $-0.449$ &  $-0.781$ & $-0.611$      \\ \cmidrule(rl){3-10} 
  &  Significance $(\%)$  &   $>99.99$      &  $50.5$    &   $79.8$     & $96.2$     & $>99.9$          & $68.6$   & $99.3$    &  $91.2$ \\ \hline\hline
  \multirow{6}{*}{\rotatebox[origin=c]{90}{\parbox[c]{1cm}{\centering $\Lambda(\theta_0, \delta)$}}}
  &   $\epsilon_{nl}$ ($10^{-4}$) & $14.2^{+33.0}_{-9.8}$   &  $17^{+42.9}_{-14.2}$   &   $15.4^{+39.5}_{-12.2}$   & $13.6^{+34.1}_{-4.8}$  &  $21.1^{+35.2}_{-12.6}$   & $15.1^{+38.9}_{-12.4}$  & $17.3^{+34.0}_{-11.4}$ & $13.3^{+36.5}_{-10.3}$  \\
  &  $\theta_0$ (deg)          &  $76^{+9}_{-7}$    &  $49^{+27}_{-36}$ &   $66^{+16}_{-41}$ & $78^{+9}_{-21}$ &  $58^{+2}_{-3}$    & $58^{+21}_{-40}$ &  $61^{+3}_{-4}$ & $75^{+11}_{-43}$ \\
  &   $\delta$ (deg)            &    $7^{+16}_{-5}$  &  $6^{+19}_{-5}$   &   $6^{+17}_{-4}$   & $5^{+15}_{-4}$  &  $10^{+16}_{-6}$    & $6^{+17}_{-5}$ & $8^{+16}_{-5}$ & $5^{+16}_{-4}$        \\ \cmidrule(rl){3-10} 
  &   $\ln(P(\boldsymbol{O} \vert \mathrm{M}_{AR}$)  &   $-0.698$         &  $-0.296$  &   $-0.492$   & $-0.699$   & $-0.930$        & $-0.427$ &  $-0.847$ & $-0.639$      \\    \cmidrule(rl){3-10} 
  &  Significance $(\%)$  &   $>99.99$      &  $51.8$    &   $79.6$     & $96.0$     & $>99.99$          & $69.0$   & $99.4$    &  $91.0$ \\ \hline\hline
  \multirow{6}{*}{\rotatebox[origin=c]{90}{\parbox[c]{1cm}{\centering $\mathcal{N}(\theta_0, \delta)$}}}
  &   $\epsilon_{nl}$ ($10^{-4}$) & $3.7^{+16.0}_{-2.2}$   &  $15.1^{+42.8}_{-13.5}$   &   $12.2^{+39.5}_{-10.5}$   & $6.6^{+24.3}_{-5.0}$  &  $13.3^{+24.1}_{-7.0}$   & $12.6^{+41.0}_{-11.0}$  & $6.6^{+24.3}_{-5}$ & $8.9^{+35.4}_{-7.5}$  \\
  &  $\theta_0$ (deg)          &  $73^{+11}_{-10}$    &  $48^{+28}_{-34}$ &   $63^{+18}_{-38}$ & $75^{+11}_{-27}$ &  $56.6^{+3}_{-7}$    & $55^{+23}_{-37}$ &  $75^{+11}_{-27}$ & $72^{+14}_{-42}$ \\
  &   $\delta$ (deg)            &    $9^{+26.6}_{-8}$  &  $7^{+20}_{-6}$   &   $6^{+22}_{-5}$   & $4^{+20}_{-3}$  &  $6^{+22}_{-5}$    & $7^{+21}_{-6}$ & $4^{+20}_{-3}$ & $4^{+19}_{-3}$        \\ \cmidrule(rl){3-10} 
  &   $\ln(P(\boldsymbol{O} \vert \mathrm{M}_{AR}$)  &   $-0.708$         &  $-0.384$  &   $-0.584$   & $-0.811$   & $-0.943$        & $-0.516$ &  $-0.811$ & $-0.759$      \\   \cmidrule(rl){3-10} 
  &  Significance $(\%)$  &   $>99.99$      &  $49.6$    &   $78.1$     & $95.6$     & $>99.99$          & $67.1$   & $99.4$    &  $90.0$ \\ \hline\hline
  \multirow{6}{*}{\rotatebox[origin=c]{90}{\parbox[c]{1cm}{\centering Average}}}
  &   $\epsilon_{nl}$ ($10^{-4}$) & $7.7^{+25.7}_{-5.6}$   &  $11.4^{+41.6}_{-10.0}$   &   $9.2^{+34.2}_{-7.5}$   & $8.0^{+29.4}_{-6}$  &  $15.5^{+30.7}_{-9.1}$   & $10.5^{+37.2}_{-9.0}$  & $11.8^{+31.5}_{-7.5}$ & $9.1^{+34.7}_{-7.4}$  \\
  &  $\theta_0$ (deg)          &  $75^{+10}_{-8}$    &  $53^{+25}_{-37}$ &   $68^{+14}_{-36}$ & $78^{+9}_{-14}$ &  $58^{+3}_{-4}$    & $61^{+18}_{-34}$ &  $60^{+4}_{-6}$ & $75^{+11}_{-38}$ \\
  &   $\delta$ (deg)            &    $7^{+22}_{-6}$  &  $5^{+18}_{-4}$   &   $5^{+18}_{-4}$   & $5^{+17}_{-4}$  &  $8^{+18}_{-6}$    & $6^{+18}_{-5}$ & $7^{+19}_{-6}$ & $4^{+17}_{-3}$        \\ \cmidrule(rl){3-10} 
  &   $\ln(P(\boldsymbol{O} \vert \mathrm{M}_{AR}$)  &   $-0.697$         &  $-0.342$  &   $-0.519$   & $-0.722$   & $-0.917$        & $-0.464$ &  $-0.824$ & $-0.670$      \\    \cmidrule(rl){3-10} 
  &  Significance $(\%)$  &   $>99.99$      &  $50.6$    &   $79.2$     & $96.0$     & $>99.99$          & $68.2$   & $99.4$    &  $90.7$ \\\bottomrule
\end{tabular}
\end{center}  
\end{table*}

\section{Analysis of solar data} \label{sec:results:Sun}

The Sun has a well-known 11-year activity cycle that makes it ideal for testing further the accuracy of the method. The Sun activity cycle is analysed at two instants, highlighted in Figure \ref{fig:Butterfly} and for which high-quality helioseismic data are available: (a) the maximum of activity between January 1999 and January 2002; and (b) the minimum of activity between January 2006 and January 2009.  
Data are from the Variability of Solar Irradiance and Gravity Oscillations instrument aboard SOHO spacecraft \citep{Frohlich1997} and presents very few gaps. 
 
    \subsection{a-coefficients of the Sun} \label{sec:results:Sun:aj}
    
 As for the simulated data, the analysis consists in fitting the power spectrum for cases (a) and (b), to measure the $a_1$, $a_2$ and $a_4$ coefficients using the Bayesian modelling and the MCMC method described in the Appendix \ref{appendix:spectrum_analysis}. {Figures of the best fits and their discussions are provided in Appendix \ref{appendix:best_fits}}. The goal being to evaluate the accuracy of the inference of the activity zone from a-coefficients, the presented results are for a stellar inclination  fixed to $90^\circ$, instead of having it as a free parameter. This value correspond approximately to the stellar inclination as seen by the SOHO satellite. 
 This alleviates biases on $a_j$ coefficients that may arise due to the systematic underestimation of $i$ when it is close to 90 degrees. However, carried tests with a free inclination during the maximum of activity\footnote{The measured stellar inclination is $i=74\pm3$ degrees at the maximum of activity of 1999-2002.} lead to a difference of $-12$nHz in $a_2$ and to $-1.3$nHz in $a_4$. This is consistent with the expected difference from the bias map of Figure \ref{fig:bias:hnr30:Eq}. In agreement with the discussion of Section \ref{sec:bias}, the difference does not have a significant impact on the activity inference because the uncertainties for those parameters are larger than the observed measurement shift.
 
\begin{figure}
 \begin{center}
\includegraphics[angle=0, scale=0.32]{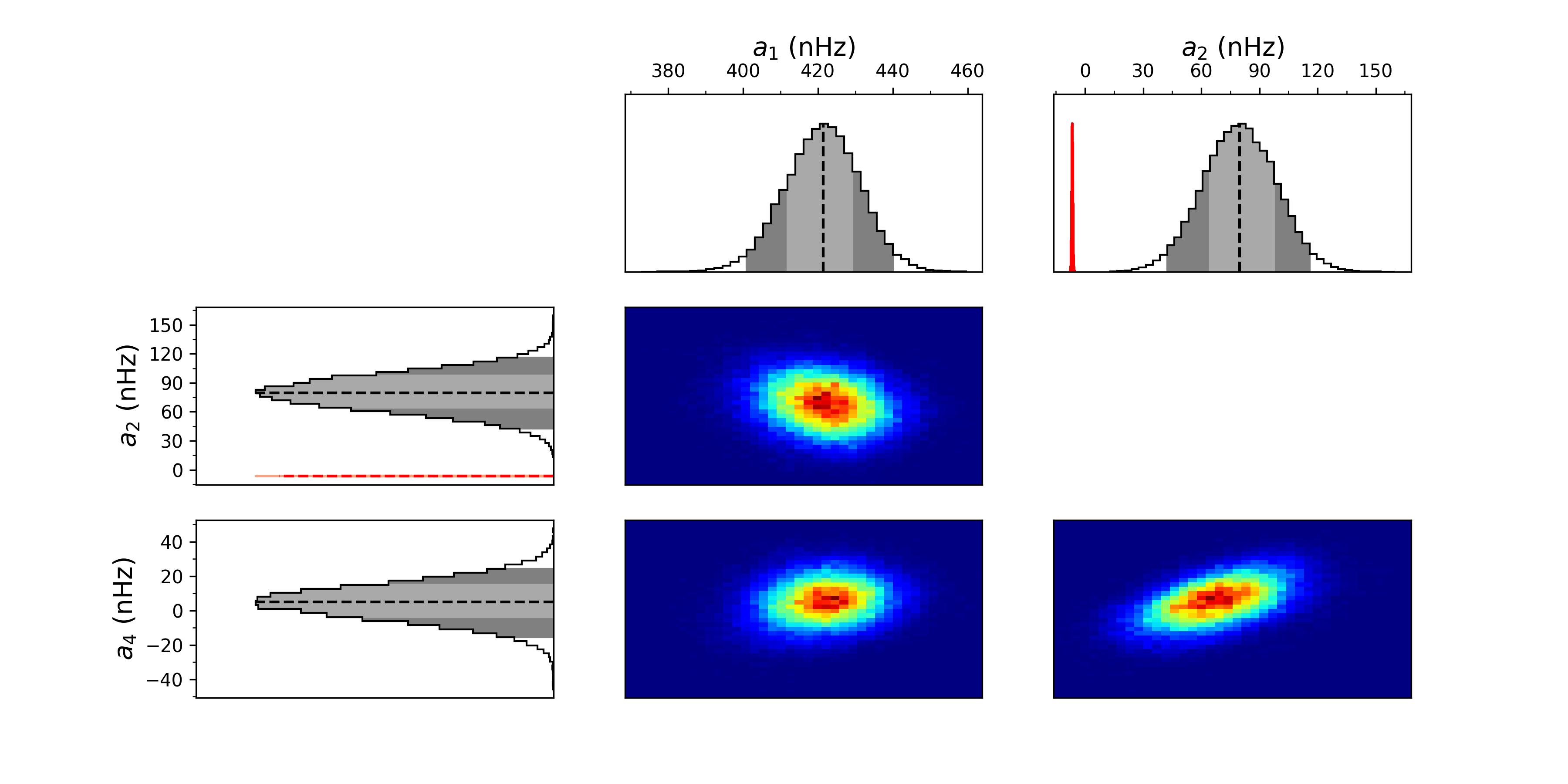} \\
\includegraphics[angle=0, scale=0.32]{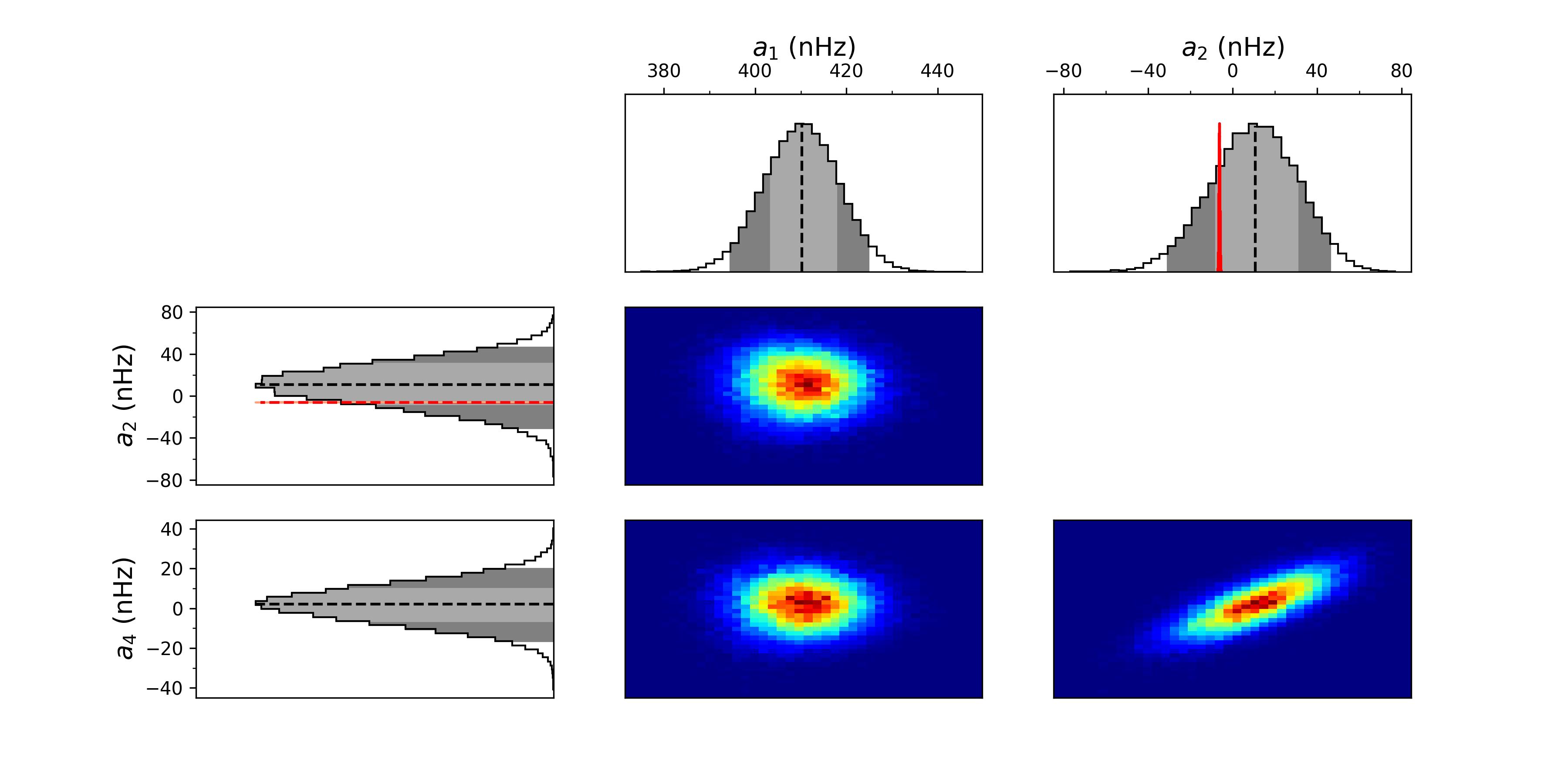}
\caption{Probability Density Functions and their correlations, obtained by MCMC for coefficients $a_1$, $a_2$, $a_3$, $a_4$ for the Sun at Maximum of activity (1999-2002, left) and at minimum of activity (2006-2009, right).
The red curve is the expected $a_2^{CF}$ coefficient for a pure centrifugal distortion. The light and dark gray PDF filling is for the $1\sigma$ and $2\sigma$ confidence interval, respectively.}
\label{fig:Sun:acoefs}
\end{center}
\end{figure}

 Figure \ref{fig:Sun:acoefs} shows the measured probability distribution function of the relevant parameters along with their correlations during the maximum of solar activity (left) and the minimum (right). The $a_2^{(CF)}$ distribution in red represents the expected centrifugal effect on $a_2$ as derived from frequency shifts $\delta^{(CF)}_{nlm}$ of equation (\ref{Eq:8}). The distributions are Gaussians and show a weak correlation, reflecting the quality of the data. Between the maximum and the minimum of activity, the $a_2$ coefficient drifted significantly, from $a_2 = 80 \pm 19$ nHz to $a_2= 11 \pm 21$ falling within the  $1\sigma$ confidence interval of the centrifugal term. Although $a_4$ may have changed, the effect is below uncertainty levels and remain close to 0. As shown in Figure \ref{fig:Sun:acoefs:20062011}, other time intervals may lead to $a_4$ departing from 0.
 The figures demonstrate that a comparison of $a_2$ and $a_2^{(CF)}$ may reveal the activity of a Sun-like star. At the maximum of activity $a_2^{(CF)}$ is inconsistent with $a_2$ at $4.5\sigma$, but it is in agreement at $1\sigma$ during the minimum of activity. {Interestingly, \cite{Chaplin2003} also studied in details the frequency asymmetry of $l=2$ modes using BiSON and GOLF data with a different methodology, and over period that encompasses the maximum of 1999-2002. Their measure consider only $T_{n22}$ so a direct comparison is not straightforward. However, we note that with a 844 day-long timeseries starting in Feb 1999\footnote{{To be compared to our timeseries of 1095 days starting in Jan 1999.}}, they detect a frequency shift $T_{n22} \simeq 190$ nHz at a similar significance ($\simeq 4 \sigma$) than us when averaging over all modes between $\simeq 2000-3300\, \mu$Hz (to be compared to our range of $\simeq 2300 - 3600\, \mu$Hz). Their Figure 8 also show that the global effect of the activity cycle between 1994 and 2000 is evident on the averaged $T_{n22}$ while the frequency dependence of $T_{n22}$ may have too large uncertainties to ascertain a frequency-trend. This is in line with our own findings (see our Section \ref{sec:3.3}).}
 
 {A} rigorous statistical evaluation of {our} significance requires the joint use of $a_2$ and $a_4$. This is discussed in Section \ref{sec:results:Sun:aj:1}, along with other activity results. Finally note that for inclination of $\simeq 90$ degrees, it is not possible to determine $a_3$, because the amplitudes of the azimuthal components for $l=1,2$, are not favourable \citep{Gizon2004}.  
 
    \subsection{Activity of the Sun} \label{sec:results:Sun:aj:1}
\begin{figure}
\includegraphics[angle=0, scale=0.32]{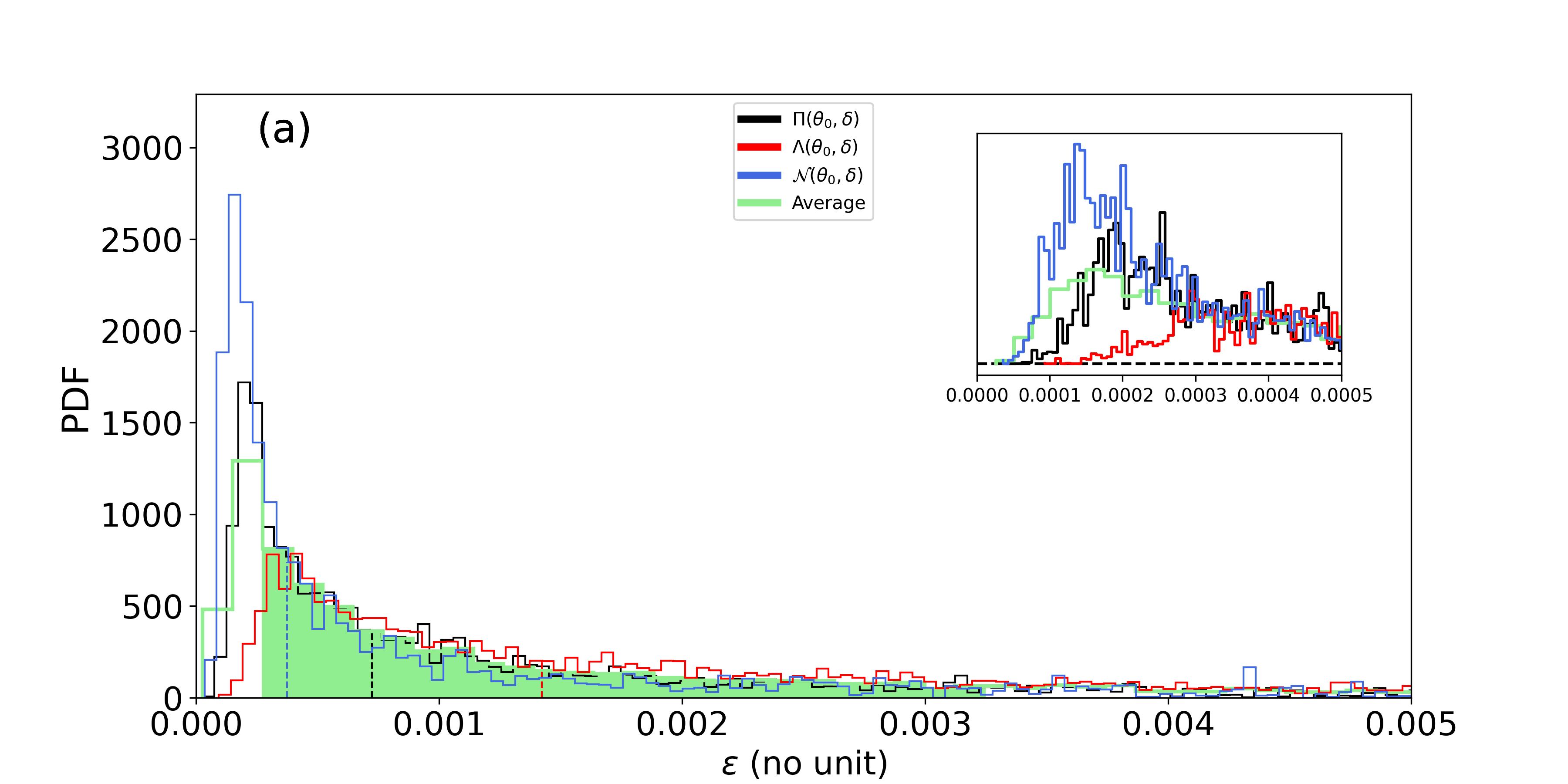}\\
\includegraphics[angle=0, scale=0.32]{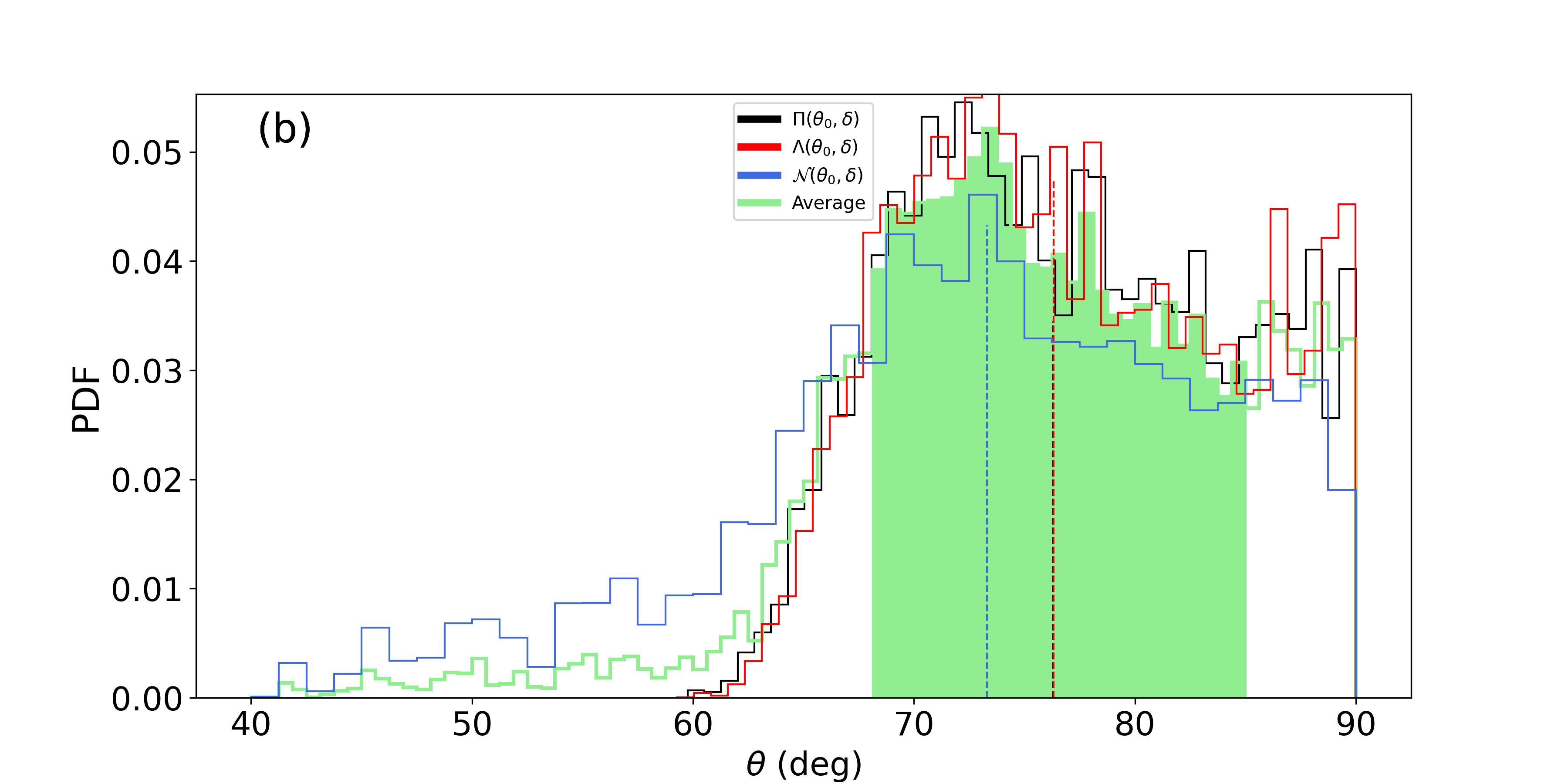}\\
\includegraphics[angle=0, scale=0.32]{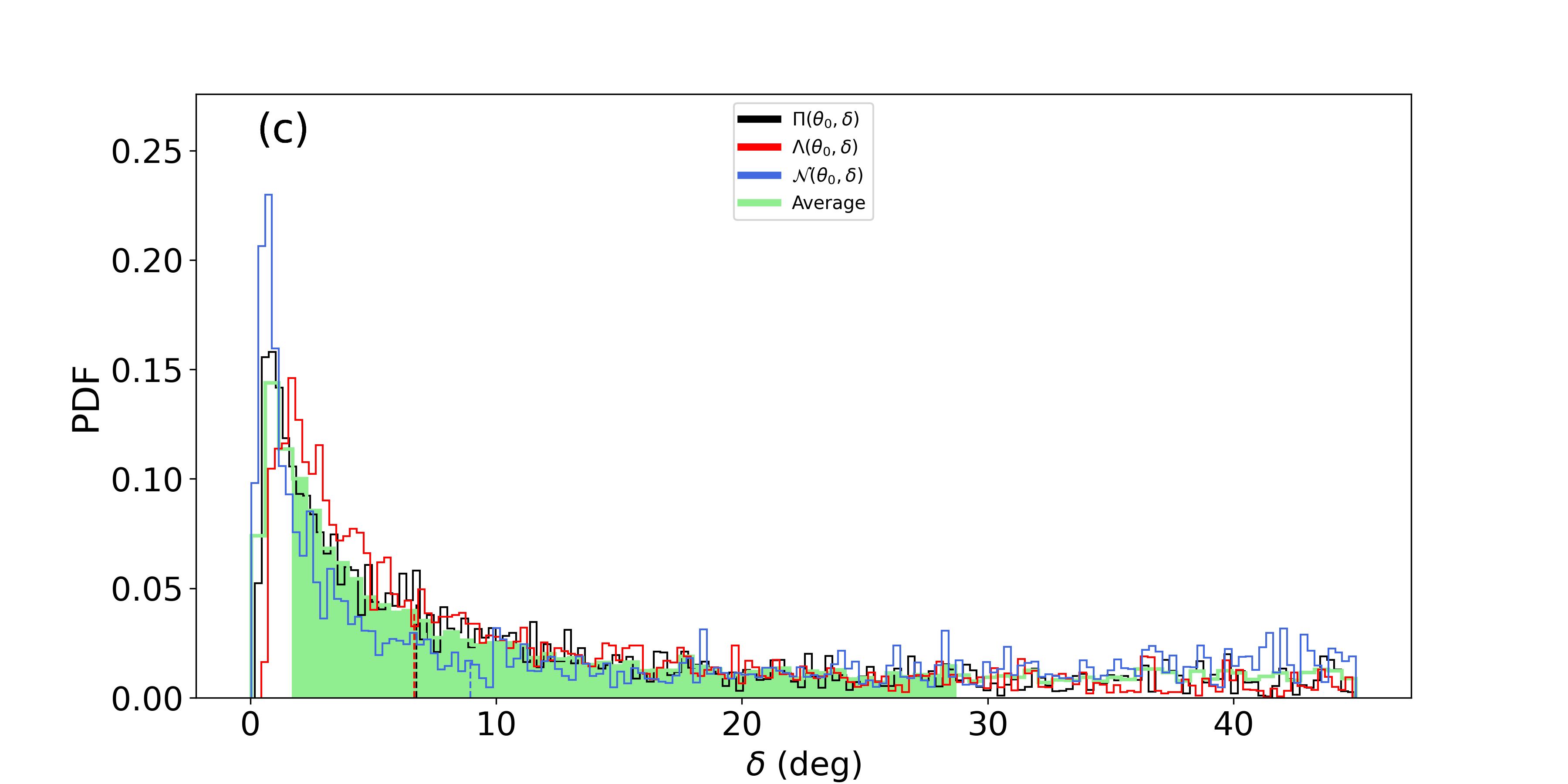}
\caption{Inferred pdf for (a) $\epsilon$, (b) $\theta_0$ and (c) $\delta$, during the Sun maximum activity of 1999-2002. The inset of (a) is a zoom into the near-zero $\epsilon$ values with smaller binning.} \label{fig:inference:Sun:max}
\end{figure}

\begin{figure}
\begin{center}
\includegraphics[angle=0, scale=0.32]{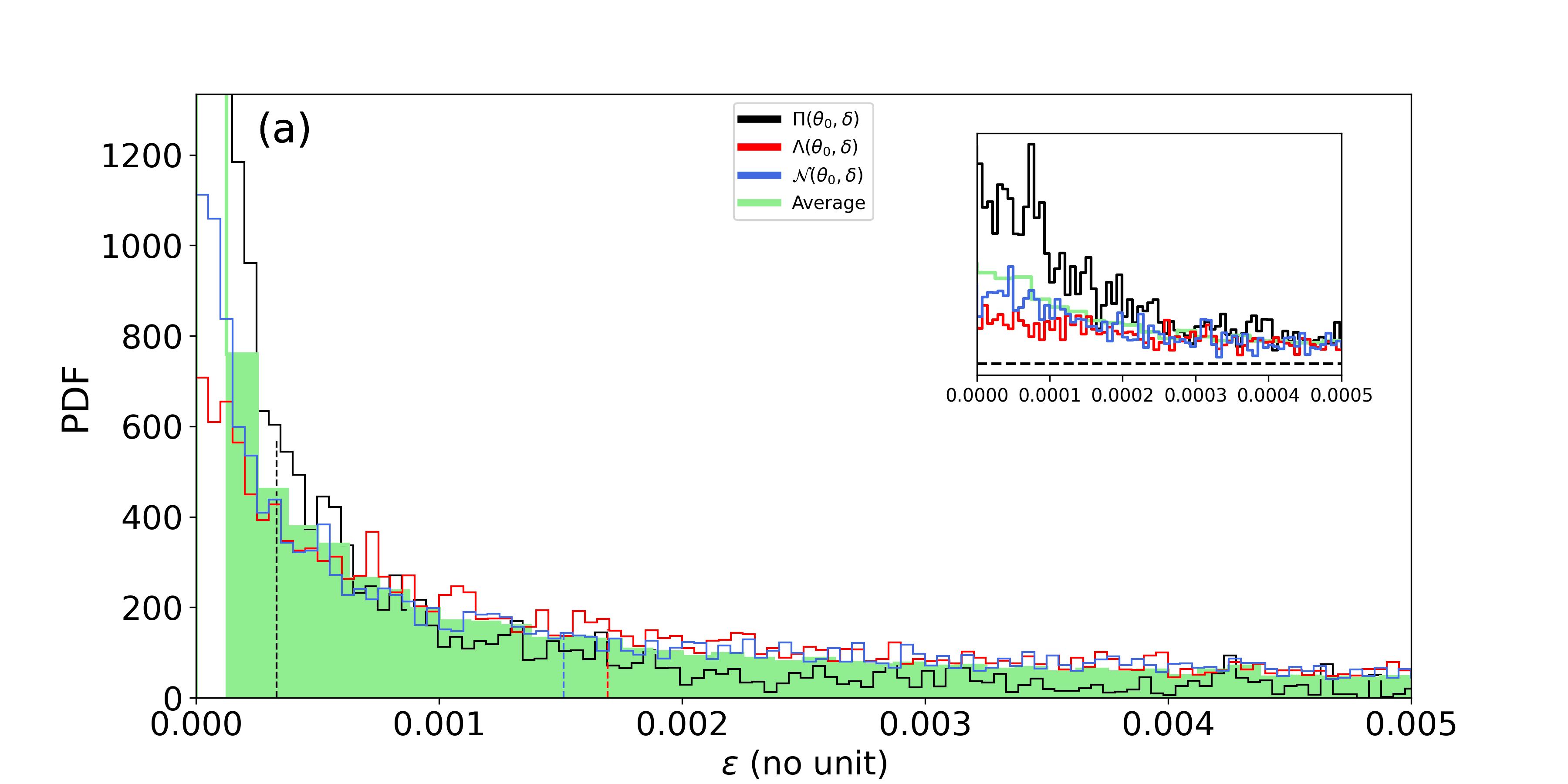}\\
\includegraphics[angle=0, scale=0.32]{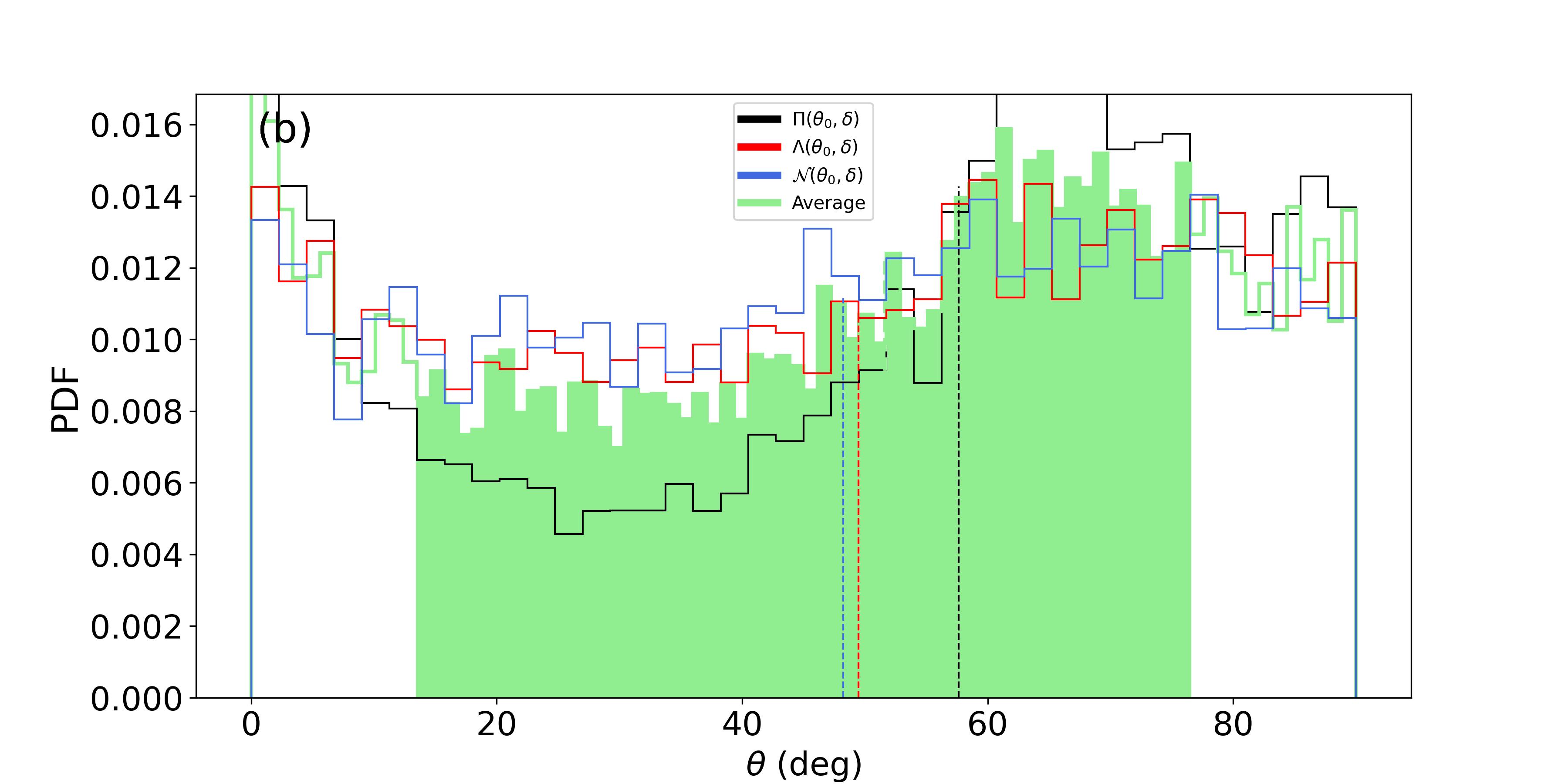}\\
\includegraphics[angle=0, scale=0.32]{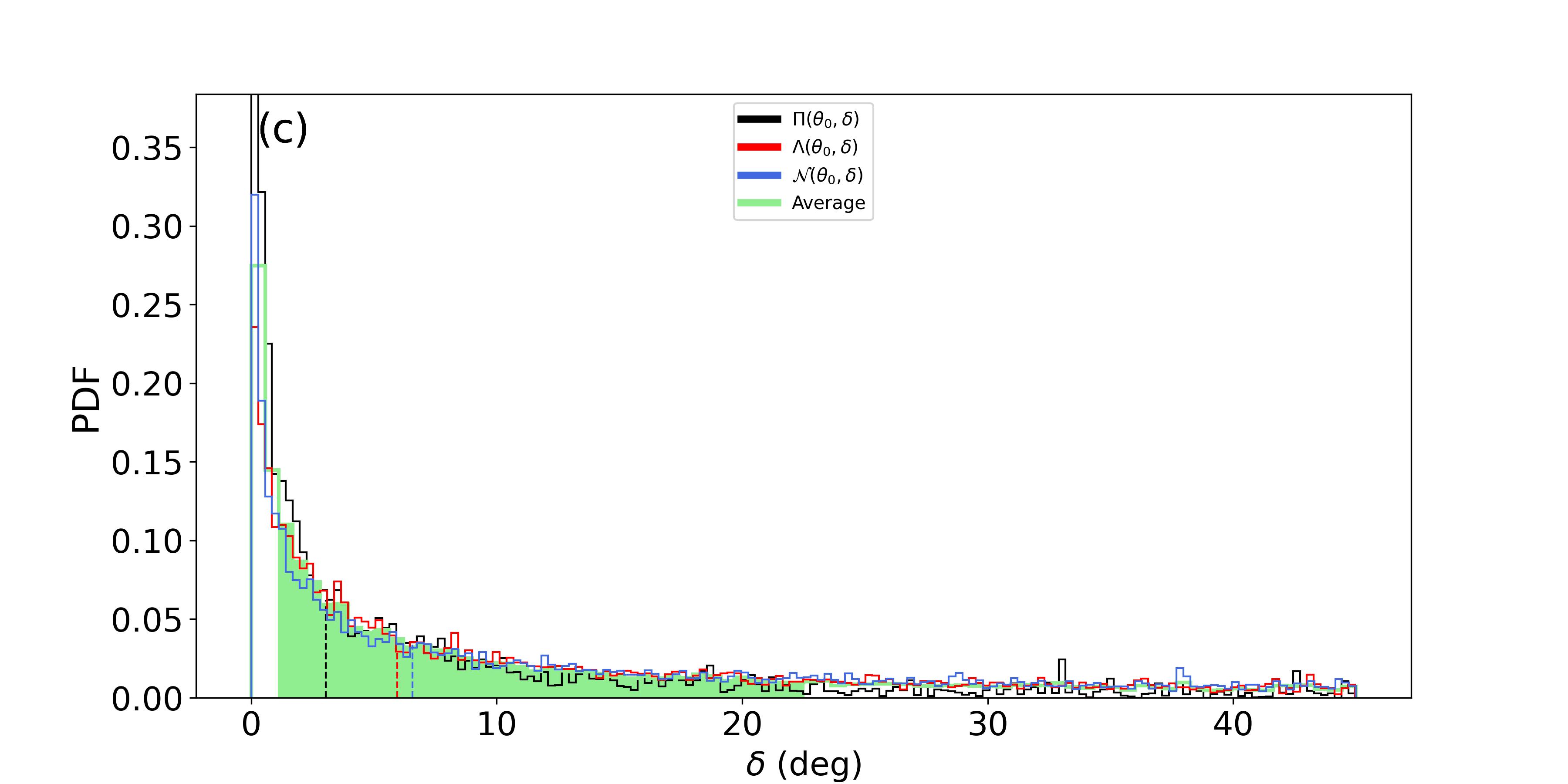}
\caption{Inferred pdf for (a) $\epsilon$, (b) $\theta_0$ and (c) $\delta$, during the Sun minimum activity of 2006-2009. The inset of (a) is a zoom into the near-zero $\epsilon$ values with smaller binning.} \label{fig:inference:Sun:min}
\end{center}
\end{figure}

The activity intensity and its latitudinal coverage is derived from a fit of the $a^{(AR)}_2$ and $a^{(AR)}_4$ coefficients. Technical details are in Appendix \ref{appendix:aj_analysis}. Statistical summary of the results are listed in Table \ref{tab:real_vals:inference}. The statistical significance of the activity exceeds $99.99\%$ (highly significant) at the maximum of activity, but is 
{around} $50.6\%$ (not significant) during the minimum of activity, demonstrating the possibility of detecting activity in Sun-like stars. {The choice of the function $F$ describing the active region changes the detection significance by only a few percents. The three explored models fit equivalently the data implying that the shape of the active region cannot be determined with currently available $a_2$ and $a_4$ constraints.}

Figure \ref{fig:inference:Sun:max} and \ref{fig:inference:Sun:min} show $\epsilon_{nl}=\epsilon$, $\theta_0$ and $\delta$ for the Sun at its maximum of activity (1999-2002) and its minimum (2006-2009), respectively. {There is no major differences between the three activity profiles, although it is noted that uncertainties are larger in the case of a triangular and a Gaussian activity zone. } During the maximum of activity, $\epsilon \simeq 0$ is clearly excluded (see inset). {With $F=\Pi$,} we observe a log-normal distribution, with median $7.6.10^{-4}$, consistent with the value reported by \cite{Gizon2002AN}. However, the uncertainty is large, {suggesting that only the order of magnitude can be constrained in Sun-like stars}. At the minimum of activity, the log-normal distribution morphs into a $1/x$ law, similar to the Jeffreys prior.  This is a sign of weaker statistical significance for the activity.

The co-latitude $\theta_0 \simeq 75$$^\circ$ is consistent with the butterfly diagram, which suggests $\theta_0 \simeq 80^\circ$. At the minimum of activity and despite a non-significant detection, the probability distribution of $\theta_0$ shows a weak indication of activity at mid and high co-latitudes.
Despite the high significance of the detection during the maximum of activity, $\delta$ is poorly constrained, showing that this parameter is challenging to {measure} on the Sun and on other Sun-like stars.

\section{ Analysis of 16 Cyb A and B} \label{sec:16CygAB}

Due to their brightness (magnitudes V=5.95 and 6.20) 16 Cyg A and B have modes with the highest HNR among all of Sun-like stars observed asteroseismically so far. They constitute ideal candidates to evaluate the activity. The two stars are  wide binaries, with a confirmed planet around 16 Cyg B \citep{Cochran1997} and were extensively studied \citep[e.g.][]{Neckel1986A&A,King1997AJ,Deliyannis2000AJ,Schuler2011ApJ,Yoichi2005PASJ,Metcalfe2012, Lund2014,Verma2014,Buldgen2015, Deal2015,Metcalfe2016,Roxburgh2017,Bellinger2017,Maia2019,Bazot2019, Bazot2020, Farnir2020, Morel2021, Buldgen2022,Nsamba2022MNRAS}. The 2.5 years ( 13 September 2010 to 8 March 2013) observation by the Kepler space-borne instrument is used in this section to measure pulsation parameters. The data are the same as those used in \cite{Bazot2019} for which instrumental issues (outliers, jumps, trends) and the quarter stitching is performed using the procedure described in \cite{Garcia2011}.
The binary system has precisely measured angular diameters, making it two of the few Sun-like stars with known interferometric radii. Their measured radius are  $1.22 \pm 0.02$ \(R_\odot\) and $1.12 \pm 0.02$ \(R_\odot\) for 16 Cyg A and B, respectively \citep{White2013}. Their spectroscopic parameters are very close to those of the Sun: Their effective temperatures are $T_{\mathrm{eff}} = 5825 \pm 50$ K, $T_{\mathrm{eff}} = 5750 \pm 50$ K and their metallicity are $[M/H] = 0.10 \pm 0.09$, $[M/H] = 0.05 \pm 0.06$ \citep{Ramirez2009} for 16 Cyg A and B, respectively. With an estimated age of around $7$ Gyrs for both stars \citep[e.g.]{Metcalfe2016, Bazot2020}, they are significantly older than the Sun.

    \subsection{Seismic constraints for 16 Cyg A} \label{sec:16CygA:acoefs}
    Earlier studies of 16 Cyg A revealed around 60 modes of pulsations with significance in the power spectrum\footnote{The exact value may differ from author to author, due to different choice for the significance.}. The asteroseismic analysis of \cite{Davies2015} found a stellar inclination of $56^{+6}_{-5}$ and a rotation $23.8^{+1.5}_{-1.8}$ ($\langle\delta\nu_{nlm}/m\rangle_{nl}\simeq 486^{+40}_{-29}$nHz). Using a refined power spectrum modelling that account for $a_1$, $a_3$ and parameterises the cavity asphericity $(R_{\mathrm{eq}} - R_{\mathrm{pol}})/R_{\mathrm{eq}}$, \cite{Bazot2019} reported values ($i=58.5\pm6.8$, $a_1=464 \pm 43$nHz), consistent with \cite{Davies2015}. 
    
\begin{figure*}
\begin{center}
\includegraphics[angle=0, scale=0.5]{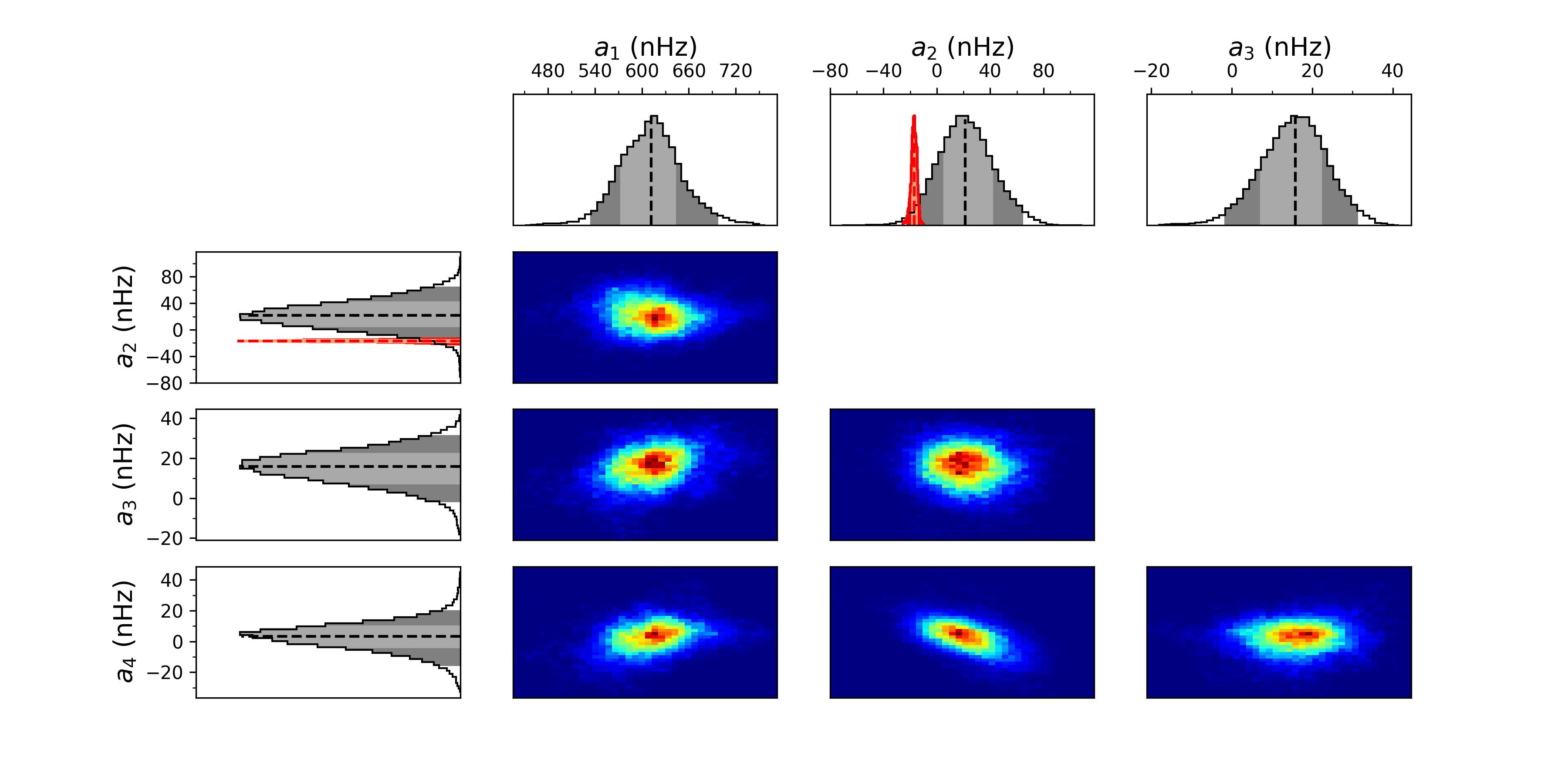}
\caption{Probability Density Functions and their correlations, obtained by MCMC of the power spectrum of 16 Cyg A and for coefficients $a_1$, $a_2$, $a_3$, $a_4$ for 16 Cyg A. The red curve shows the expected $a_2^{CF}$ coefficient for a pure centrifugal distortion of the star. The light and dark gray PDF filling is for the $1\sigma$ and $2\sigma$ confidence interval, respectively.} \label{fig:16CygA:acoefs}
\end{center}
\end{figure*}
    
\begin{figure*}
\includegraphics[angle=0, scale=0.215]{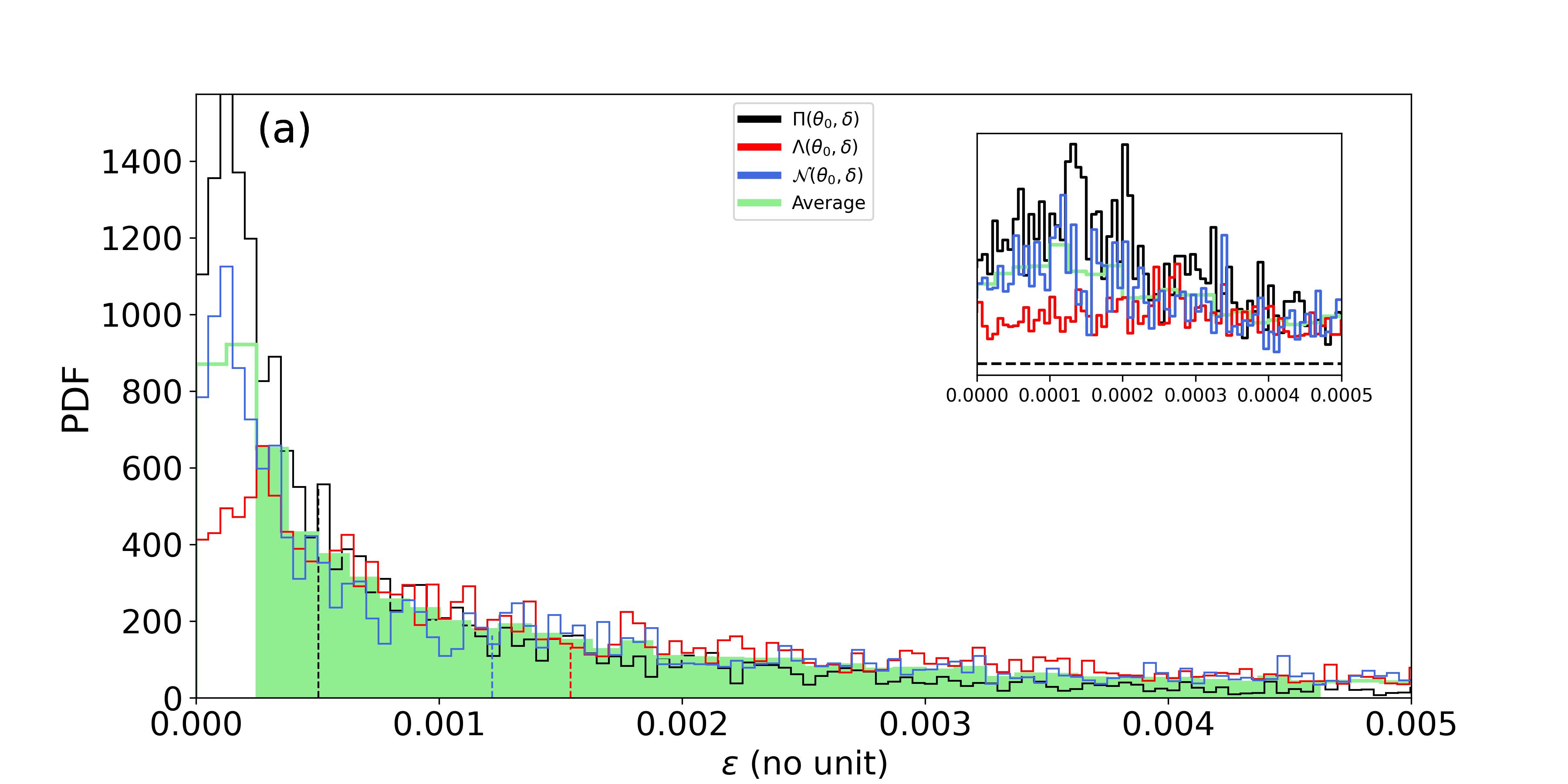}
\includegraphics[angle=0, scale=0.215]{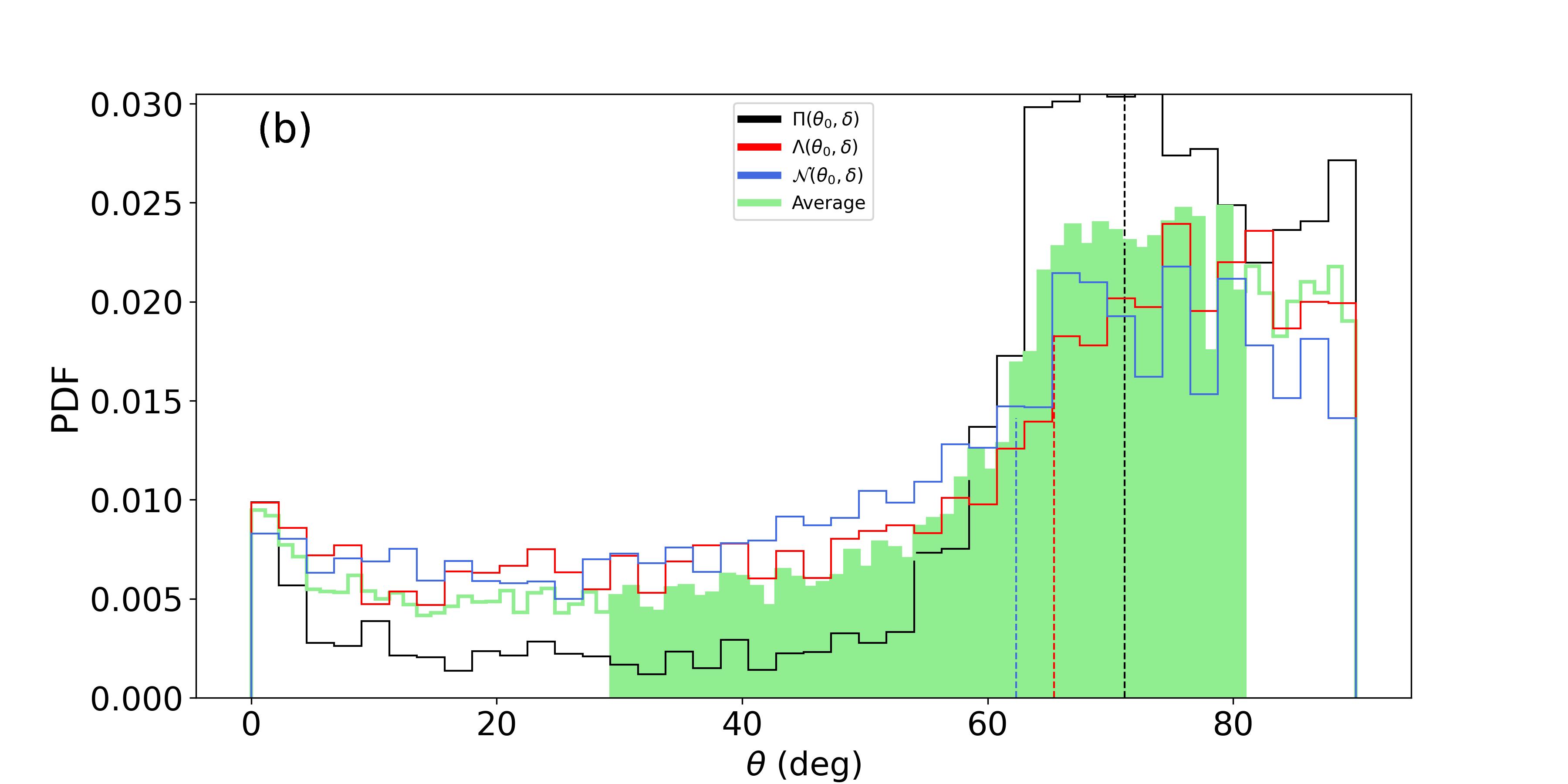} 
\includegraphics[angle=0, scale=0.215]{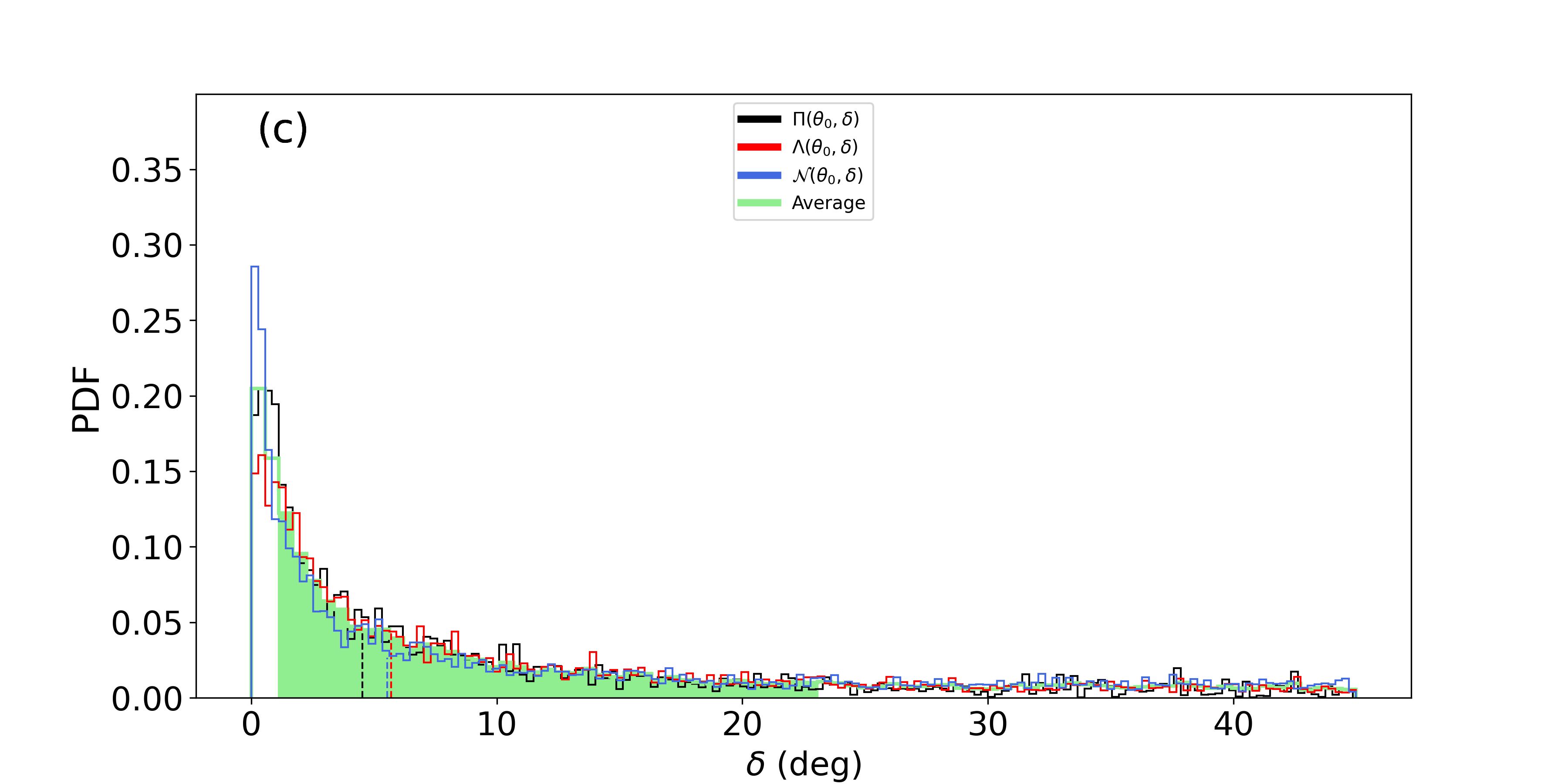}\\
\includegraphics[angle=0, scale=0.215]{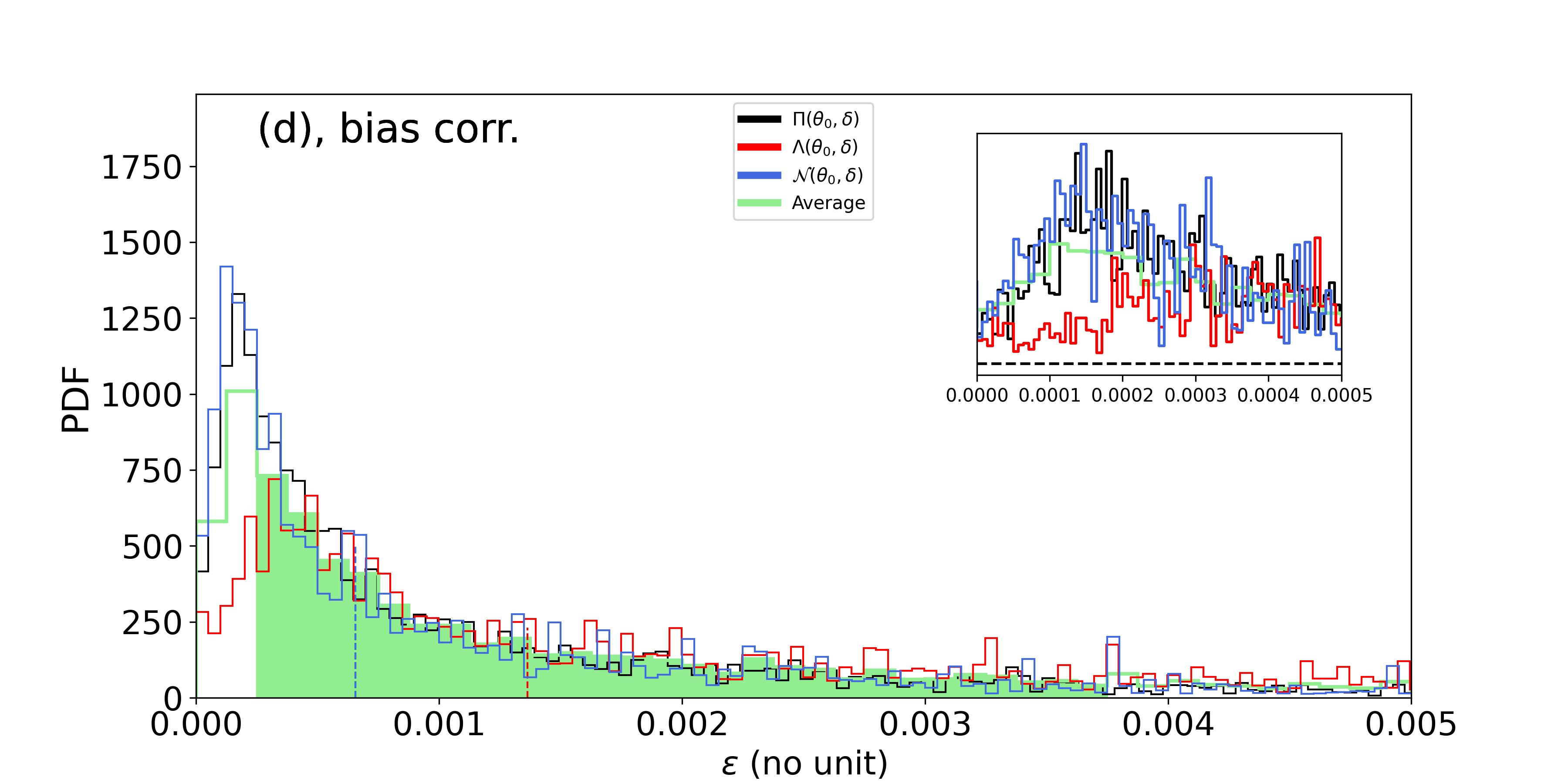} 
\includegraphics[angle=0, scale=0.215]{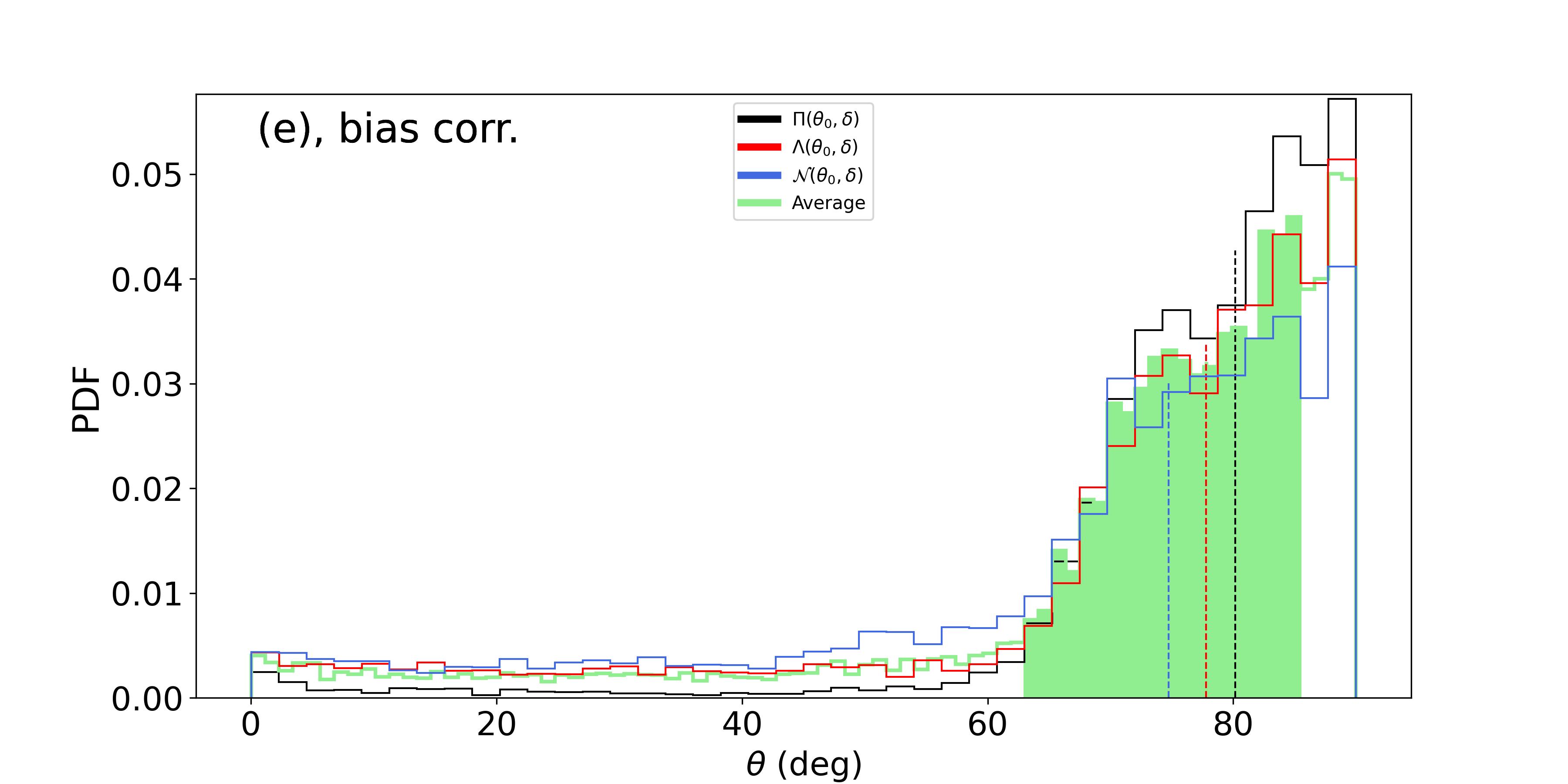}
\includegraphics[angle=0, scale=0.215]{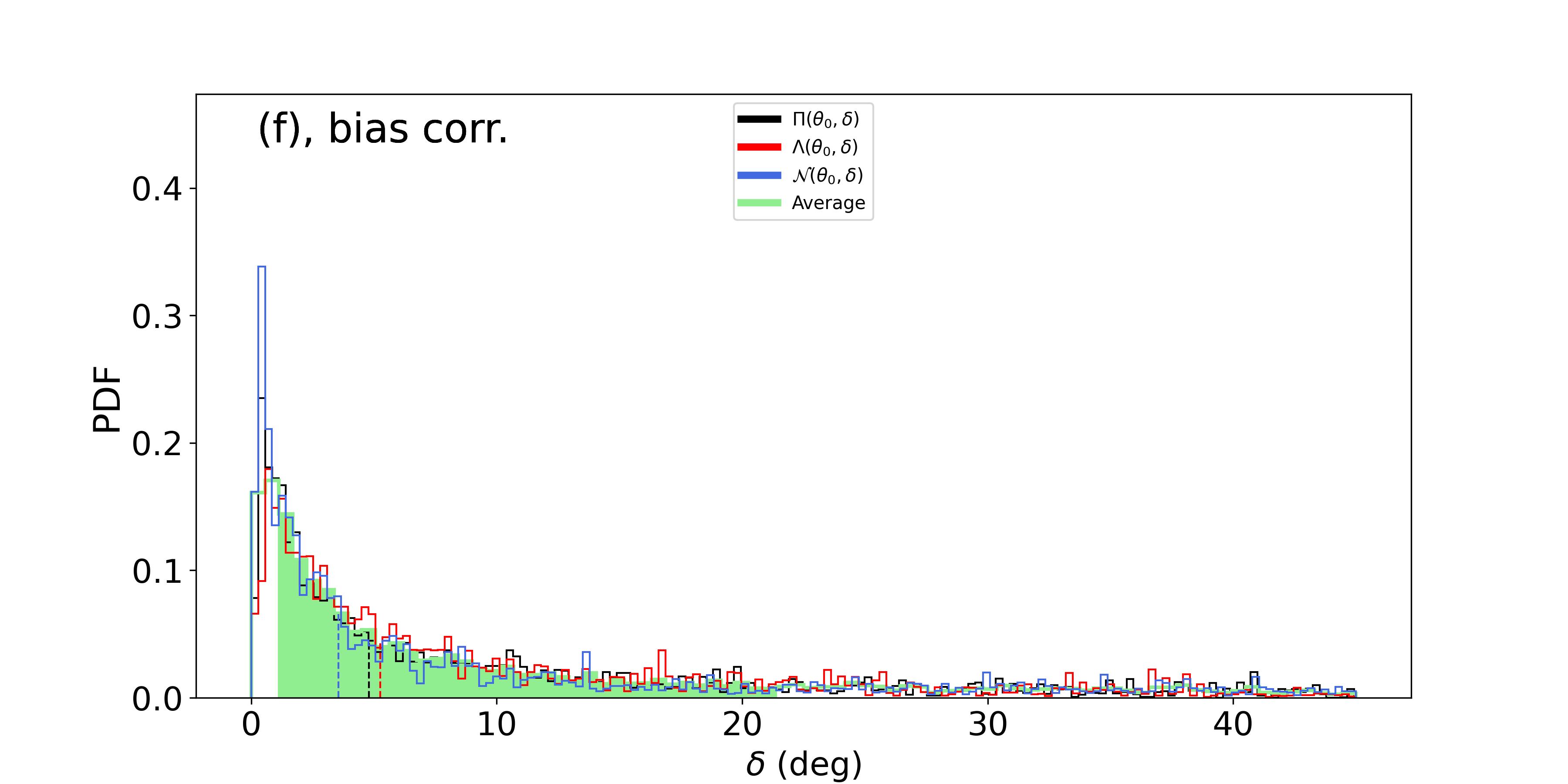}
\caption{Inferred pdf for 16 Cyg A for {raw measures (a,b,c) or with bias correction (d,e,f). The green curve is for the average pdf with $F=\Pi,\,\Lambda$ and $\mathcal{N}$. The shaded area is its} $1\sigma$ confidence interval. The inset of {(a) and (d)} is a zoom into the near-zero $\epsilon$ values with smaller binning.} \label{fig:inference:16cygA}
\end{figure*}

Figure \ref{fig:16CygA:acoefs} shows the probability density function for $a_1$, $a_2$, $a_3$ and $a_4$ for 16 Cyg A, along with their correlation and with $a^{(CF)}_2$ superimposed in the $a_2$ quadrant.
Table \ref{tab:real_vals} synthesises the inferred values of the coefficients. The probability density functions are near-Gaussian and thus the quadratic mean of the asymmetrical uncertainties is reported in the table.
The $a_1$ coefficient is significantly lower than past estimates but remain consistent at a $2\sigma$ confidence level. The difference may be due to the lower stellar inclination $i=45\pm4$, that is also only consistent with earlier determination at $2\sigma$. The $a_3$ coefficient is marginally greater but with smaller uncertainty than \cite{Bazot2019} (they reported $a_3 = 11.15\pm10.95$nHz).
    
More importantly, $a_2$ is positive at $1\sigma$, which is consistent with the assertion of star prolateness from \cite{Bazot2019} and is inconsistent with the centrifugal distortion term, $a_2^{(CF)}$. Finally, $a_4$, includes zero within $1\sigma$.
    
    With respect to Figures \ref{fig:bias:hnr30:Eq}-\ref{fig:bias:hnr30:Pol}, $a_1/\Gamma_{\nu_{max}} = 0.45 \pm 0.04$ and $i=45\pm4^\circ$ place 16 Cyg A in a parameter space where the expected magnitude of the bias for $a_1$, $a_2$ and $a_4$ is of the order of $0.5\sigma$, $0.5\sigma$ and $0.8\sigma$ respectively. This remains accurate even when accounting for the slightly higher estimates of inclination from previous publications.  
    Translated into absolute units, this corresponds to approximately $b(a_1) \simeq +10$nHz (overestimation), $b(a_2) \simeq -10$nHz (underestimation), $b(a_4) \simeq -10$nHz (underestimation). These fiducial values are used in Section \ref{sec:16CygA:activity} to evaluate the effect of the potential bias on the estimates of the activity zone.

    \subsection{Seismic constraints for 16 Cyg B}\label{sec:16CygB:acoefs}

    Similarly to 16 Cyg A, 16 Cyg B has around 60 observed modes. However, the reported precision for the past determination of the seismic parameters is less accurate than for 16 Cyg A, despite a similar HNR. \cite{Davies2015} and \cite{Bazot2019} both note a large degeneracy between the stellar inclination and the average rotation rate, with even a clear bi-modality in the distributions obtained by \cite{Bazot2019}. Their global solution of $i=36^{+17}_{-7}$ is associated with two separate solutions for $a_1$, centred around $\simeq 300$nHz and $\simeq 550$nHz, that they use to infer the latitudinal differential rotation profile.
    Meanwhile, as for 16 Cyg A, the star is found to be prolate, indicating a surface activity.
    
\begin{figure*}
\begin{center}
\includegraphics[angle=0, scale=0.5]{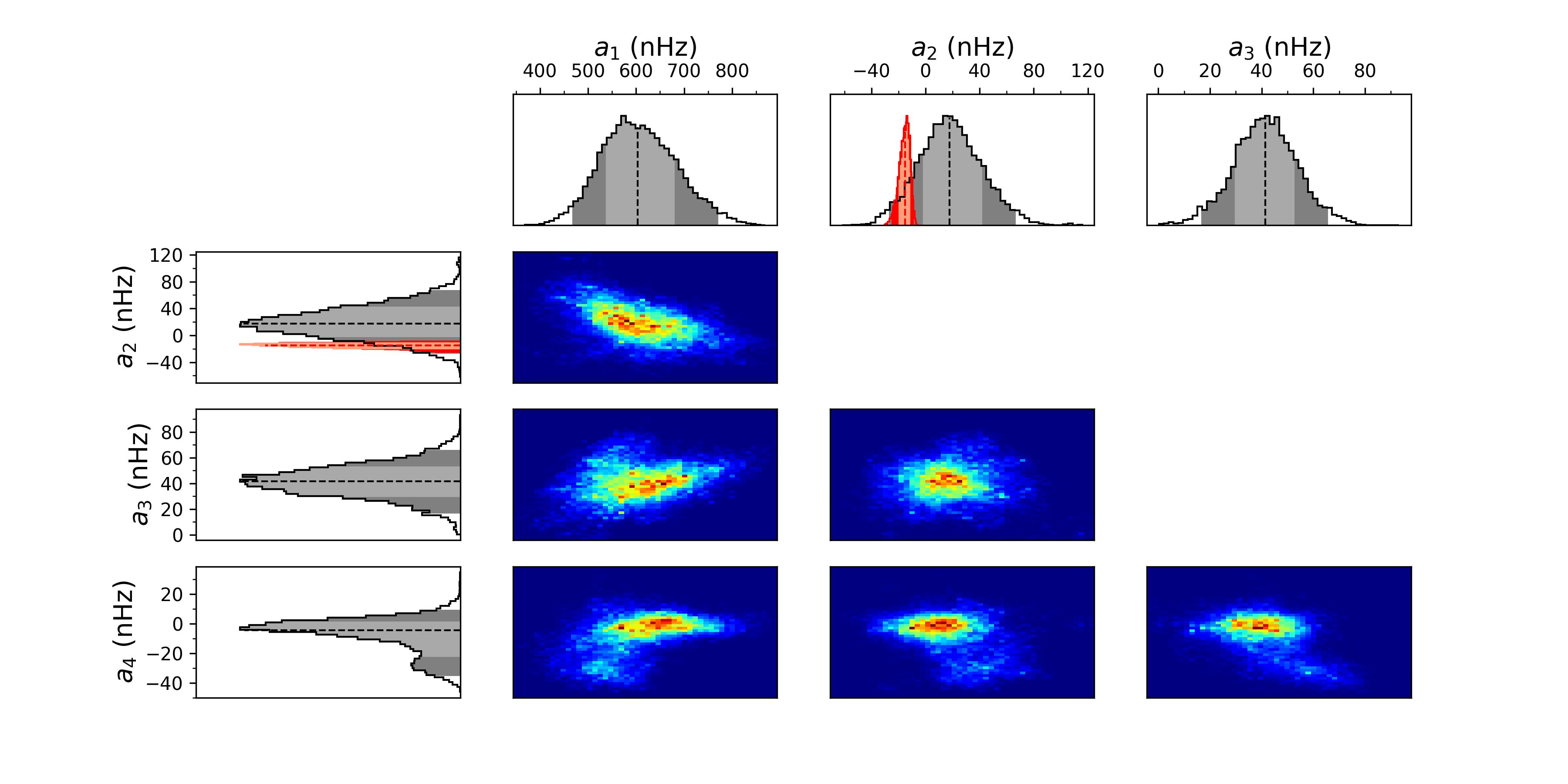}
\caption{Probability Density Functions and their correlations, obtained by MCMC of the power spectrum of 16 Cyg B and for coefficients $a_1$, $a_2$, $a_3$, $a_4$ for 16 Cyg A. The red curve shows the expected $a_2^{CF}$ coefficient for a pure centrifugal distorsion of the star. The light and dark gray PDF filling is for the $1\sigma$ and $2\sigma$ confidence interval, respectively.} \label{fig:16CygB:acoefs}
\end{center}
\end{figure*}

    Figure \ref{fig:16CygB:acoefs} shows the measured a-coefficients. It appears that including the $a_4$ coefficient remove the degeneracy issue observed by previous publications. The stellar inclination, $a_1$ and $a_2$ are precisely determined and have approximately Gaussian probability distributions and these are reported as such in Table \ref{tab:real_vals}. The found rotation rate $a_1$ correspond to the higher solution of rotational splitting of \cite{Davies2015} and \cite{Bazot2019} (see their Figures of probability density functions). Meanwhile, the degeneracy observed in $a_1$ in earlier studies is moved to $a_4$: It exhibits two solutions. As our method of activity inference assumes Gaussian distributions, the bi-modality of $a_4$ requires to separate the two apparent solutions. Using Gaussian mixture modelling \citep{Bishop1995,Bazot2019}, the mean and standard deviation of both solutions are measured, enabling their separate analysis. 
    The lowest estimate ($a_4=-27.9\pm6.5$nHz) weights $30\%$ and the highest estimate ($a_4=-1.0\pm6.7$nHz) is more significant as it weights $70\%$. The inference of the activity from the two solutions is discussed in Section \ref{sec:16CygB:activity}.
    Note that $a_3=45^{+13}_{-14}$nHz is significantly higher than the reported values from \cite{Bazot2019} ($a_3=13.89\pm13.95$nHz) and may have an impact on the rotational profile.
    
    \subsection{Activity inference for 16 Cyg A} \label{sec:16CygA:activity}
Figure \ref{fig:inference:16cygA} shows the results from the inference of the activity parameters and a statistical summary is in Table \ref{tab:real_vals:inference}. 
{Triangular and Gaussian descriptions of active latitudes give larger uncertainties than the simple gate model. However, they all suggest a near-equatorial activity, with a similar detection significance level of $\simeq 79\%$ (without bias correction) or $\simeq 96\%$ with fiducial correction. As for the Sun and because the shape of the active region is \emph{a priori} unknown, the average distribution of the parameters are discussed.}
The activity intensity has a large uncertainty, but according to the median of $\epsilon$, may be between the maximum and the minimum of solar activity.

The posterior probability distribution of $\delta$ does not allow us to precisely constrain the extension of the activity region. As already noted in the case of the Sun, this parameter requires stringent constraint on both the $a_2$ and $a_4$ coefficient in order to inform us about the size of the active region. 

    \subsection{Activity inference for 16 Cyg B}  \label{sec:16CygB:activity}

\begin{figure*}
\includegraphics[angle=0, scale=0.215]{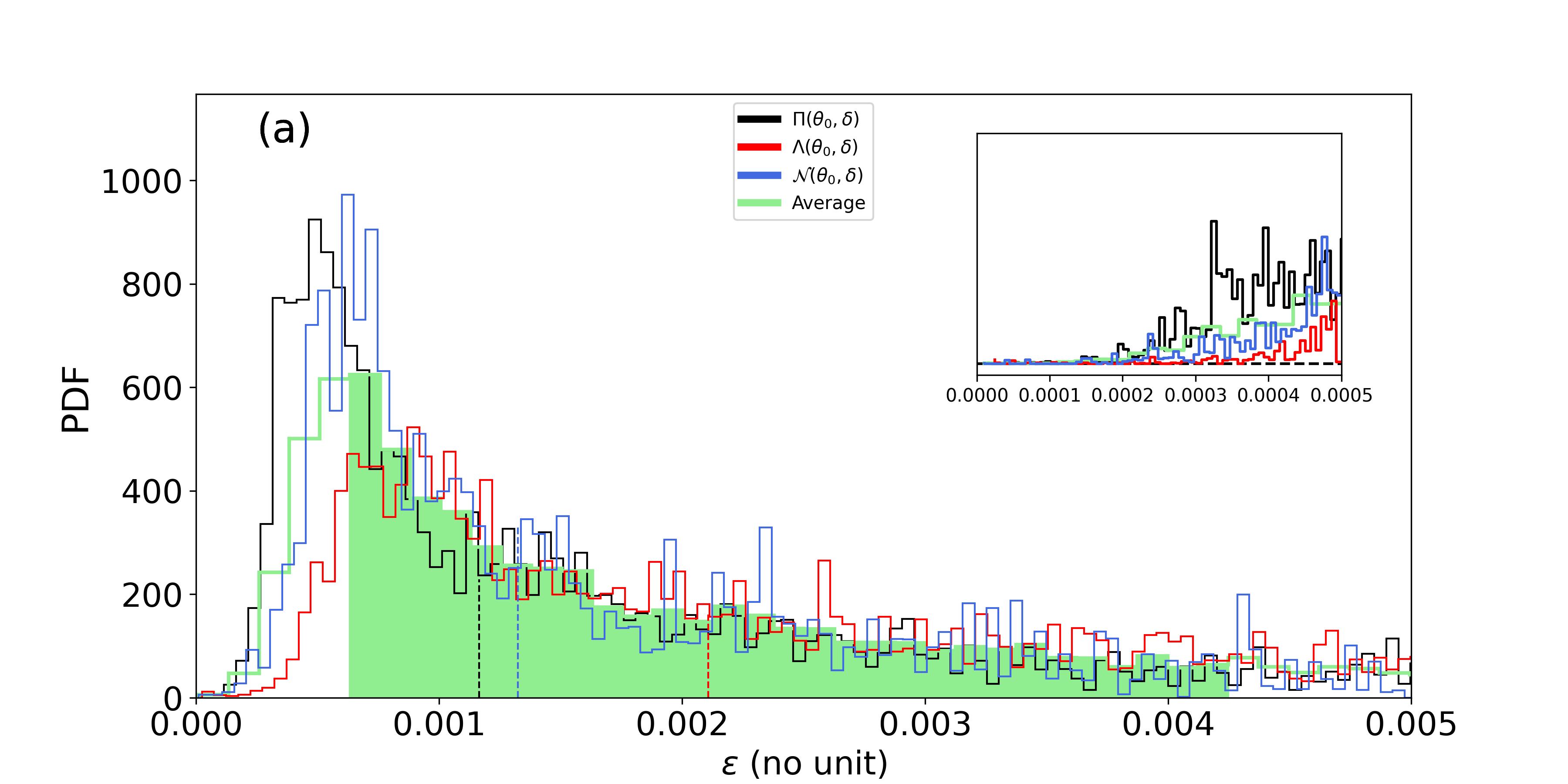}
\includegraphics[angle=0, scale=0.215]{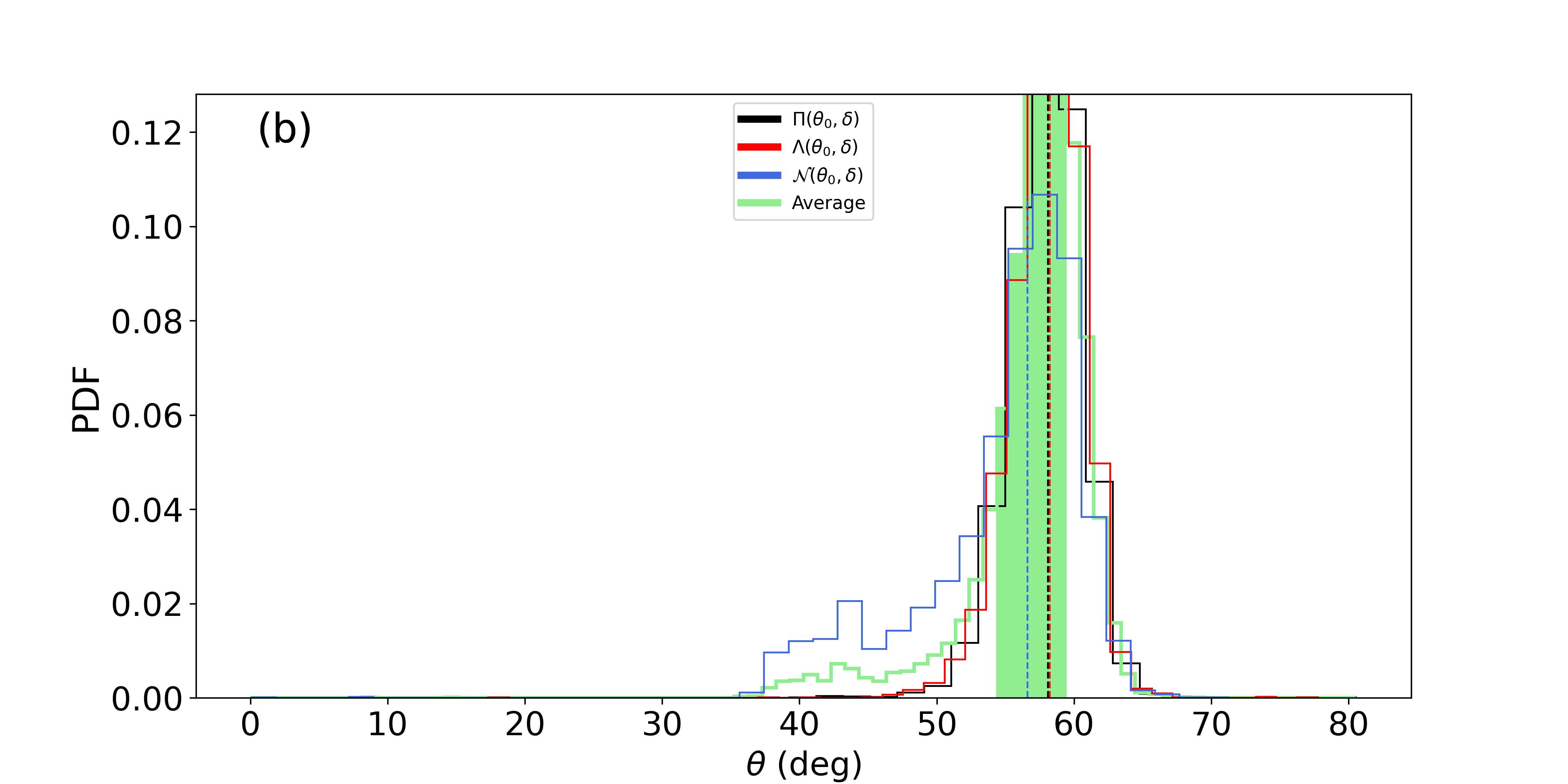}
\includegraphics[angle=0, scale=0.215]{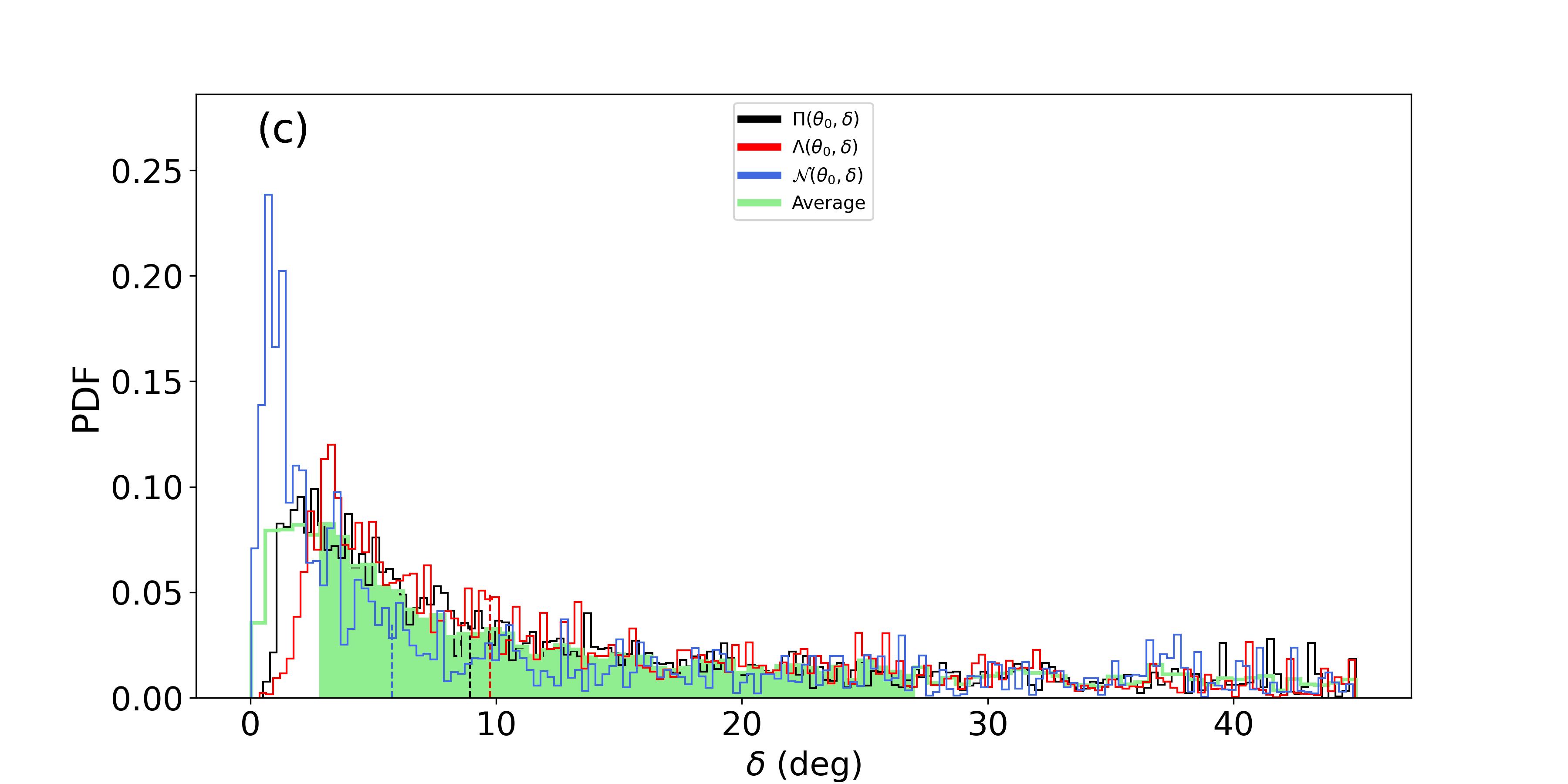}\\
\includegraphics[angle=0, scale=0.215]{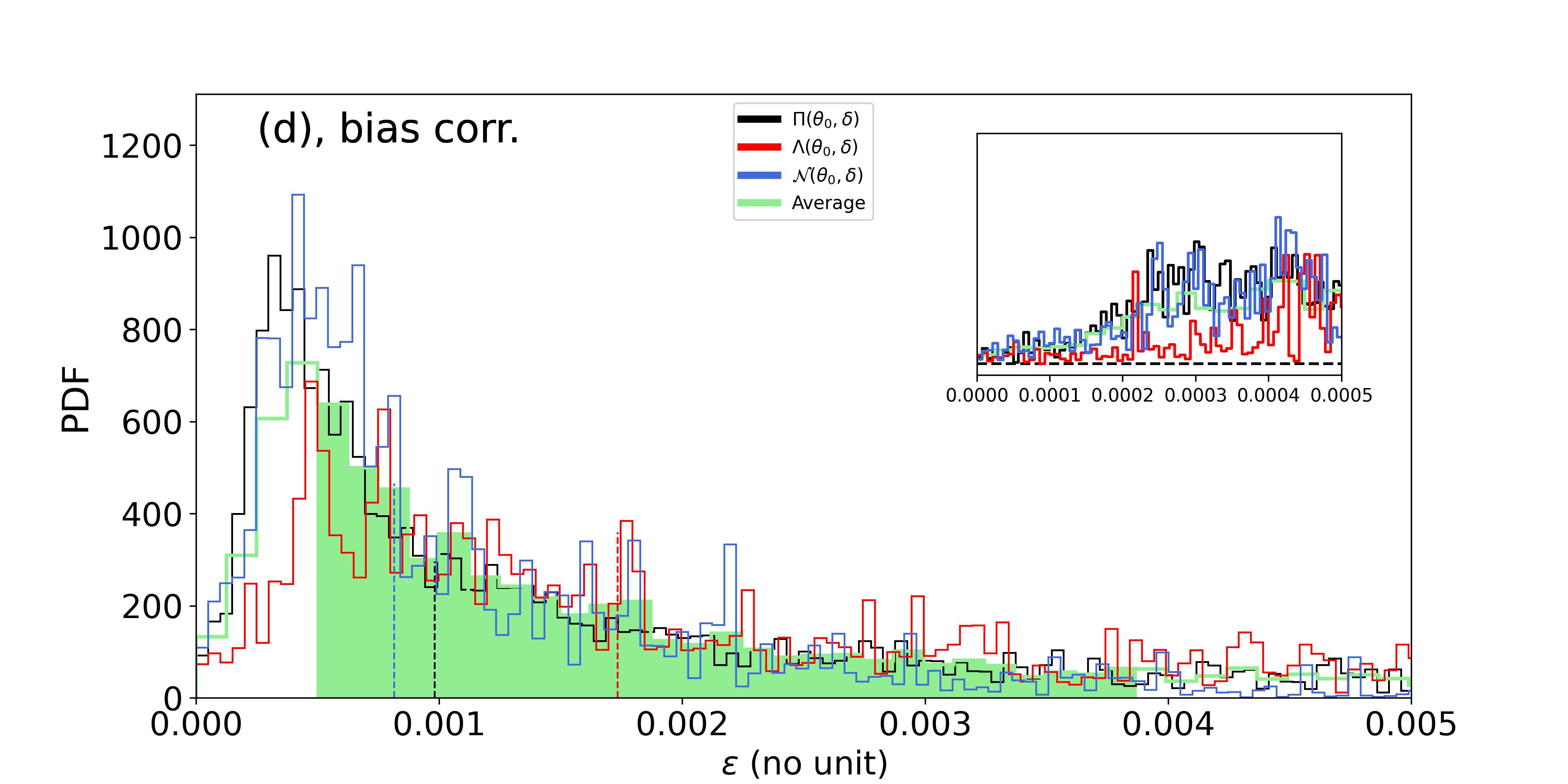}
\includegraphics[angle=0, scale=0.215]{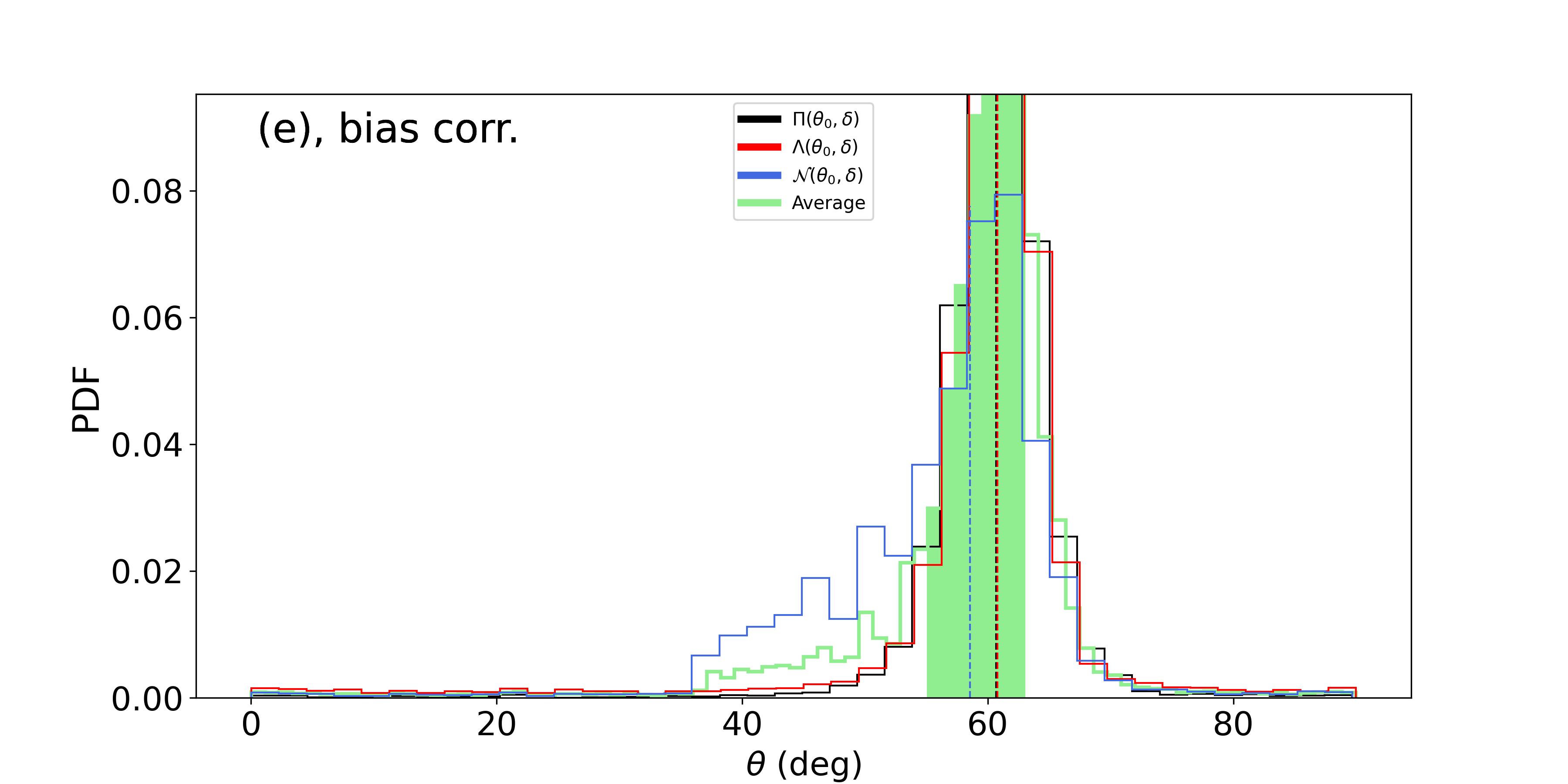} 
\includegraphics[angle=0, scale=0.215]{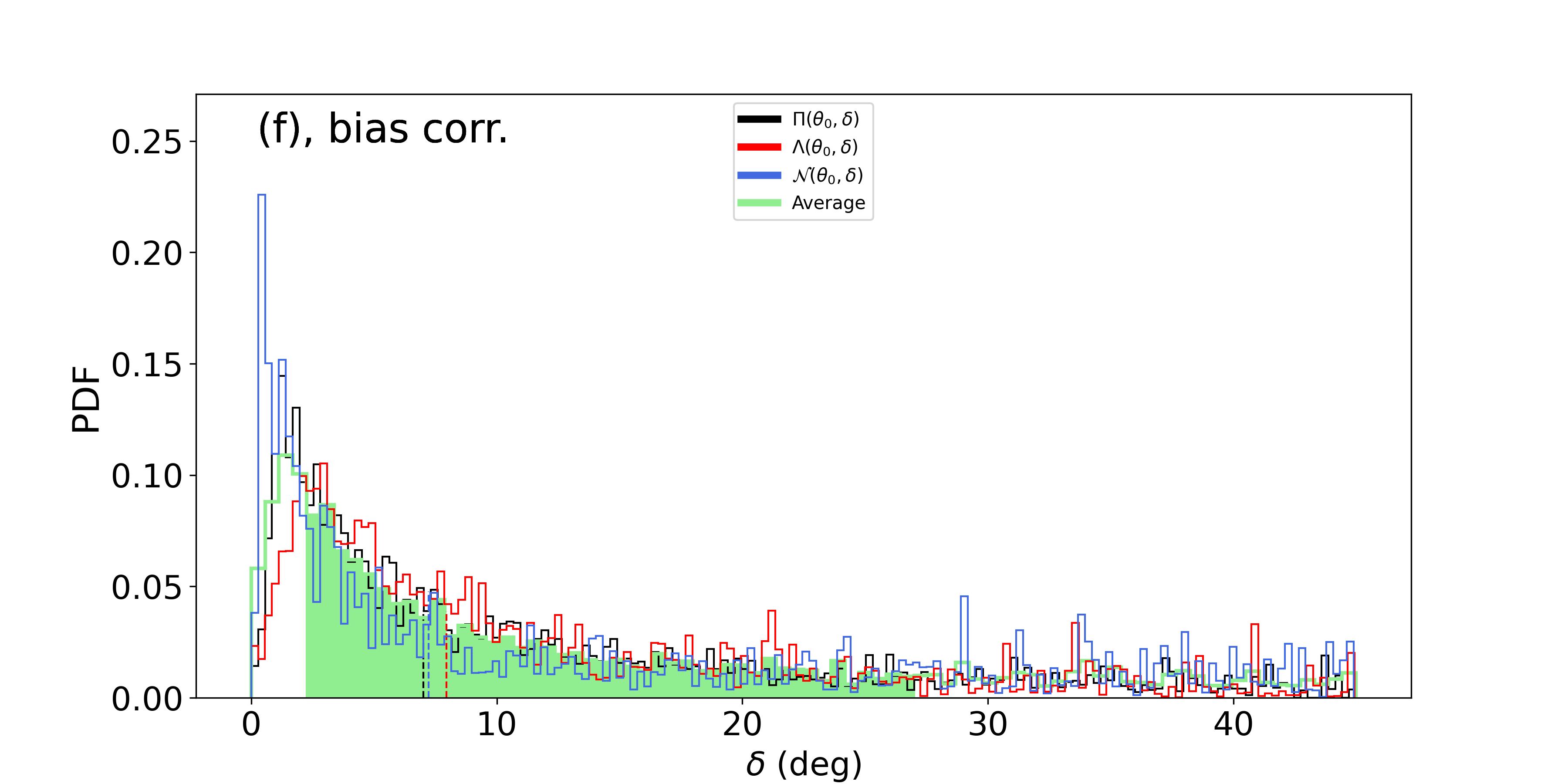}
\caption{Inferred pdf of $\epsilon$, $\theta_0$ and $\delta$ {for raw results of 16 Cyg B with the lower solution $a_4=-29.2\pm5.6$nHz, without bias correction (a, b, c) or with it (d, e, f). The green curve is for the average pdf with $F=\Pi,\,\Lambda$ and $\mathcal{N}$. The shaded area is its $1\sigma$ confidence interval. The inset of (a,d) } is a zoom into the near-zero $\epsilon$ values with smaller binning.} \label{fig:inference:16cygB}
\end{figure*}

\begin{figure*}
\includegraphics[angle=0, scale=0.215]{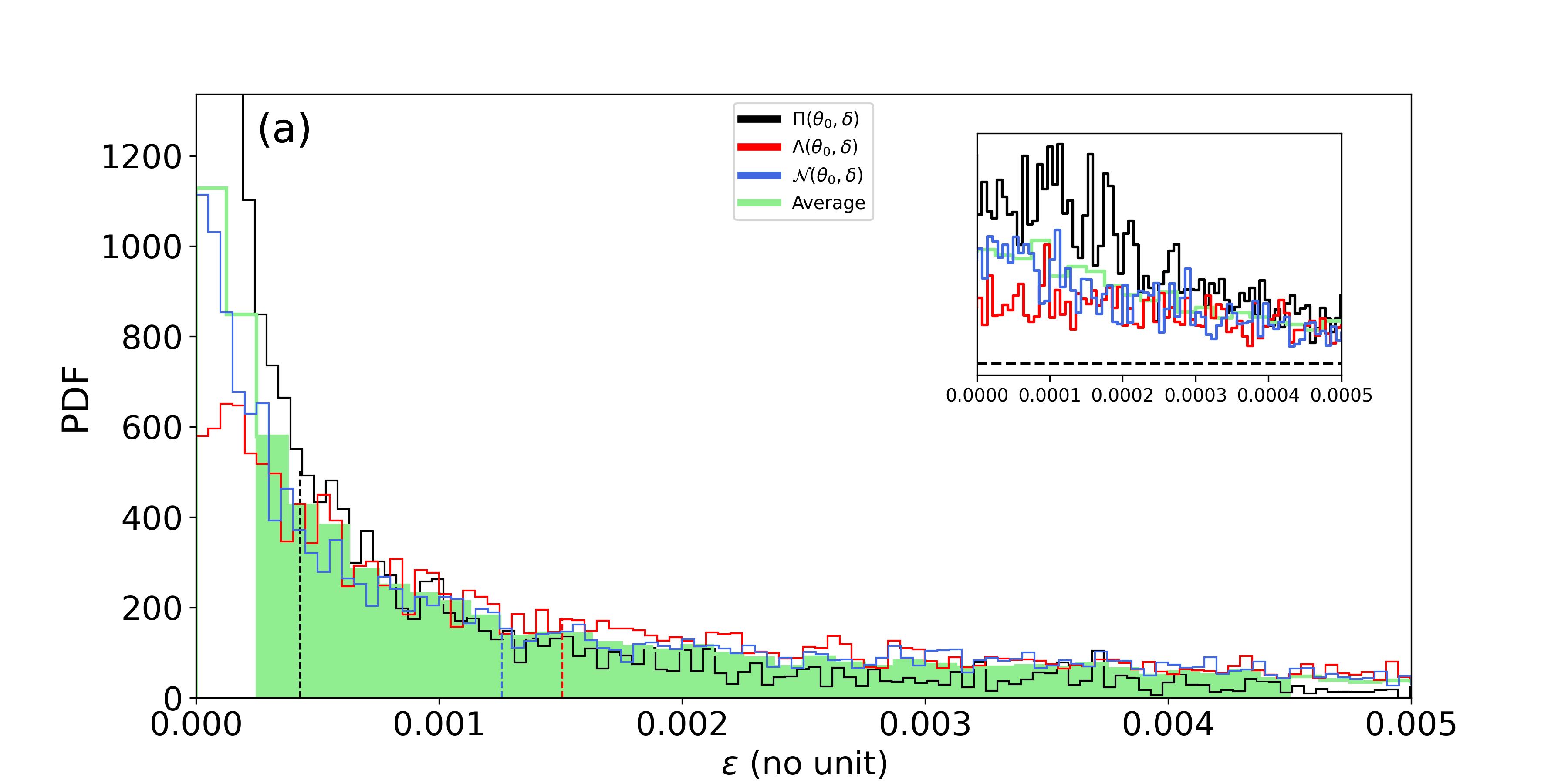}
\includegraphics[angle=0, scale=0.215]{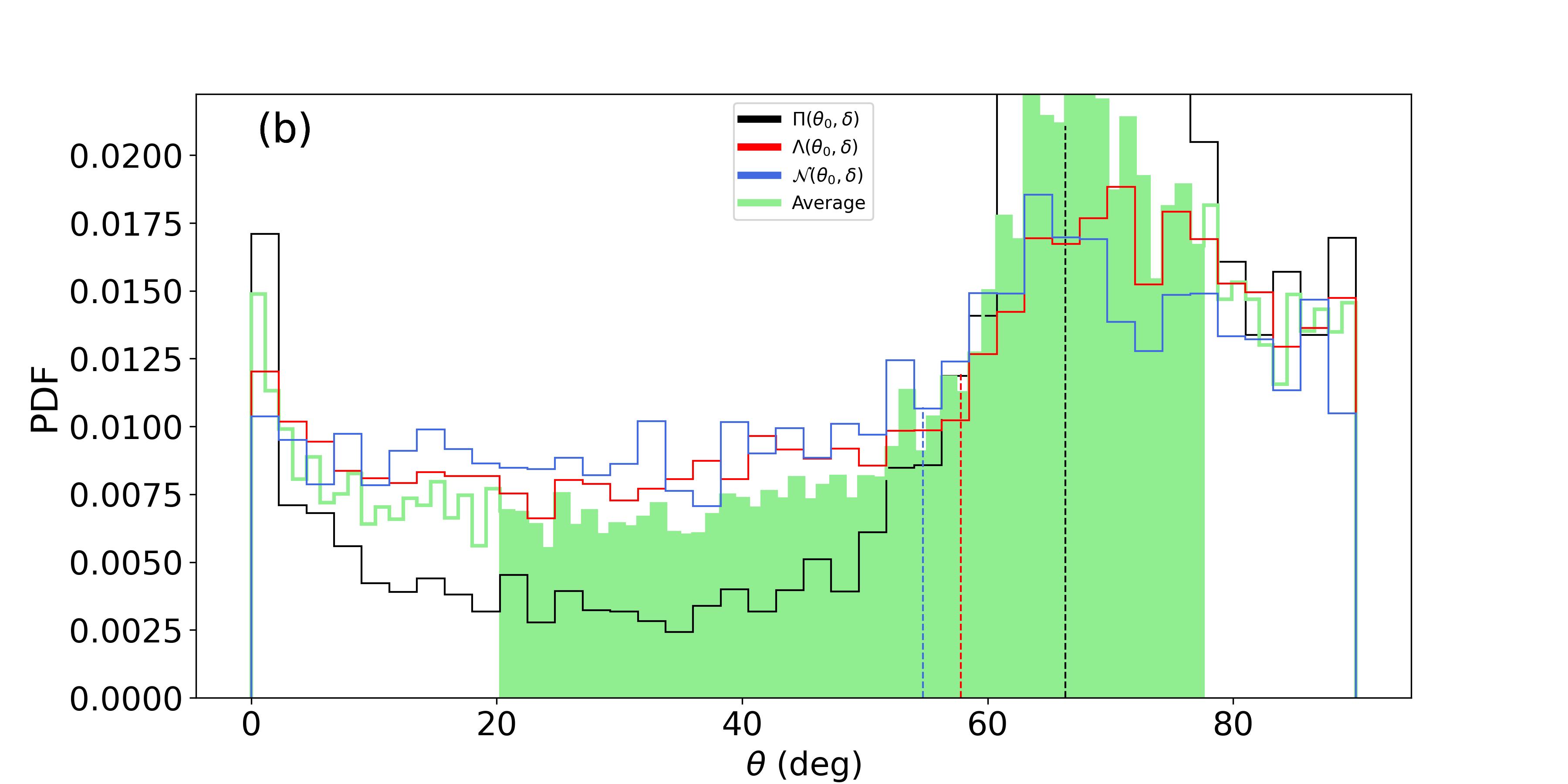}
\includegraphics[angle=0, scale=0.215]{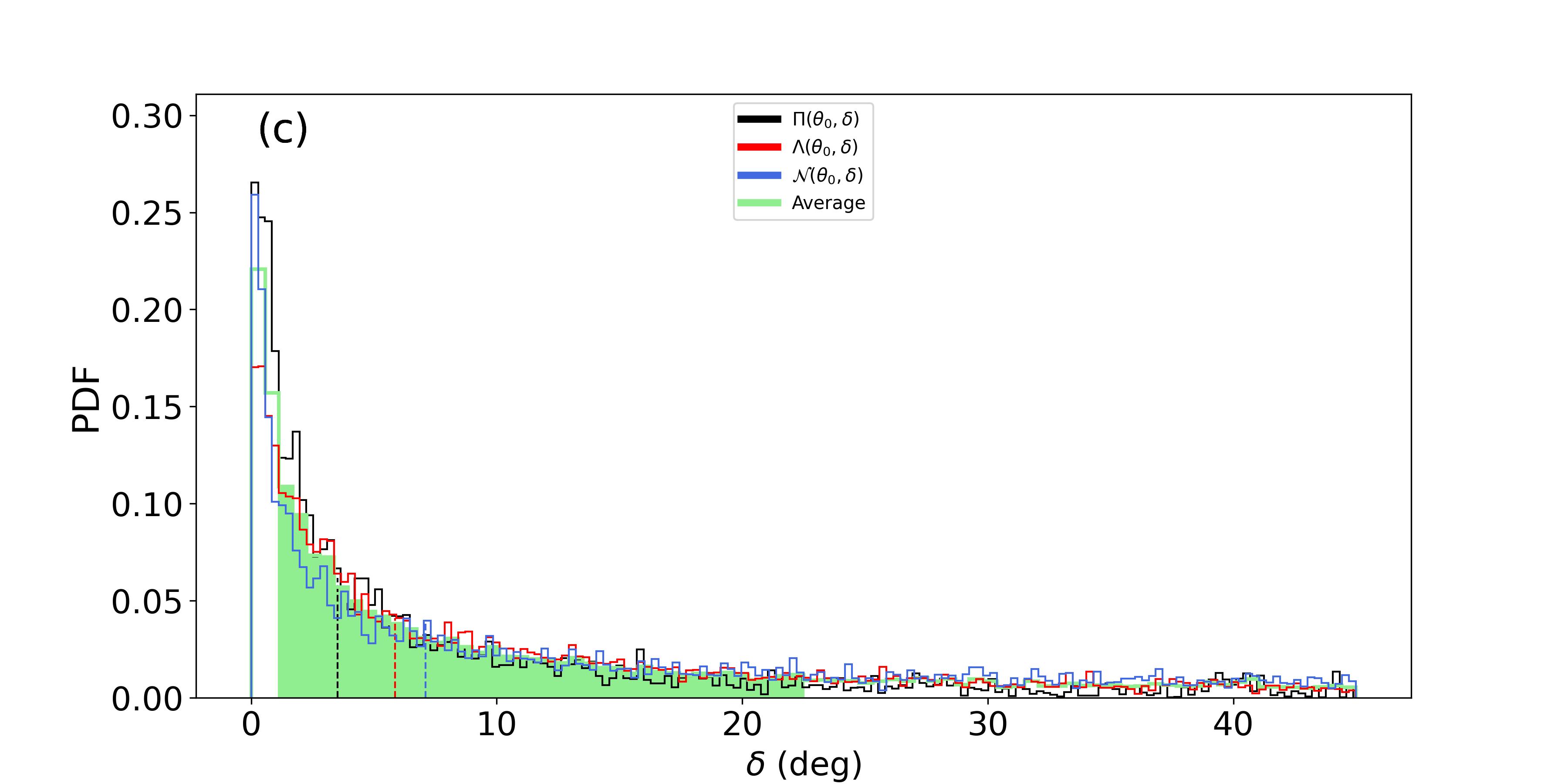} \\
\includegraphics[angle=0, scale=0.215]{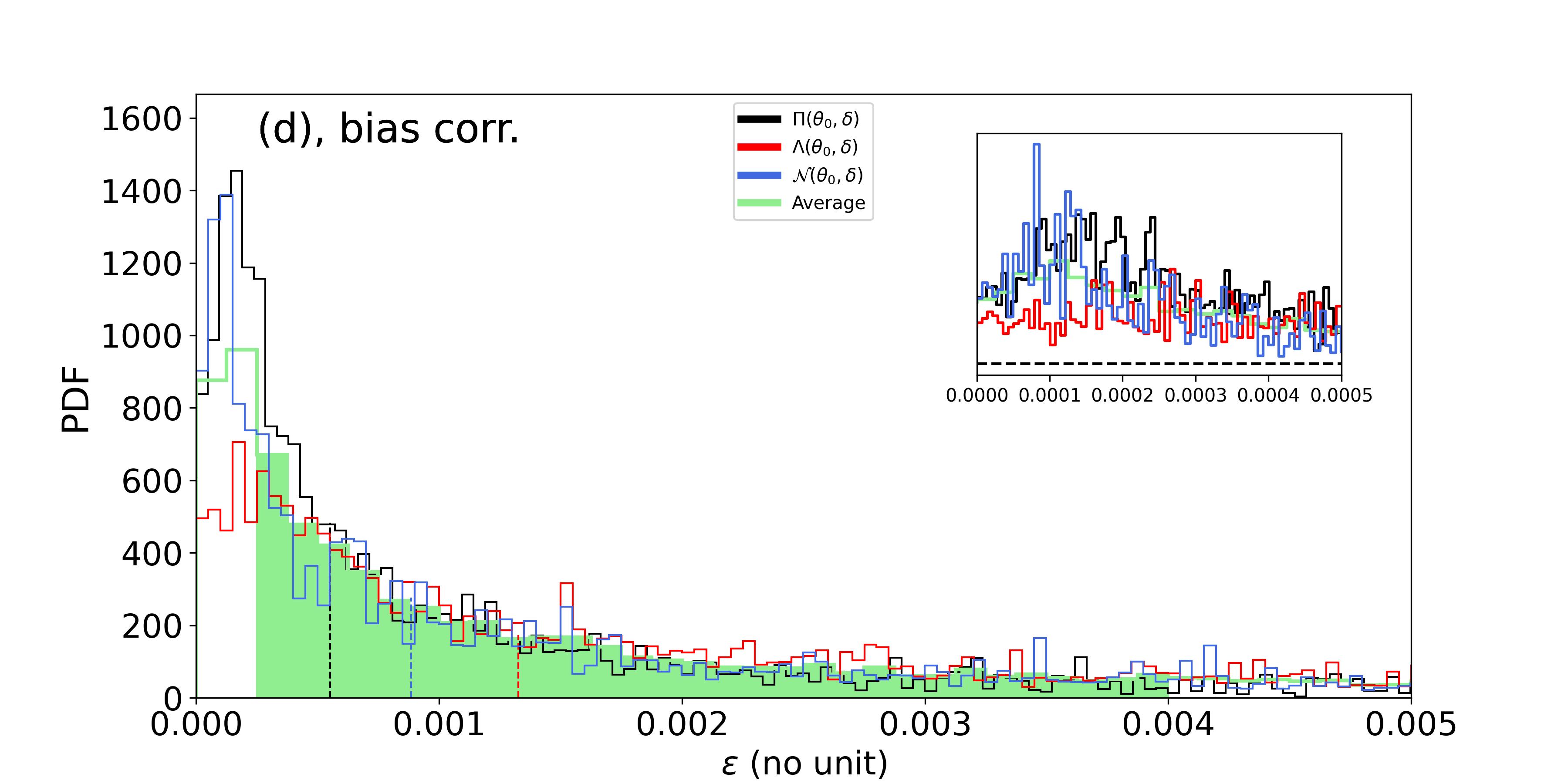}
\includegraphics[angle=0, scale=0.215]{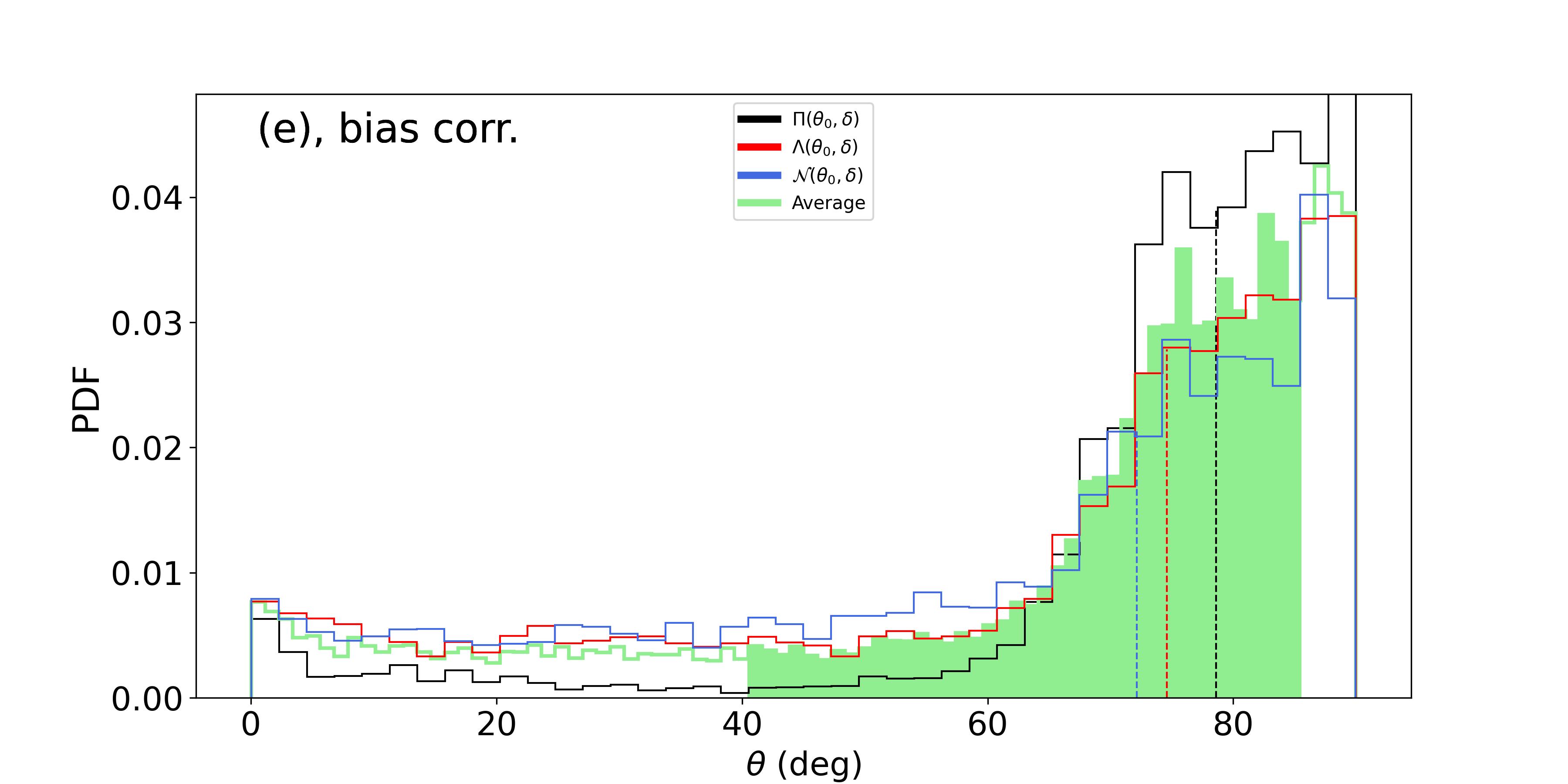}
\includegraphics[angle=0, scale=0.215]{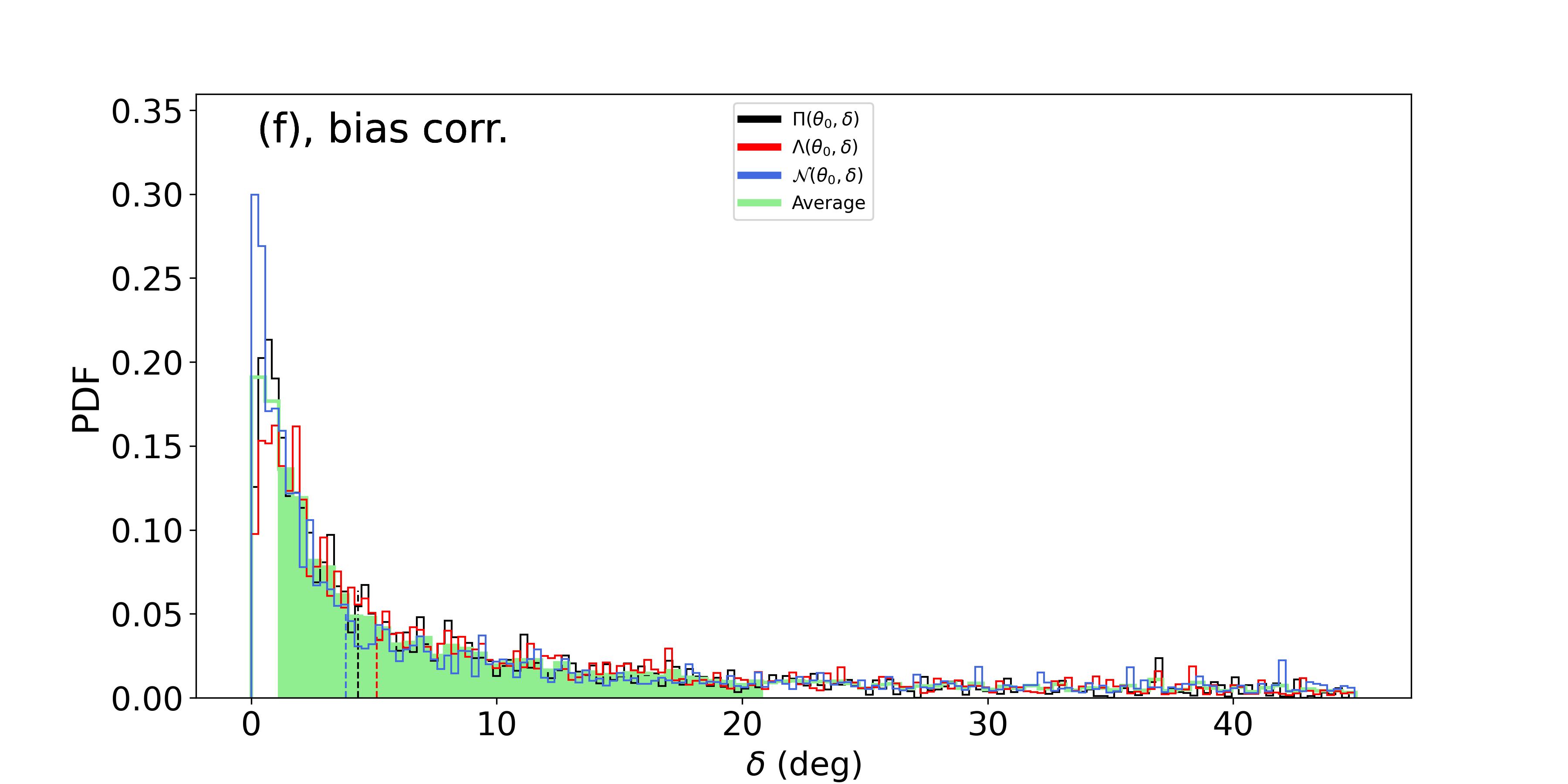}
\caption{{Same as Figure \ref{fig:inference:16cygB} but for the higher solution $a_4=-0.9\pm6.8$nHz.}} \label{fig:inference:16cygB:bias}
\end{figure*}

Figure \ref{fig:inference:16cygB} {and \ref{fig:inference:16cygB:bias}} shows the results for the activity for 16 Cyg B in the two possible scenarios of $a_4$ discussed in Section \ref{sec:16CygB:acoefs} (see Table \ref{tab:real_vals:inference} for the statistical summary). In the case of $a_4=-27.9\pm6.5$nHz
 ({Figure \ref{fig:inference:16cygB}}), the activity is highly significant (greater than $99.3\%$ with or without bias correction, {whatever is the activity zone model}) and has stronger intensity than in the case of the Sun. The active region is then located at $\theta_0 \simeq 58^\circ$ ($\theta_0 \simeq 60^\circ$, after bias correction), ie. at comparable latitudes seen during a maximum of activity of the Sun. The bias correction has negligible effect on the inferred activity latitude. Due to $a_4$ significantly departing from 0, the extension of active region is better constrained than in 16 Cyg A or the Sun, but remains weakly informative.

The second more likely solution (probability of $70\%$) corresponds to $a_4=-1.0\pm6.7$nHz ({Figure \ref{fig:inference:16cygB:bias}}). It is associated to an activity at co-latitudes above $\simeq40^\circ$ and of lower activity. Its statistical significance is low, and the activity intensity may be of the same order or lower than 16 Cyg A. {In fact, it looks similar to the solar case when approaching its minimum of activity.} Accounting for the fiducial bias, the solution is more concentrated to the equatorial region and differs significantly from the lower probability $a_4$ solution. 
Here, the uncertainty on $a_2$ and $a_4$ is again too large to provide a stringent constraint on the extension of the activity zone $\delta$.

\section{Discussion and Conclusion} \label{sec:discuss}
The stellar activity is a complex phenomenon emerging from the interplay between the stellar plasma, rotation and magnetism. The activity distorts the shape of mode cavity, perturbing the pulsation frequencies. This perturbation is here observationally evaluated using a a-coefficient decomposition. This allows us to separate observables (the a-coefficients) from their physical interpretation (the activity modelling). The separation enables us to demonstrate that the measurements of the average low a-coefficient, $a_2$ and $a_4$, under some observational conditions, are sufficient to reveal the presence of a statistically significant activity of similar intensity to the Sun and to determine its latitude. The required observational conditions are analysed using a methodology similar to \cite{Kamiaka2018}, that is, by constructing a grid of artificial power spectra, that allows us to determine the bias for $a_1$, $a_2$ and $a_4$. 
It is found that if the height-noise-ratio exceeds 10, the mode blending factor $f=a_1/\Gamma_{\nu_{max}}$ is greater than 0.4, the inclination is above $30^\circ$ and that the observation is longer than 2 years, the inaccuracy remains mild and generally smaller than the $1\sigma$ uncertainty. The uncertainty and/or the inaccuracy may however become too large to reliably detect any activity beyond the above-specified conditions. In particular and in agreement with \cite{Kamiaka2018}, the stellar inclination is a decisive variable to ensure the accuracy of the measurement. Below $30^\circ$, $a_1$ is often wrong by a factor 2. This is likely due to the fit mistakenly identifying $l=2, m=\pm1$ as $l=2, m=\pm2$ and vice-versa. Other a-coefficients are then severely inaccurate. Therefore, such a bias analysis suggests that rotation studies of stellar ensembles require a careful star selection.

A method that uses the average $a_2$ and $a_4$ coefficients is proposed to perform a subsequent analysis of the activity, considering a geometrical model of the activity effect on the pulsations frequencies and accounting for the stellar asphericity due to the centrifugal effects. This two-step approach is tested on the case of the Sun and shows that it is effectively able to detect the change of activity between the solar maximum of activity around 1999-2002 and the minimum of activity around 2006-2009. Although the use of averaged a-coefficients makes it difficult to evaluate the extension of the activity zone, the model successfully retrieves the mean latitude of activity during the maximum of the solar cycle.

The method is then applied to the brightest stars observed during the initial observational phase of the Kepler space instrument, 16 Cyg A and B. These stars were selected as a test-bed due to the fact that they are well studied and present the highest mode signal-to-noise ratio of all the currently known main-sequence stars. \cite{Davies2015} suggested that these stars have mild to no activity. However, \cite{Bazot2019}, using a parametric model for describing the asphericity $(R_{\mathrm{eq}} - R_{\mathrm{pol}})/R_{\mathrm{eq}}$ found an asphericity significant at $1\sigma$. Our current analysis, performed using the same data set, confirms this asphericity and found a mild (relatively to the Sun) to moderate activity for both stars. It is found that 16 Cyg A has a near-equatorial band of activity during the period of observation
(13 September 2010 to 8 March 2013), with a significance of the detection greater than $79.8\%$. 

\begin{figure}
\begin{center}
\includegraphics[angle=0, scale=0.32]{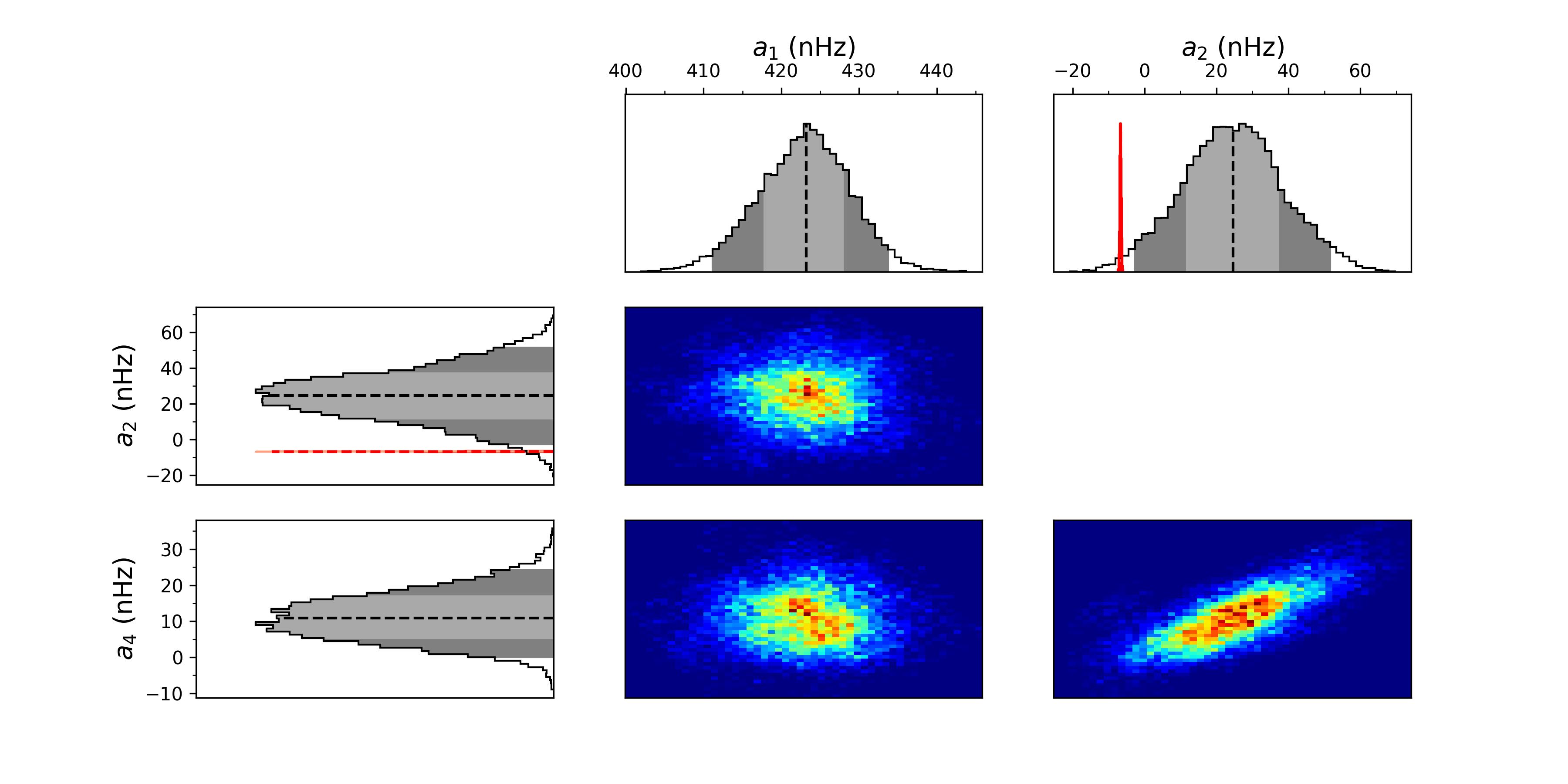}
\caption{Probability Density Functions and their correlations, obtained by MCMC for coefficients $a_1$, $a_2$, $a_3$, $a_4$ for the Sun in-between two activity cycles (2006-2011).
The red curve is the expected $a_2^{CF}$ coefficient for a pure centrifugal distortion. The light and dark gray PDF filling is for the $1\sigma$ and $2\sigma$ confidence interval, respectively.}
\label{fig:Sun:acoefs:20062011}
\end{center}
\end{figure}

The case of 16 Cyg B is more ambiguous. A bi-modality on the average splitting $\langle\delta\nu_{nlm}/m\rangle \simeq a_1$ and on the stellar inclination is already reported in \cite{Davies2015} and \cite{Bazot2019}. Our refined model suggests that this bi-modality is in fact related to the $l=2$ modes. Indeed, as we account for $a_1$, $a_2$, $a_3$ and $a_4$, we note that the bi-modality previously seen on $a_1$ in earlier studies is displaced to $a_4$. The solutions of $a_4$ are separated using a Gaussian process algorithm, which found that the weight (or importance) of the highest solution, close to 0 nHz is of $70\%$. The lower solution, close to -28nHz has a weight of $30\%$. The associated probability distribution for stellar inclination $i=35\pm3$ is unique (instead of being bi-modal in past studies). 
The two solutions of $a_4$ are analysed independently. The lower solution (with lower weight) is associated to an overall activity that is stronger than the Sun, localised at latitudes of approximately {$32^\circ$} (with an uncertainty of $\simeq 3^\circ$). The higher solution is linked to a lower activity and is weakly significant (probability greater than $67.1\%$). Although the uncertainty is large, the study suggests an activity closer to the equatorial region.
 In the Sun and as evident in its butterfly diagram, the quiet phase is associated to a magnetic activity closer to the equator, while the transition to the active Sun is abrupt and characterised by the appearance of magnetic spots at latitudes of $30$-$40^\circ$. 
 In that context and although it is not possible to rule out the possibility of a statistical fluke, an interpretation of the bi-modality is that the star was transitioning from a period of low activity to a more active period during the observation time of the Kepler instrument. 
To evaluate that hypothesis, we selected solar-data between Jan 2006 and Jan 2011, including the end of a cycle and the start of a new one. The measured a-coefficients for that analysis are shown in Figure \ref{fig:Sun:acoefs:20062011}. There is no visible bi-modality on either $a_2$ or $a_4$. Because 16 Cyg B is evidently different than the Sun, this does not refute the hypothesis, but weakens it. A more firm verification would require a follow-up observation of the star, ie with PLATO \citep{PLATO2014}. Or extensive simulations in order to attempt to reproduce the bi-modality.
 
 This work demonstrates that it is possible to determine the latitude and intensity of the activity for Sun-like stars, when this activity is similar to, or exceed the one of the Sun. The limited observed bias suggests that the analysis of the stellar activity could also be extended to a larger set of stars. The Kepler LEGACY stars \citep{Lund2017} are ideal candidates as these generally satisfy the criteria of reliability that are discussed in this conclusion and detailed in Section \ref{sec:bias}.
 
 Although the a-coefficient analysis has the benefit to allow to simplify the bias studies and to make the analysis faster, a further axis of improvement would consist in using a single analysis step, ie, fitting directly the power spectrum with the model of activity. Contrary to the two-step analysis, such an approach does not require assumptions on the properties of probability distribution of the a-coefficients as it is currently the case. It would also make use of the full set of a-coefficients (not only their average), which may lead to smaller uncertainties. However, the two approaches would certainly be needed in tandem, as the stability of the solution (and it's accuracy) is harder to ascertain when performing a single step fitting approach.
 

\begin{acknowledgements}
O.B. and L.G. acknowledge partial support from NYUAD Institute Grant G1502 "Center for Space Science". 
L.G. is supported in part by the
European Research Council (ERC) Synergy Grant WHOLE SUN 810218 and the  Max Planck Society grant "Preparations for PLATO Science".  We thank Saskia Hekker for comments and advice.
\end{acknowledgements}

{All data presented in this paper were obtained from the Mikulski Archive for Space Telescopes (MAST) at the Space Telescope Science Institute. The specific observations analysed can be accessed via \url{https://doi.org/10.17909/T9059R}.}

%

\vspace{5mm}



\begin{appendix}

\section{Polynomials, a-coefficients and splittings}

    \subsection{The first six used polynomials} \label{appendix:Plm}

    The $\mathcal{P}^{(l)}_j(m)$
    coefficients, which are introduced by \cite{Schou1994}, can be obtained by normalising the $H_j^m(l)$ coefficients by \cite{Ritzwoller1991} so that $\mathcal{P}^{(l)}_j(m=l) = l$. 

    \begin{align}
        \mathcal{P}_{j}^{(l)}(m) &= \frac{l(2l-j)!}{(2l)!} H_j^m(l) \nonumber \\
        &= (-1)^{-l+m} \frac{l\sqrt{(2l-j)!(2l+j+1)!}}{(2l)!}
        \begin{pmatrix}
            j & l & l \\
            0 & m & -m
        \end{pmatrix}
        \label{eq:Pjlm}
    \end{align}
in which the last factor means the Wigner $3j$ symbol.

{They can also be obtained by recurrence using equation (\ref{Eq:4}) and (\ref{Eq:4b}), together with the normalisation condition, $\mathcal{P}_j^l(m)=l$ and starting with $\mathcal{P}^l_0 (m) = l$.}
    
The first six polynomials are as follows:
    \begin{equation}
        \mathcal{P}^{(l)}_1(m) = m
    \end{equation}
    \begin{equation}
        \mathcal{P}^{(l)}_2(m) = \frac{3m^2 -l(l+1)}{2l-1}
    \end{equation}
    \begin{equation}
        \mathcal{P}^{(l)}_3(m) = \frac{5m^3 - (3l(l+1)-1)m}{(l-1)(2l-1)}
    \end{equation}
    \begin{equation}
        \mathcal{P}^{(l)}_4(m) = \frac{(35m^4 - 5(6l(l+1)-5)m^2) + 3l(l+1)(l(l+1)-2)}{2(l-1)(2l-1)(2l-3)}
    \end{equation}
    \begin{equation}
        \mathcal{P}^{(l)}_5(m) = \frac{252m^5 -140(2L-3)m^3 + (20L(3L-10) + 48)m}{8(4l^4 - 20l^3 + 35l^2 - 25l + 6)}
    \end{equation}

    \begin{align}
        \mathcal{P}^{(l)}_6(m) &= \frac{924m^6 - 420m^4(3L-7) + 84m^2(5L^2 - 25L + 14)}{N} \nonumber \\
        &\quad - \frac{20L(L^2 - 8L + 12)}{N}
    \end{align}
    where $L=l(l+1)$ and $N=64l^5 - 480l^4 + 1360l^3 - 1800l^2 + 1096l - 240$.
    
    \subsection{Relationship between a-coefficients and splittings} \label{appendix:splittings}

The relationships between a-coefficients and symmetric/anti-symmetric splittings can be derived using equations (\ref{Eq:5:Snlm-Tnlm-1}) and (\ref{Eq:5:Snlm-Tnlm-2}). These are used to convert model frequencies $\nu_{nlm}$ into a-coefficients.

    \subsubsection{Explicit form for low a-coefficients and l=1}
    \begin{equation}
        a_1(n,l=1) = S_{n11} = \frac{\nu_{n,1,1}- \nu_{n,1,-1}}{2}
        \label{Eq:Appendix:a11:T}
    \end{equation}
    \begin{equation}
        a_2(n,l=1) = T_{n11}/3 = \frac{(\nu_{n,l,-1} + \nu_{n,l,1})/2 - \nu_{n,l,0}}{3}
        \label{Eq:Appendix:a21:S}
    \end{equation}
     \subsubsection{Explicit form for low a-coefficients and l=2}
    \begin{equation}
        a_1(n,l=2) =\frac{S_{n21} +4S_{n22}}{5}
        \label{Eq:Appendix:a12:T}
    \end{equation}
    \begin{equation}
        a_2(n,l=2) =\frac{2T_{n22} - T_{n21}}{7}
        \label{Eq:Appendix:a22:S}
    \end{equation}
    \begin{equation}
        a_3(n,l=2) =\frac{S_{n22} - S_{n21}}{5}
        \label{Eq:Appendix:a32:T}
    \end{equation}
    \begin{equation}
        a_4(n,l=2) =\frac{T_{n22} - 4T_{n21}}{70}
        \label{Eq:Appendix:a42:S}
    \end{equation}
     \subsubsection{Explicit form for low a-coefficients and l=3}
    \begin{equation}
        a_1(n,l=3)=\frac{S_{n31} + 4S_{n32} + 9S_{n33}}{14}
        \label{Eq:Appendix:a13:T}
    \end{equation}
    \begin{equation}
        a_2(n,l=3)=\frac{-15T_{n31}  + 25T_{n33}}{126}
        \label{Eq:Appendix:a23:S}
    \end{equation}
    \begin{equation}
        a_3(n,l=3)=\frac{-S_{n31} - 2S_{n32} + 3S_{n33}}{9}
        \label{Eq:Appendix:a33:T}
    \end{equation}
    \begin{equation}
        a_4(n,l=3) =\frac{T_{n31} - 7T_{n32} + 3T_{n33}}{77}
        \label{Eq:Appendix:a43:S}
    \end{equation}
    \begin{equation}
        a_5(n,l=3) = \frac{5S_{n31} - 8S_{n32} + 3S_{n33}}{126}
        \label{Eq:Appendix:a53:T}
    \end{equation}
    \begin{equation}
        a_6(n,l=3)=\frac{15T_{n31}  -6T_{n32} + T_{n33}}{1386}
        \label{Eq:Appendix:a63:T}
    \end{equation}
    
 \section{Spectrum analysis} \label{appendix:spectrum_analysis}
    \subsection{Acoustic-spectrum modelling}

The analysis of the asteroseismic data is very often performed by fitting the power spectrum \citep[eg.][]{Appourchaux1998,Appourchaux2008}. This requires to devise a likelihood function and a model for the observed spectrum. As our fitting involves a Bayesian approach, the priors are also required. 

\subsubsection{Likelihood}
The likelihood function is determined by the noise statistics of the power spectrum, which is a $\chi$-squared with two degree of freedom. It is also based on the assumption that the frequency bins are independent and uncorrelated. This implies that the observation duration is assumed to be much greater than the typical lifetime of the pulsation modes and that the duty cycle is sufficient to allow us to neglect any leakage induced by the window function. The observation timeframe for the LEGACY sample is of the order of years, which is significantly longer than the mode lifetime. Furthermore, the duty cycle is above $95\%$, which was shown to be sufficient to neglect leakage \citep{Stahn2010PhD}. Under these conditions, the likelihood function is \citep{Anderson1990},
\begin{equation}
    L_1(S(\nu_i)|\boldsymbol{X}, M)=\sum^{N-1}_{i=0} \frac{1}{S(\nu_i)} \exp{\left(-\frac{S(\nu_i)}{M(\nu_i,\boldsymbol{X})}\right)}
\end{equation}
where $S(\nu_i)$ is the power at the central frequency $\nu_i$ of the $i^{th}$ bin. Here, $N$ is the total number of bins and $M(\nu_i, \boldsymbol{X})$ is the model of the power spectrum with the set of variable $\boldsymbol{X}$ for the model $M$.

From Bayes theorem, the posterior probability density function is defined as,
\begin{equation}
    \pi_1(\boldsymbol{X} \vert S(\nu), M) = \frac{ \pi_1(\boldsymbol{X} \vert M)\,L_1(S(\nu) \vert \boldsymbol{X}, M)}{\pi_1(S(\nu) \vert M)}
\end{equation}
where $\pi_1(\boldsymbol{X}|M)$ is the prior knowledge on the parameters $\boldsymbol{X}$. $\pi_1(S(\nu)|M)$ is a normalisation constant, essential only for comparing the significance of models \citep[eg.][]{Gregory2005,Benomar2009}. Variables are assumed to be independent from each others.
This implies that the joint prior is the product of individual priors.
The posterior is evaluated using the Tempered Adaptive MCMC code\footnote{\url{https://github.com/OthmanB/Benomar2022/Programs/TAMCMC-1.81}} described in \cite{Atchade2006a} and implemented by \cite{Benomar2008}.

\subsubsection{Acoustic spectrum model} \label{method:model}
The model used for the power spectrum fitting, involves a sum of asymmetrical Lorentzian, superimposed to a monotonically decreasing function of frequency (pink noise). The asymmetric Lorentzian is commonly used in helioseismology \citep{Duvall1993, Nigam1998, Georgobiani2000, Toutain1998} and in asteroseismology for Sun-like stars \citep{Benomar2018ApJ},
 \begin{equation}
    M(\nu) = \sum^{n_{max}}_{n=n_0} H_{nlm} \frac{(1 + B_{nlm}\,z)^2 + B^2_{nlm}}{1 + z^2} + N(\nu),
\end{equation}

with $z=2(\nu -\nu_{nlm})/\Gamma_{nlm}$. Each asymmetrical Lorentzian is defined by a height $H_{nlm}$, width $\Gamma_{nlm}$, a central frequency $\nu_{nlm}$ and an asymmetry $B_{nlm}$. 
As explained by \cite{Gizon2006,Benomar2018ApJ}, the asymmetry coefficient depends on the mode width and frequency. The normalised asymmetry coefficient $\chi_{nlm}= 2 \frac{B_{nlm}\nu_{nlm}}{\Gamma_{nlm}}$ is fitted instead as it is nearly constant over the range of fitted modes. 

Several prescriptions exist for describing the noise background $N(\nu)$. Here, it is assumed to be a sum of two generalised Lorentzian, sometimes referred as Harvey-like profiles \citep{harvey1985}, and of a white noise,
\begin{equation} \label{Eq:Noise}
    N(\nu)= \frac{A_1}{1 + (\tau_1 \nu)^{p_1}}  + \frac{A_2}{1 + (\tau_2 \nu)^{p_2}} +N_0.
\end{equation}
Here $N_0$ is the white noise and $A_k$ is the maximum heights of the $k^{th}$ generalised Lorentzian. The $\tau_k$ parameter is timescale that is the inverse of the full width at half maximum of the Lorentzians, and $p_k$ is a power exponent.

The fit is performed globally over all the statistically significant peaks visible in the power spectrum of a star. This can lead to a very large number of fitted parameters (up to a few thousands), which is practically unsuitable. A model simplification is therefore preferred, similar to eg. \cite{Appourchaux2008,Benomar2009,Campante2011, Handberg2011} with main simplifications recalled hereafter,
    \begin{itemize}
        \item[-] The m dependence on heights is controlled by its relationship with the stellar inclination \citep{Gizon2003}. This saves several hundreds of parameters.
        \item[-] The relative height of the different degree  $l$ is constant across the fitted range and for a given $l$, hence $V^2_{l} = H_{n,l}/H_{n,l=0} = const$. Only the $l=0$ heights $H_{n,l=0}$ are variables. The mode visibility $V^2_{l}$ replaces the $H_{n,l>0}$ as a variable.
        \item[-] Given a degree $l$, $\Gamma_{n,l,m} = \Gamma_{n,l}$ is imposed. This is justified by the fact that the width depends weakly on the frequency: all split components are assumed to have the same width.
        \item[-] Because $\Gamma_{n,l}$ is nearly independent of the degree, it is possible to fit $\Gamma(\nu) = \Gamma_{n, l=0}$ and interpolate it to the frequencies of the modes with degree $l>0$.
    \end{itemize}
These assumptions reduce the number of variables to a few tens in the case of CoRoT, Kepler or TESS observations.
The frequencies of the modes follow equation (\ref{Eq:9}) in Section \ref{sec:splittings:3}. 

    \subsection{Priors} \label{appendix:priors}
Priors are fundamental of a Bayesian method. This section explicits $\pi_1(\boldsymbol{X} \vert M, I)$, the prior used during the power spectrum fitting. Parameters of the vector $\boldsymbol{X}$ are assumed independent to each others, such that the product rule is used to the define $\pi_1(\boldsymbol{X} \vert M, I)$.

\begin{table*}[htp]
    \centering
        \caption{Mode priors used for the fit of the power spectrum. Prior parameters on frequencies are visually determined, see Section \ref{appendix:priors}.}
    \begin{tabular}{|c|c|c|c|c|c|c|}
        \multicolumn{7}{|c|}{Main mode parameters} \\ \hline
        Height ($ppm^2/\mu$Hz)    & Width ($\mu$Hz)    & Frequency ($\mu$Hz)   & Asymmetry ($\vert \chi \vert$) & $V^2_{l=1}$ & $V^2_{l=2}$ & inclination \\ \hline
        $\mathcal{J}$(1,1000)  & $\mathcal{J}$(0.1, 45) & $\mathcal{U}$($\nu_{\mathrm{min}}$, $\nu_{\mathrm{max}}$)        & $\mathcal{J}$(5, 100) & $\mathcal{G}$(1.5, 0.15) & $\mathcal{G}$(0.53, 0.053) & $\mathcal{U}$(0,90) \\ \hline \hline
    \end{tabular}
    \label{tab:priors}
\end{table*}

Heights, widths, frequencies, asymmetry and inclination use non-informative priors. These are either Jeffreys priors for scale parameters, noted $\mathcal{J}(x_{\mathrm{min}}, x_{\mathrm{max}})$ or uniform prior, noted $\mathcal{U}(x_{\mathrm{min}}, x_{max})$. 
Priors on frequencies are uniform and require a visual inspection of the power spectrum in order to assign the lower and upper bound of the prior for each mode that follow the expected pattern for main sequence Sun-like stars (equally spaced p modes) and that show an excess of power relative to the background that exceeds $80\%$. The excess of power is determined using a smooth spectrum for which the noise statistics is derived by \cite{Appourchaux2003}.
Mode visibilities are defined by Gaussian priors (noted $\mathcal{G}(x_0, \sigma)$), with mean $x_0$ set as the solar value \citep[see][]{ballot2011} and the standard deviation $\sigma$ is $10\%$ of the mean. Table \ref{tab:priors} lists the type and the prior characteristic values that are used for the parameters of the modes.

The prior on $a_1$ is uniform between 0 and 1500 nHz. For 16 Cyg A and B, an uniform fixed prior is set on $\vert a_3 \vert$ over the range [0, 100] nHz. However at each iteration of the optimisation process, $\vert a_3 \vert$ is not allowed to exceed $20\%$ of $a_1$. Preventing extremely large relative value of $a_3/a_1$, improves the fit stability and ensure a faster convergence of the algorithm. For the Sun analysis and because it is not possible to measure $a_3$ for $i\simeq 90^\circ$ (due to the lack of amplitude of $l=2, m=\pm 1$ at that inclination), $a_3$ is fixed to 0.

The priors on $a_2$ and $a_4$ are also uniform. The range is defined by using the maximum range of Figure \ref{fig:aj_mean} (showing $a^{(AR)}_2$ and $a^{(AR)}_4$), increasing it by $50\%$ and adding the expected centrifugal term $a^{(CF)}_2$, assuming $a_1=400$nHz for the Sun and $a_1=600$nHz for 16 Cyg A/B. The $\Delta\nu$ reported in Table \ref{tab:real_vals} is also used. For the same reason as to $a_3$, $a_2/a_1$ cannot exceed $50\%$ and $a_4/a_1$ cannot exceed $20\%$ at each iteration step of the optimisation process.

The noise priors are obtained from a global MAP approach similar to \cite{Benomar2012a}. This provides the best fit values and $1\sigma$ uncertainties that are used as priors. The noise background of the individual model analysis is described by equation (\ref{Eq:Noise}). During the global MAP fit, the model is made of that same background model, plus a Gaussian envelope to account for the power excess due to the modes \citep[eg][]{Mathur2010,Huber2011}.

    \section{Inference of the active latitudes } \label{appendix:aj_analysis}
 The determination of the activity is performed as a second step using  as observables the marginalised posterior distribution of the average a-coefficients. This section describes the method and assumptions to determine the posterior distribution and to compute the significance of the detection.
 
    \subsection{Posterior distribution and likelihood}
 
 Similarly as to the power spectrum fitting described in Section \ref{appendix:spectrum_analysis}, the determination of the most likely latitudes for the activity and its significance necessitate the computation of a posterior distribution,
 \begin{equation} \label{Eq:posterior:activity}
    \pi_2(\boldsymbol{X} \vert \boldsymbol{O}, M) = \frac{ \pi_2(\boldsymbol{X} \vert M)\,L_2(\boldsymbol{O} \vert \boldsymbol{X}, M)}{\pi_2(\boldsymbol{O} \vert M)},
\end{equation}
with $\pi_2(\boldsymbol{X} \vert \boldsymbol{O}, M)$, $\pi_2(\boldsymbol{X} \vert M)$, $L_2(\boldsymbol{O} \vert \boldsymbol{X}, M)$ are the posterior distribution, the prior, the likelihood, respectively. The denominator $\pi_2(\boldsymbol{O} \vert M)$ is the normalisation constant used for model comparison (marginal likelihood).
 The posterior distributions of the a-coefficients $a_{2,o}$ and $a_{4,o}$ obtained by power spectrum fitting are the observables contained in $\boldsymbol{O}$. The class of the model is identified by the variable $M$.
 
 For simplicity, the observables are assumed to be distributed according to un-correlated Gaussian functions, such that $\boldsymbol{O}=\{a_{2,o}, a_{4,o}, \sigma_{2,o}, \sigma_{4,o}\}$, where  $a_{2,o}$ and $a_{4,o}$ are the mean for the distributions of $a_2$ and $a_4$ while $\sigma_{2,o}$, $\sigma_{4,o}$ are the standard deviations.
 This leads to a log-Likelihood in the form of a $\chi$-squared,
 \begin{equation} \label{Eq:Likelihood:Activity:General}
     \ln\,L_2(\boldsymbol{X} \vert \boldsymbol{O}, M) = -\frac{(a_{2,o} - a_{2,m}(\boldsymbol{X}))^2}{2 \sigma_{2,o}^2} - \frac{(a_{4,o} - a_{4,m}(\boldsymbol{X}))^2}{2 \sigma_{4,o}^2}.
 \end{equation}

{In \cite{Dziembowski2000}, a-coefficients are weighted using the inverse of the mode inertia (see their equation 4). This is due to the fact that the inertia account for most of the frequency-variations of the coefficients. It is essentially relevant if the uncertainties are small enough to observe a trend when 
modes are fitted individually. However, in the case of the Sun and with either GOLF or BiSON data, it is difficult to perceive a frequency-trend \citep{Chaplin2003}. Our own trials on VIRGO/SPM using a linear fit to describe the frequency dependence of a-coefficient did not detect a slope that is significant at more than $1\sigma$. A similar analysis on 16 Cyg A and B and on simulations showed evidence of large uncertainties when attempting to determine frequency-variations for stars with HNR typical of Kepler observations. In fact, only the average the a-coefficient over $l$ and $n$ is shown to be robustly determined (see also our discussion Section \ref{sec:3.3}) so that inertia effects are here neglected.}

The activity model $M_{AR}$ depends on the variables $\boldsymbol{X}=\{\epsilon, \theta_0, \delta\}$ and $a_{2,m}(\boldsymbol{X}$), $a_{4,m}(\boldsymbol{X})$ denote the modelled a-coefficients. The log-likelihood is then,
    \begin{align} \label{Eq:Likelihood:Activity:AR}
        \ln\,L_2(a_{2,o}, a_{4,o}, \sigma_{2,o}, \sigma_{4,o} \vert a_{2,m}, a_{4,m}, M_{AR}) = \nonumber \\
        &-\frac{(a_{2,o} - a_{2,m}(\epsilon, \theta_0, \delta))^2}{2 \sigma_{2,o}^2} \nonumber \\
        &- \frac{(a_{4,o} - a_{4,m}(\epsilon, \theta_0, \delta))^2}{2 \sigma_{4,o}^2}.
    \end{align}
    
{In our case, the observables are }the average of the fitted modes coefficients, $a_{2,o} = a_2^{(CF)} + a_2^{(AR)}$ and $a_{4,o} =a_4^{(AR)} = 0$. However, to conveniently propagate all errors on the parameters of $a_2^{(CF)}$ ($a_1$, $\Delta\nu$, $\Delta\nu_{\odot}$, $\rho_{\odot}$, $\nu_{nl}$), it is preferable to use $a_{2,o} = a_2^{(AR)}$, obtained by subtracting the $a_2^{(CF)}$ from the $a_2$ measured by power spectrum fitting. 
The distribution of $\Delta\nu$  is computed by linear fitting of the group-wise ensemble of samples $(n,l=0)$ for each measured frequencies $\nu_{n0}$. The term $a_2^{(CF)}$ is finally obtained by weighted average of all independent $a_{2}^{(CF)}(n,l)$ computed at the star's posterior frequencies $\nu_{nl}$.
The subtraction of $a_2^{(CF)}$ is then again performed using the samples of its posterior, enabling to construct of the posterior probability distribution function of $a_2^{(AR)}$, from which we deduce its mean $a_{2,o}$ and its standard deviation $\sigma_{2,o}$.  
To evaluate the relevance of the activity, two models are considered. First, a model ($M=M_{CF}$) without activity, that accounts only for the centrifugal effects and second, a model ($M=M_{AR}$) with activity. Their details and the choice of the priors is described in the following sections for each of them.

\subsection{Model without activity $M_{CF}$}

The model $M_{CF}$ has no unknown variable that require minimisation, see equation (\ref{Eq:8}). For the sake of the model comparison with the model $M_{AR}$, it is however important to determine the marginal likelihood of $M_{CF}$. 
 From that perspective, it is necessary to calculate the denominator of equation (\ref{Eq:posterior:activity}), $P(\boldsymbol{O} \vert M_{CF})$ which is defined as an integral,
 \begin{equation}
    P(\boldsymbol{O} \vert M_{CF}) = \int \pi(\boldsymbol{X} \vert M_{CF}) L_2(\boldsymbol{O}\vert a_{2,m}, a_{4,m}, M_{CF})\, d\boldsymbol{X}.
\end{equation}

 An absence of activity correspond to the limit case where the a-coefficients of the activity are exactly 0. This corresponds to setting Dirac priors on $a_{2,m}=\delta_0$ and $a_{4,m}=\delta_0$ and implies that the marginal likelihood is the local value of the likelihood at $\boldsymbol{X}=\{0,0\}$. 
 The model comparison is commonly performed in log-space such that,
 \begin{equation}
     \ln\,P(a_{2,o}, a_{4,o}, \sigma_{2,o}, \sigma_{4,o}  \vert M_{CF}) = -\frac{a_{2,o}^2}{2 \sigma_{2,o}^2} - \frac{a_{4,o}^2}{2 \sigma_{4,o}^2}.
 \end{equation}

\subsection{Model with activity $M_{AR}$}

 Regarding the model with activity of equation (\ref{Eq:Likelihood:Activity:AR}), priors must then be set on the variables subject to optimisation. The $\epsilon$ and $\delta$ parameters are indispensable intensive parameters and the adequate non-informative prior is then the (truncated) Jeffreys prior \citep{Jeffreys1961}, uniform in the log-space,
 \begin{equation}
     J(x) = \frac{ln(1+ x_{max}/x_{min})}{x + x_{min}},
 \end{equation}
 where $x_{min}$, $x_{max}$ are upper and lower
bounds such that if $x > x_{max}$ or $x < x_{min}$ then $J(x) = 0$.
This guarantees that the probability density is proper (the integral over x is finite). In case of weak information content in the observables, this prior is more weighted toward a null-value. We set $\epsilon_{min}=5.10^{-4}$ and $\epsilon_{max} = 10^{-2}$. This embraces the solar value $\epsilon \simeq 5.10^{-4}$ \citep{Gizon2002AN}.
Because at the maximum of solar activity, $\delta \simeq 10^\circ$,  we set $\delta_{\mathrm{min}}= 10^\circ$. Note also that consistently with the discussion on Figure \ref{fig:Butterfly} of Section \ref{sec:information_content}, $\delta_{\mathrm{max}}= 45^\circ$ is required.

 Finally, the location parameter $\theta_0$ has an uniform prior in the range $[0,90] (^\circ)$. 
 
 The marginal likelihood of the model $M_{AR}$ requires us to evaluate the triple integral,
\begin{equation} \label{Eq:GlobalLikelihood:activity}
    P(\boldsymbol{O}  \vert M) = \int \pi(\epsilon)\, \pi(\theta)\,\pi(\delta) L_2(\boldsymbol{O} \vert a_{2,m}, a_{4,m}, M)\, d\epsilon\, d\theta \, d\delta.
\end{equation}

The MCMC process used here involves the use of parallel Metropolis-Hasting tempered chains. The chains are mixing each other in order to enhance the sampling. As explained in \citealt[][see their Section A.3]{Benomar2009}, these parallel chains can be used to approximate the equation (\ref{Eq:GlobalLikelihood:activity}). This technique is here used with 10 parallel chains following a geometrical temperature law $T_{k}= 1.7^{k-1}$, with $k$, the chain index such that $k=1$ is the target distribution, given by equation (\ref{Eq:posterior:activity}). 

\section{Bias map for $\widehat{HNR}=10$ and $\widehat{HNR}=20$ } \label{appendix:extra_bias_maps}
\begin{figure*}[htp]
\includegraphics[angle=0, scale=0.37]{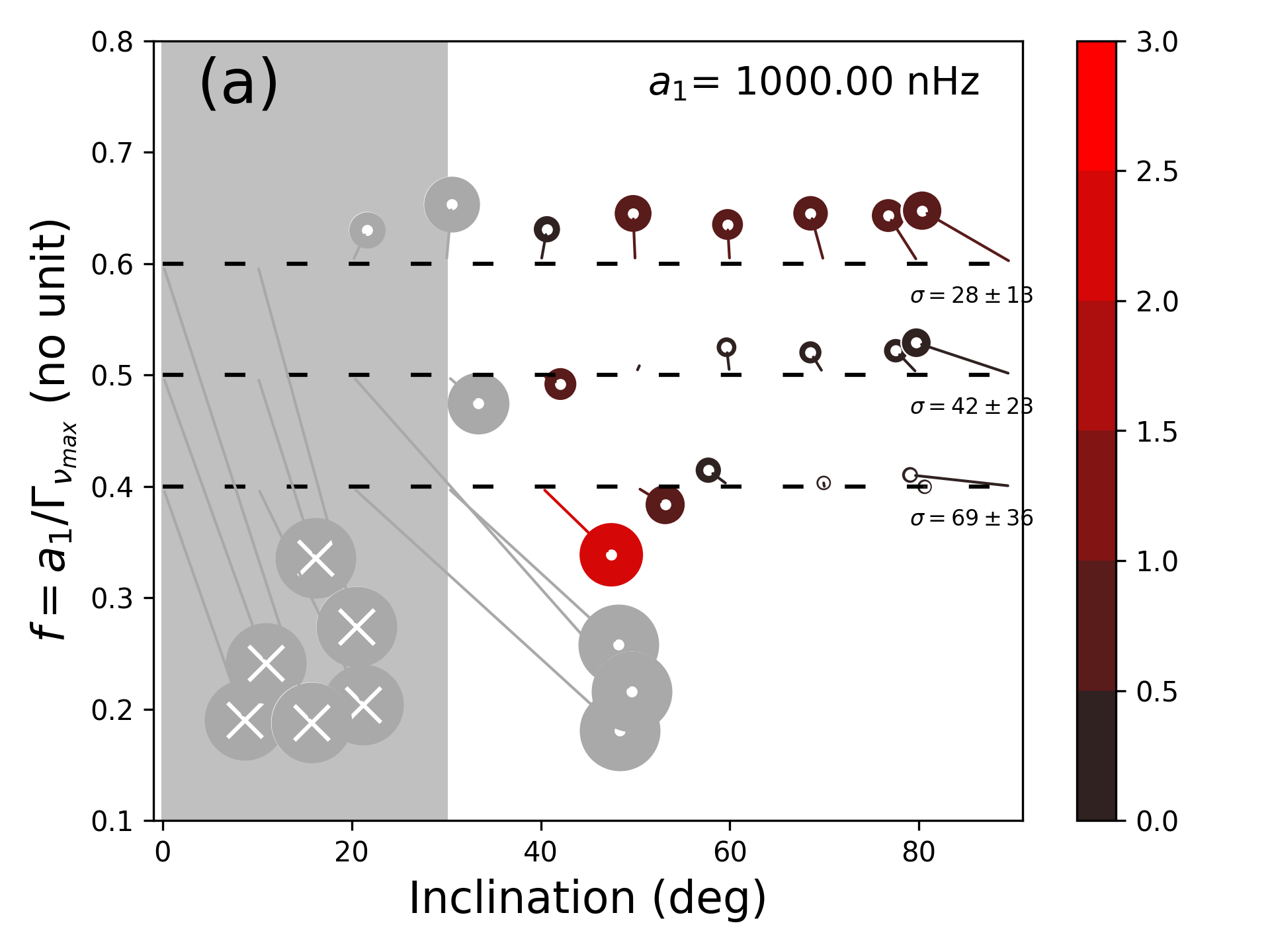}
\includegraphics[angle=0, scale=0.37]{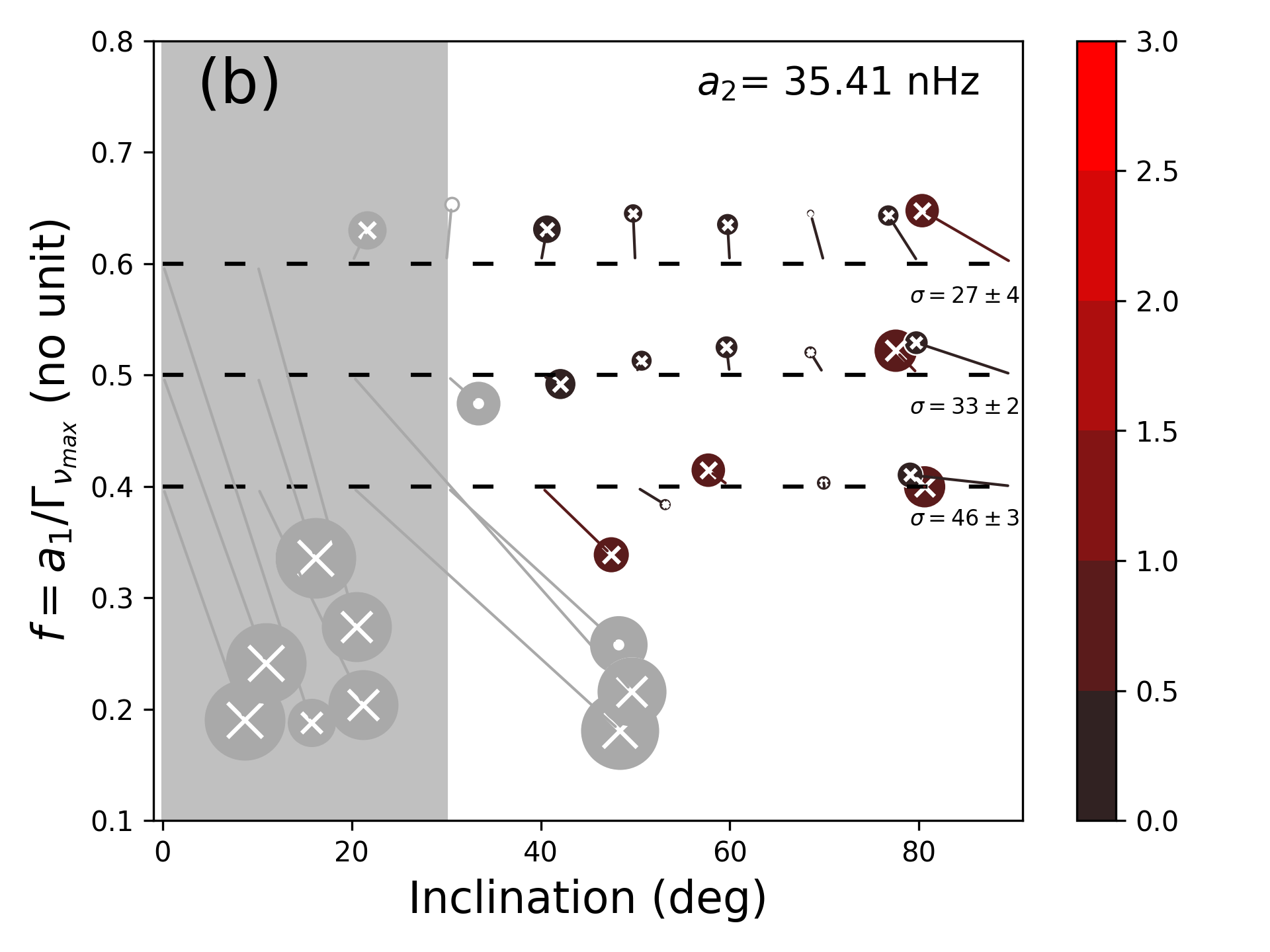} 
\includegraphics[angle=0, scale=0.37]{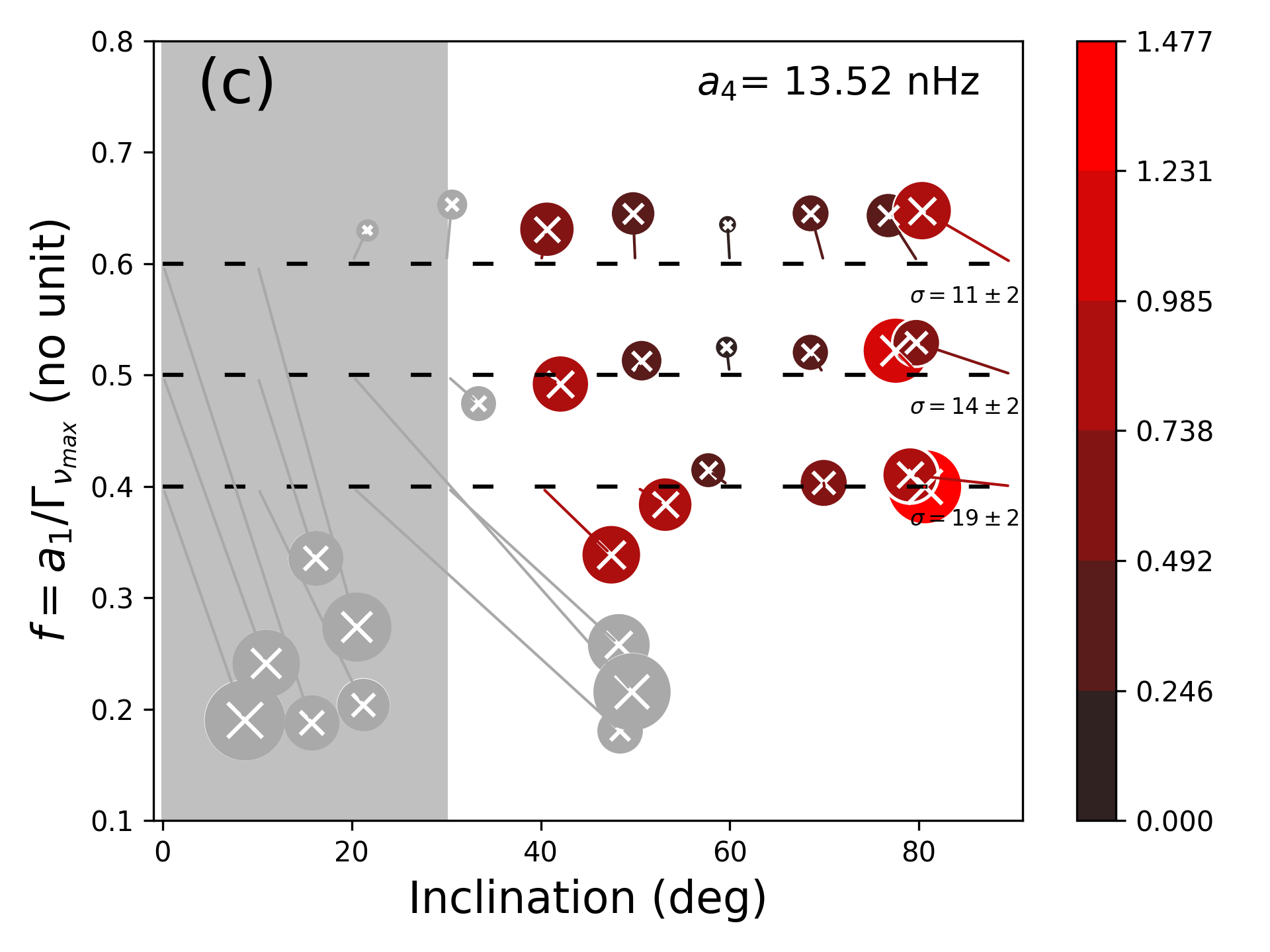} \\
\includegraphics[angle=0, scale=0.37]{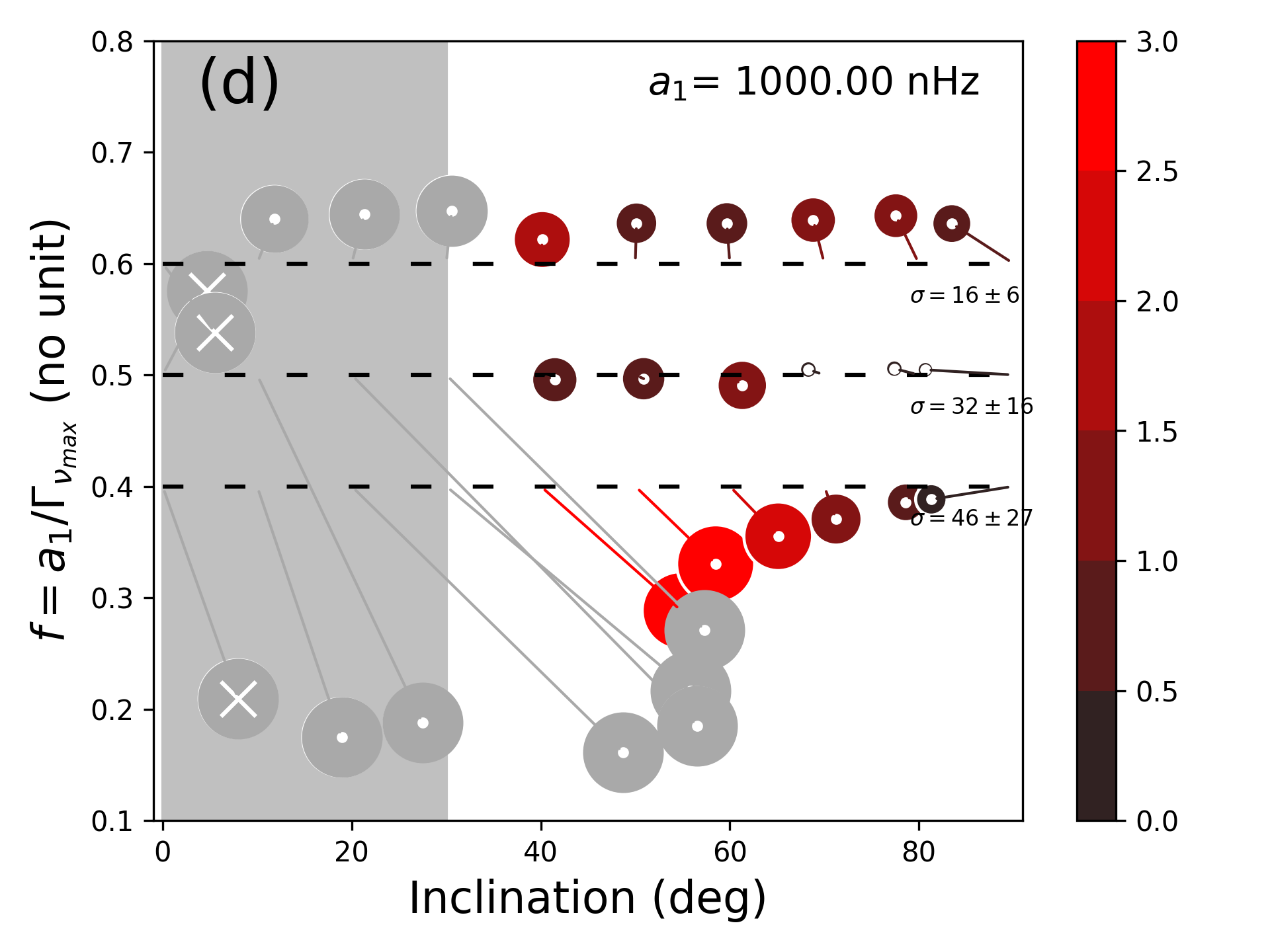} 
\includegraphics[angle=0, scale=0.37]{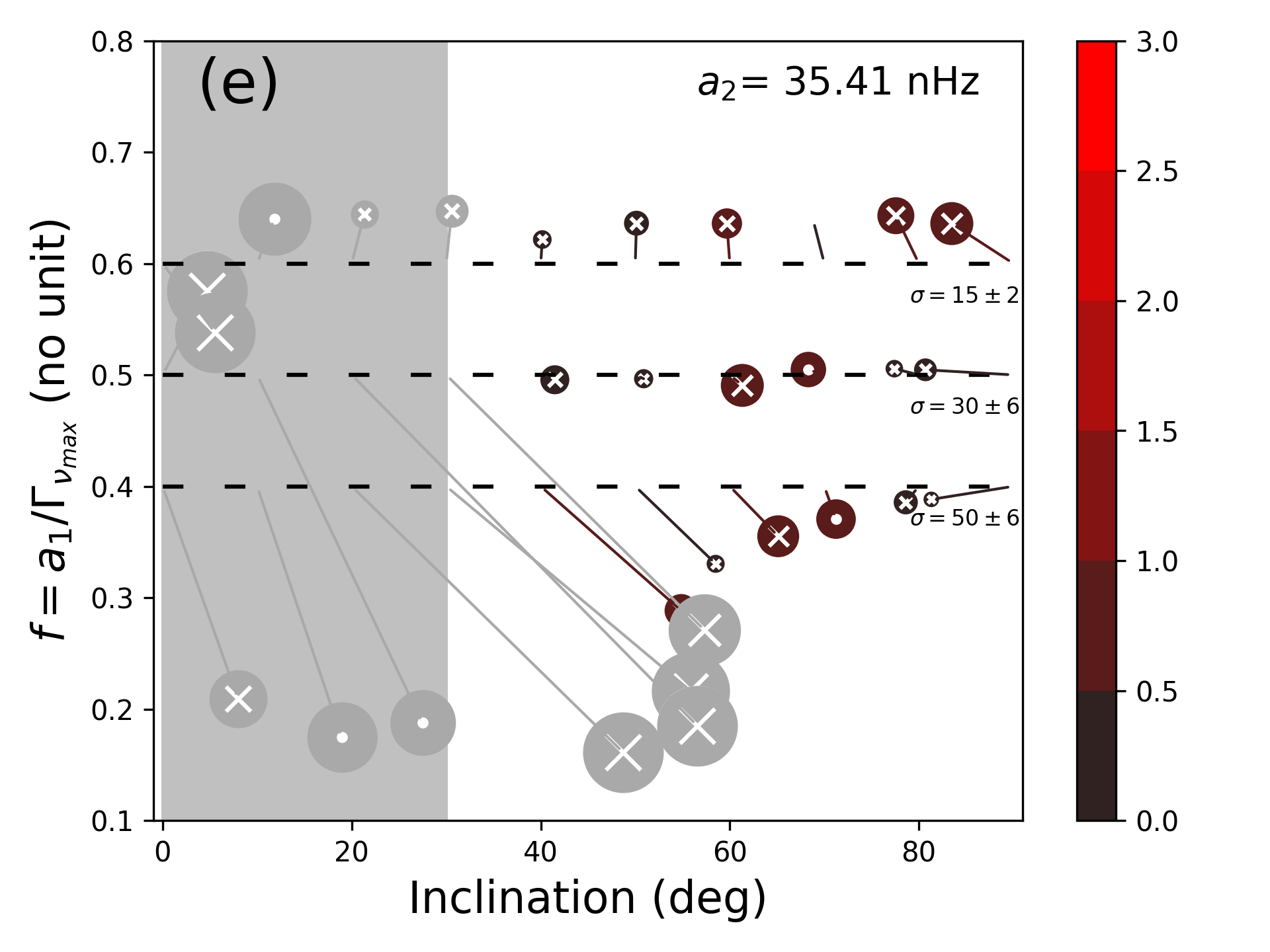}
\includegraphics[angle=0, scale=0.37]{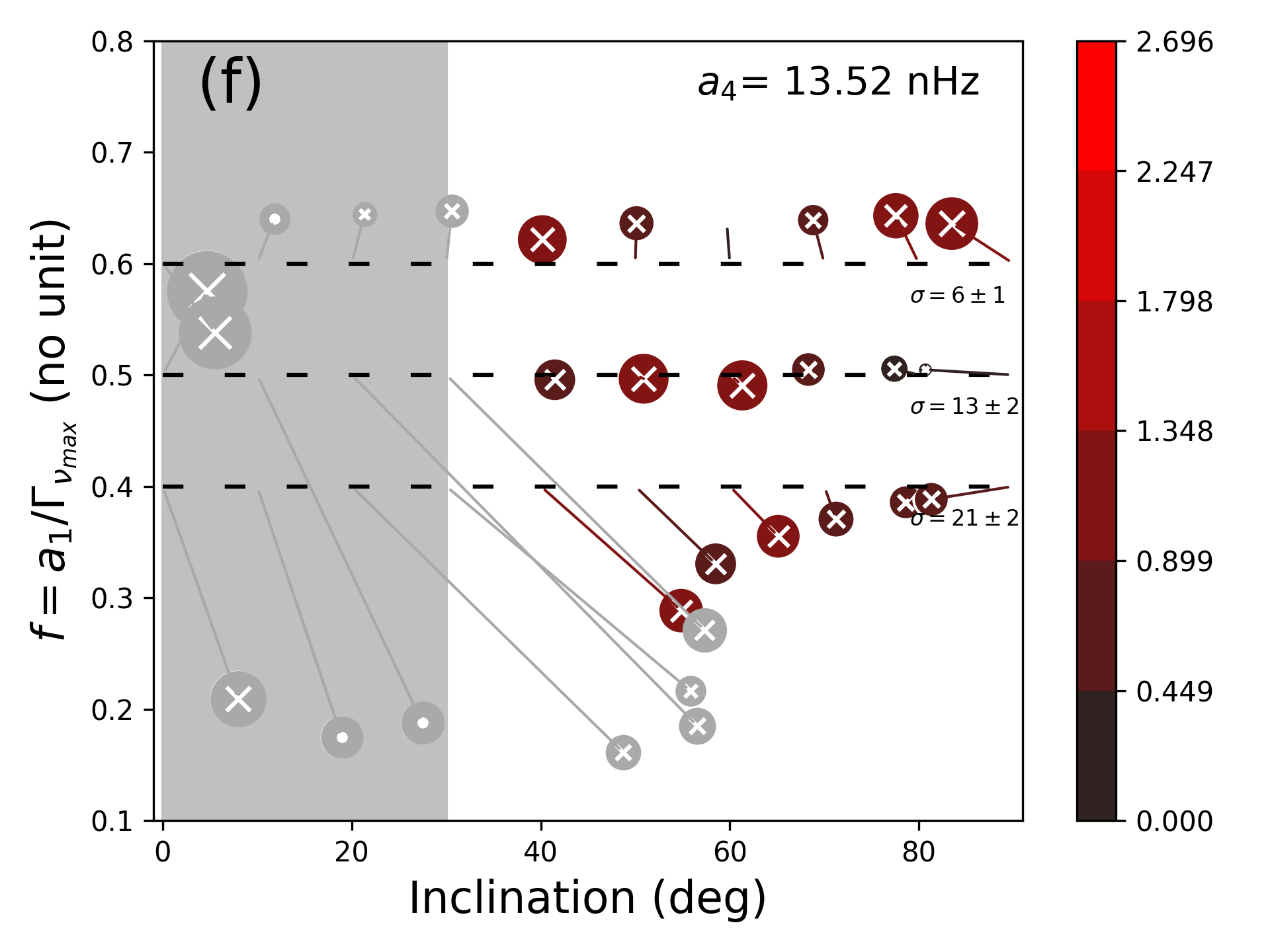}
\caption{Bias analysis for $\widehat{HNR}=10$ for an equatorial activity band ($\theta_0=85$,$\delta=10$) of similar intensity to the Sun ($\epsilon_{nl} = 5.10^{-4}$), for $T_{obs} = 2$ years (top) and $T_{obs} = 4$ years (bottom).}
\label{fig:bias:hnr10:Eq}
\end{figure*}

\begin{figure*}[htp]
\begin{center}
\includegraphics[angle=0, scale=0.37]{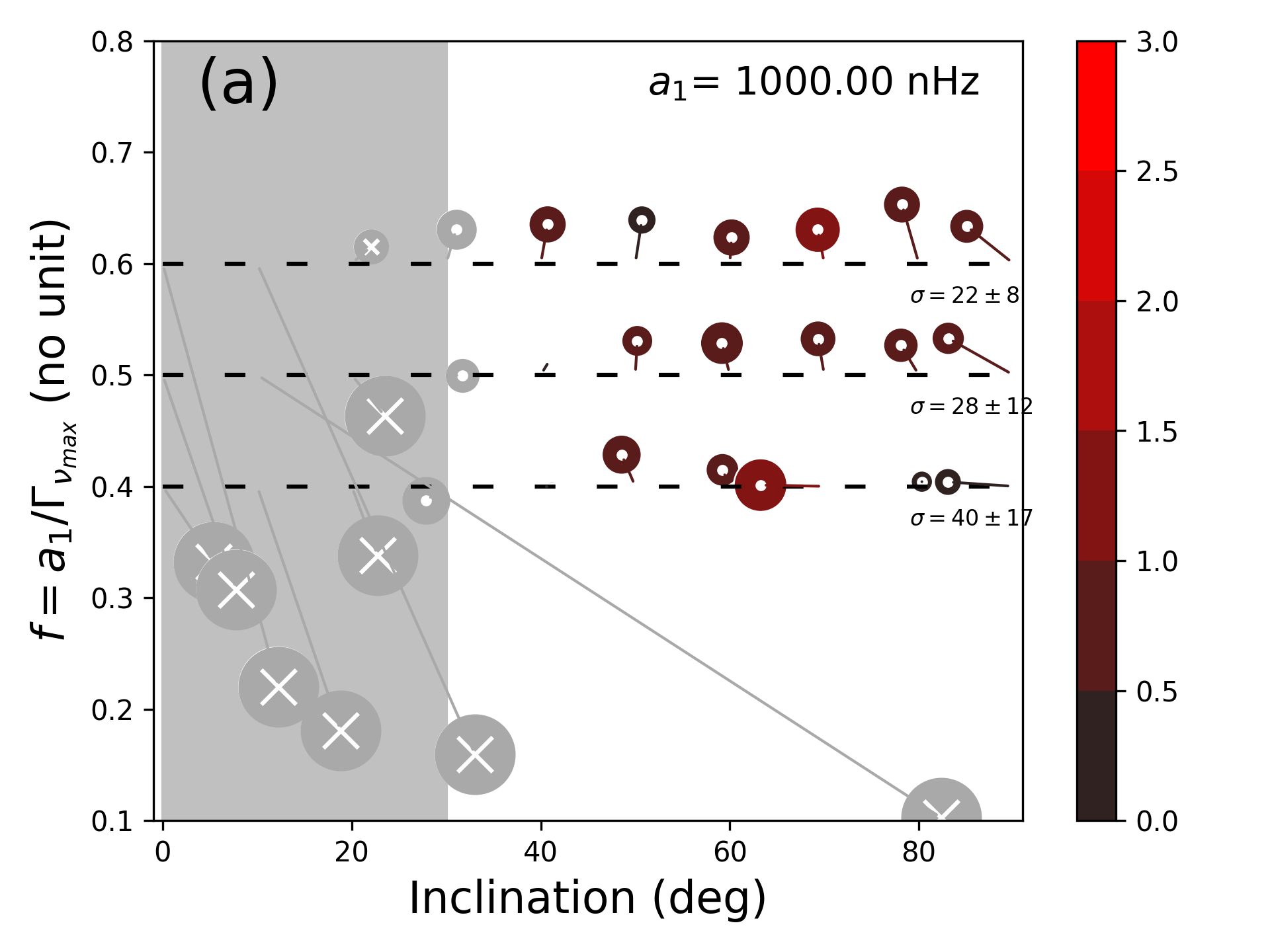}
\includegraphics[angle=0, scale=0.37]{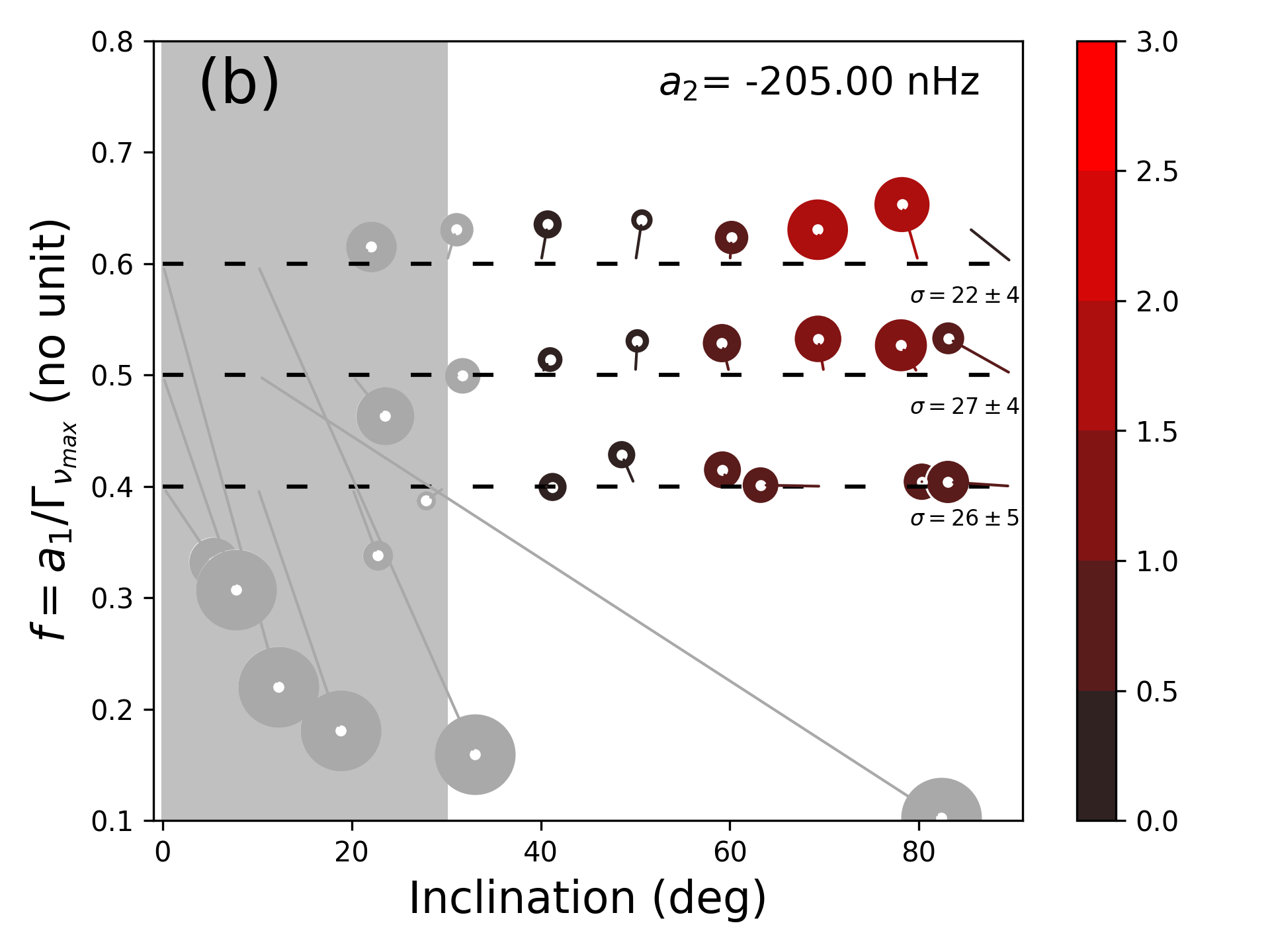} 
\includegraphics[angle=0, scale=0.37]{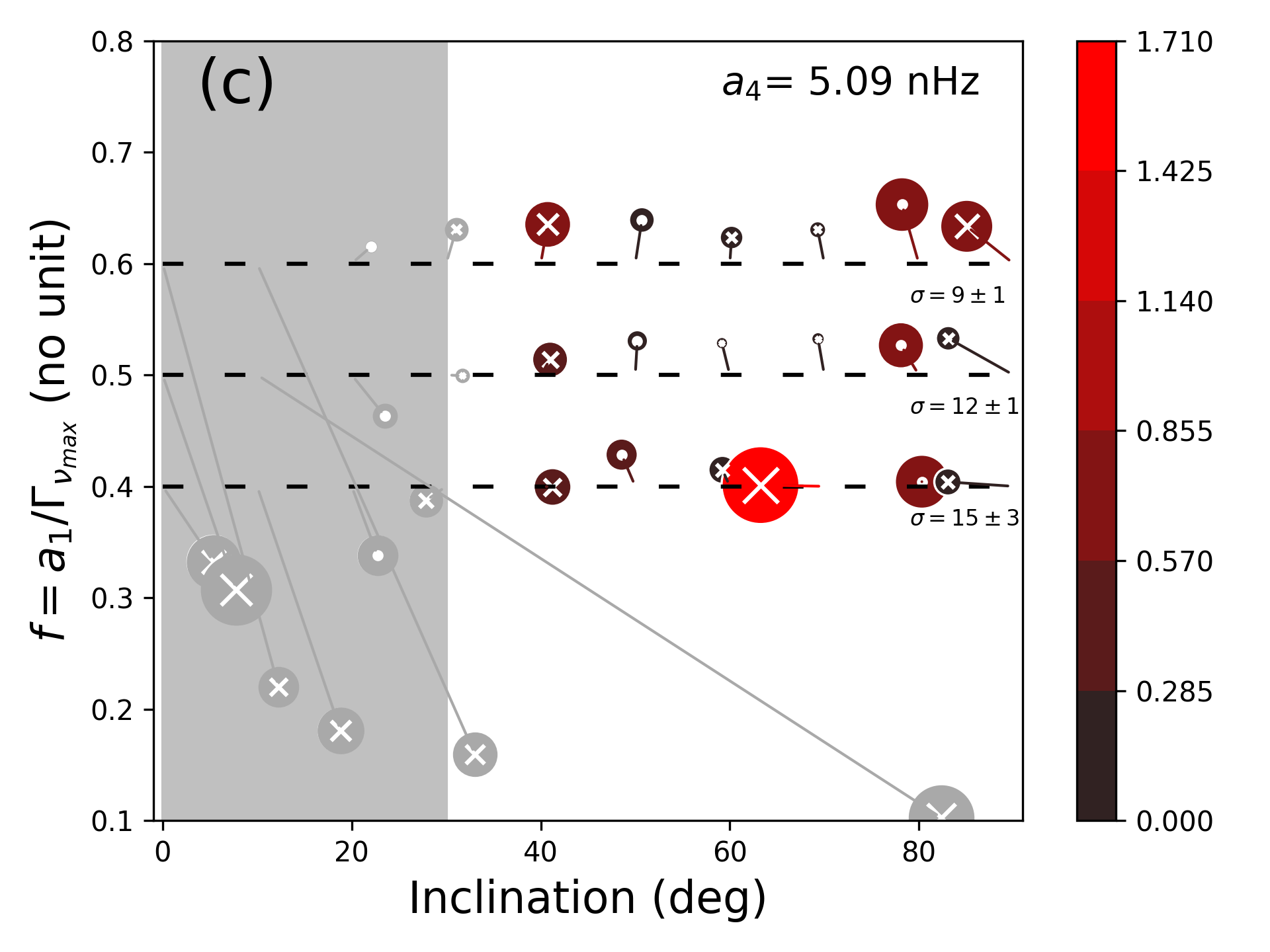} \\
\includegraphics[angle=0, scale=0.37]{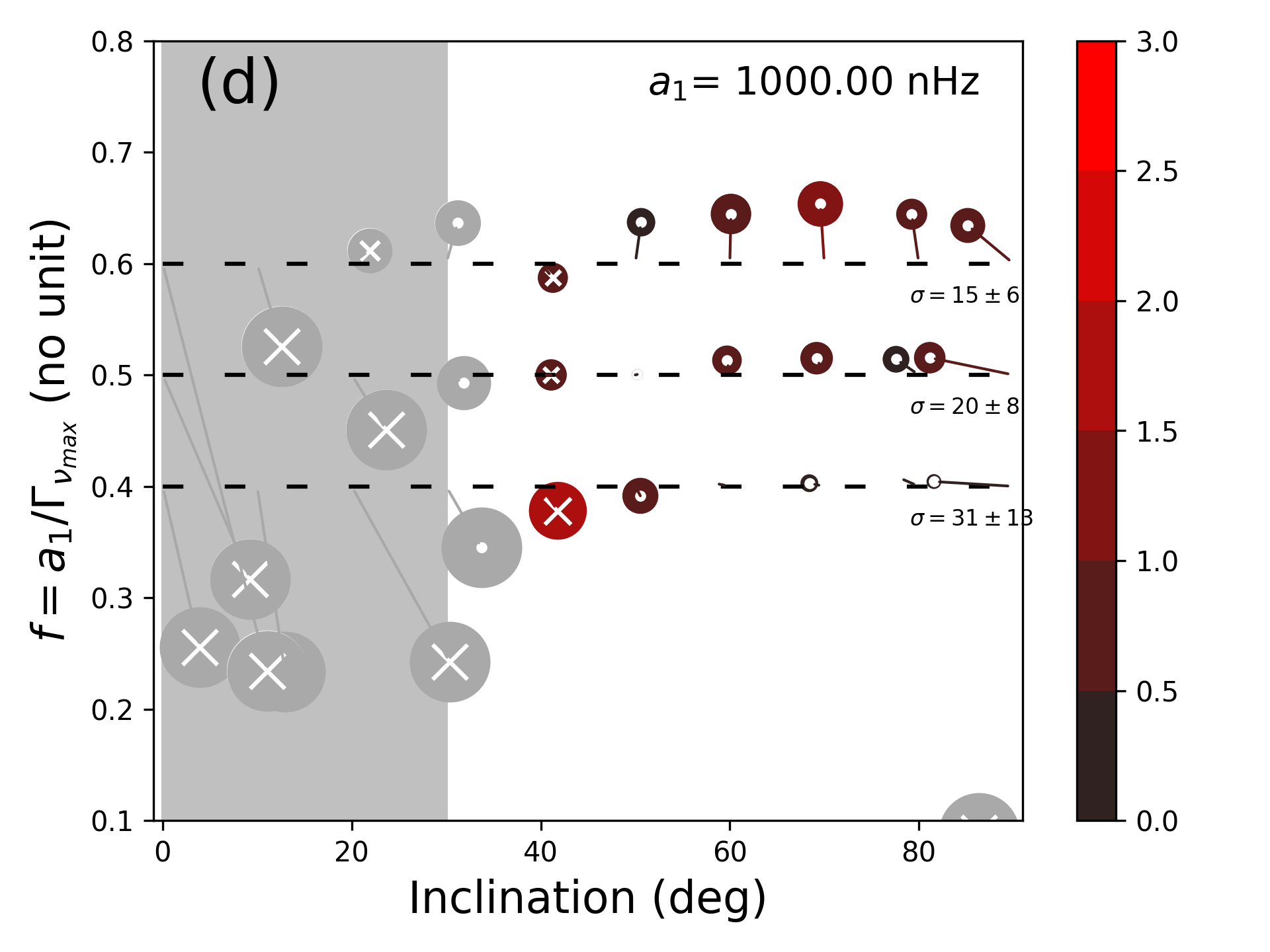} 
\includegraphics[angle=0, scale=0.37]{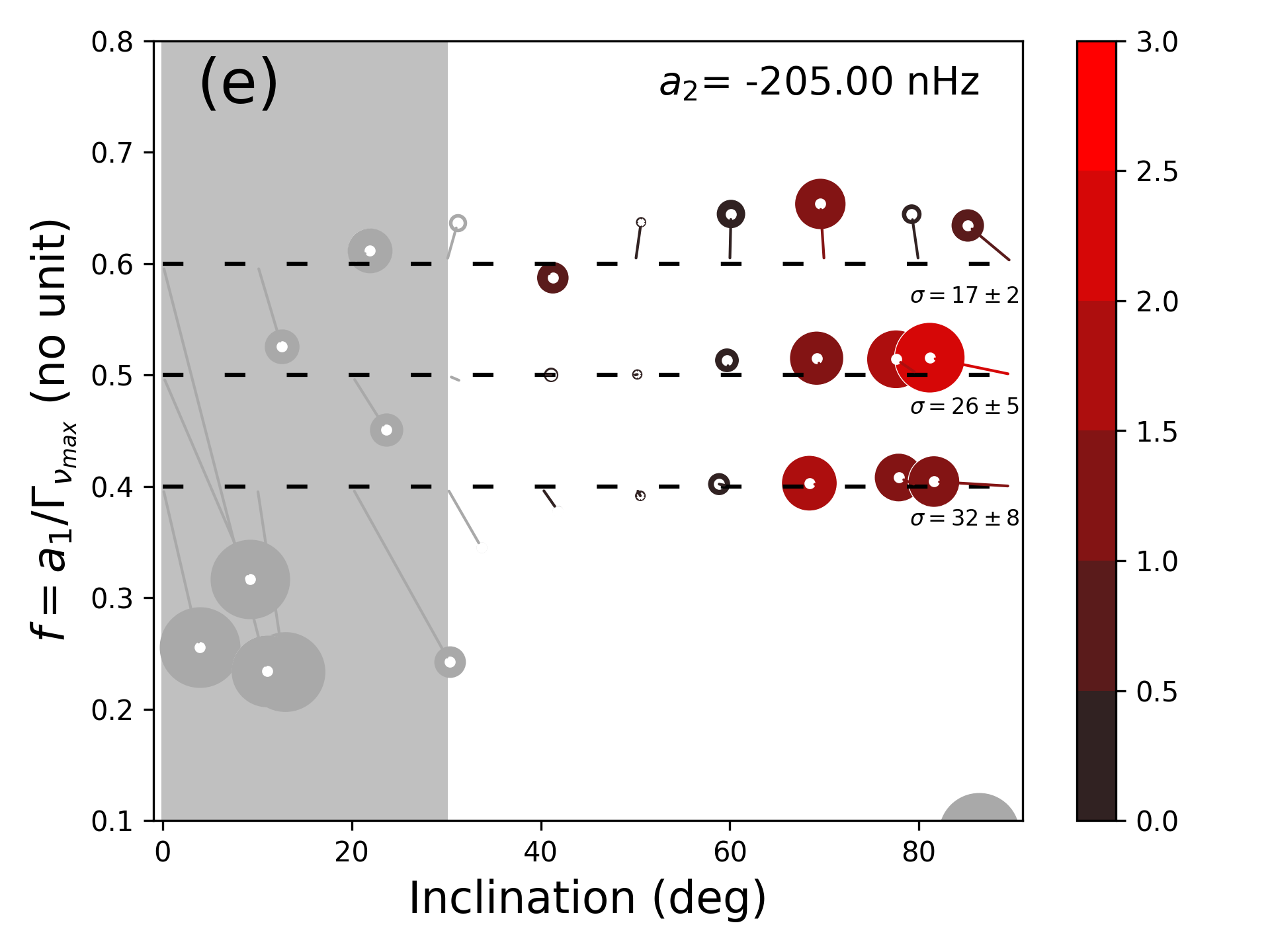}
\includegraphics[angle=0, scale=0.37]{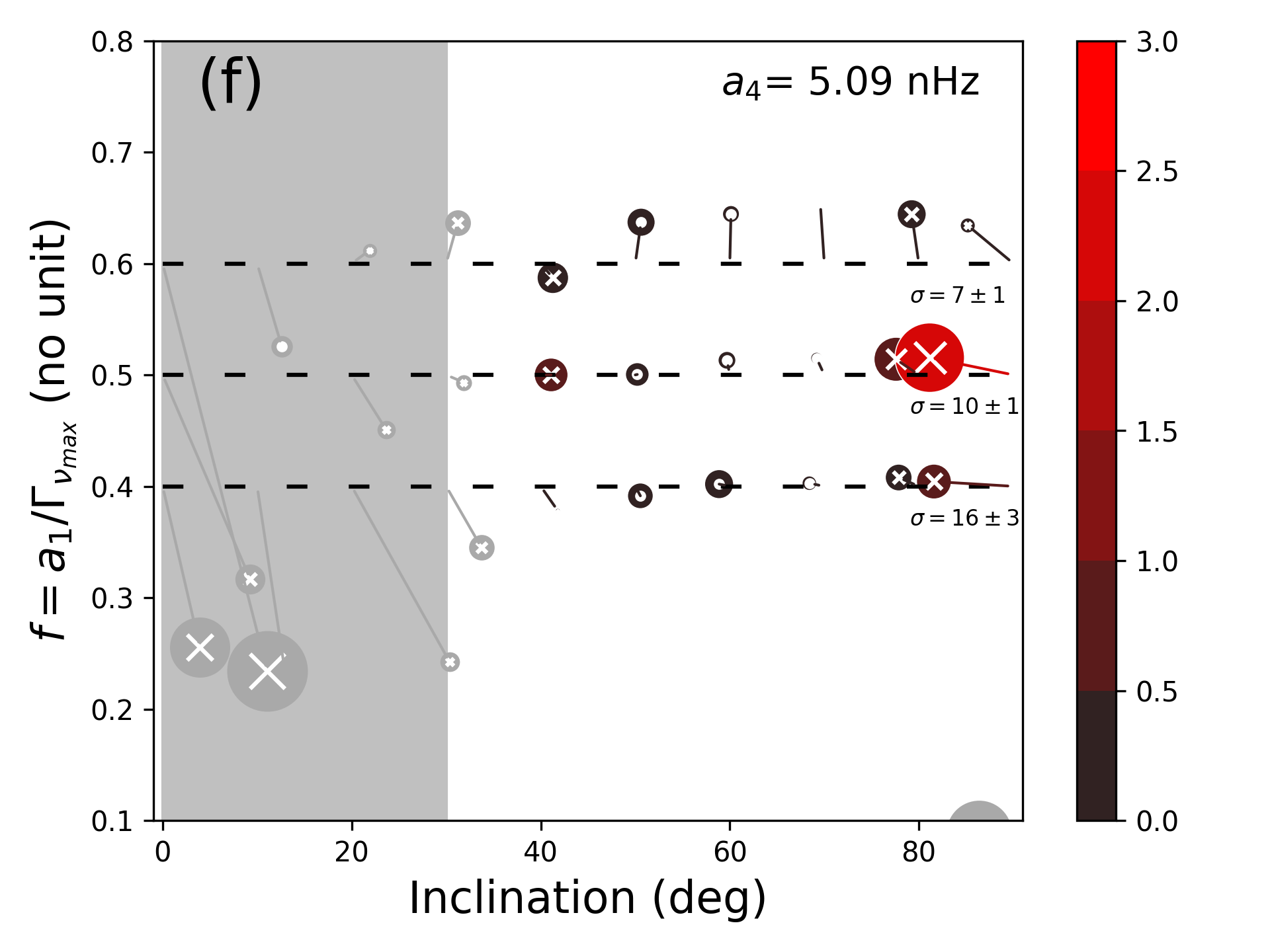}
\caption{Bias analysis for $\widehat{HNR}=10$ for a large polar activity cap ($\theta_0=22.5^\circ$,$\delta=45^\circ$) of similar intensity to the Sun ($\epsilon_{nl} = 5.10^{-4}$), for $T_{obs} = 2$ years (top) and $T_{obs} = 4$ years (bottom).}
\label{fig:bias:hnr10:Pol}
\end{center}
\end{figure*}

\begin{figure*}[ht]
\begin{center}
\includegraphics[angle=0, scale=0.37]{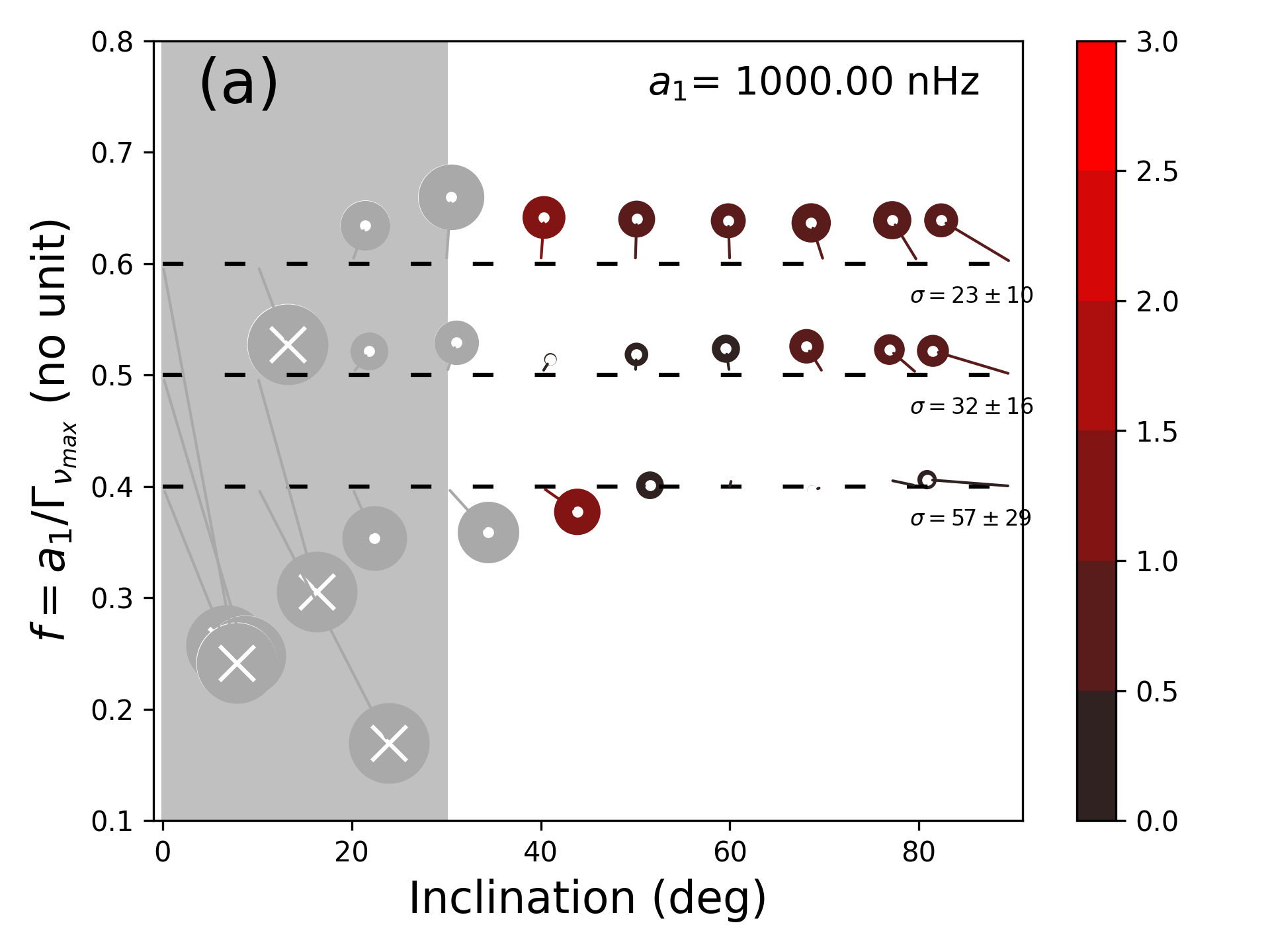}
\includegraphics[angle=0, scale=0.37]{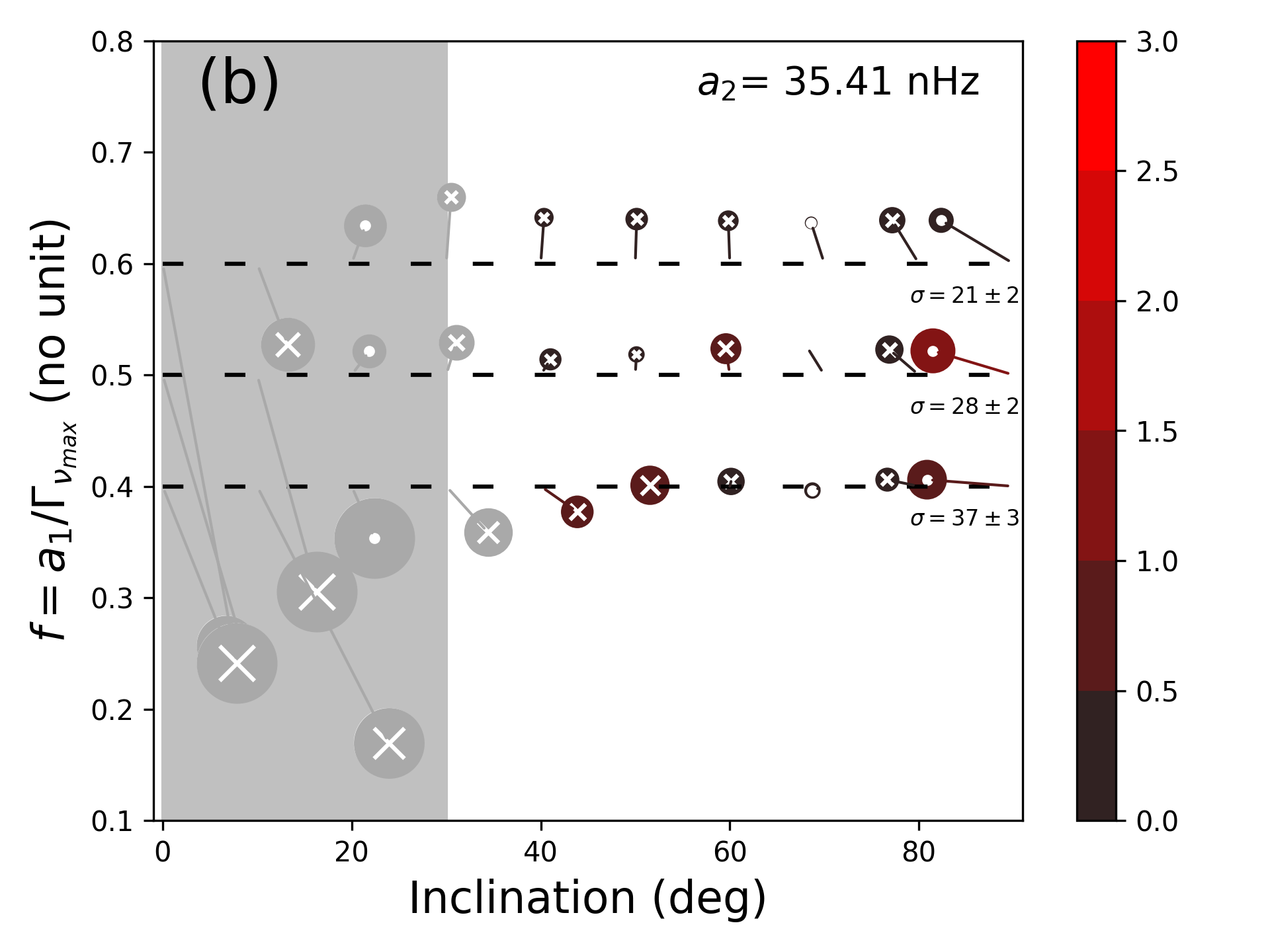}
\includegraphics[angle=0, scale=0.37]{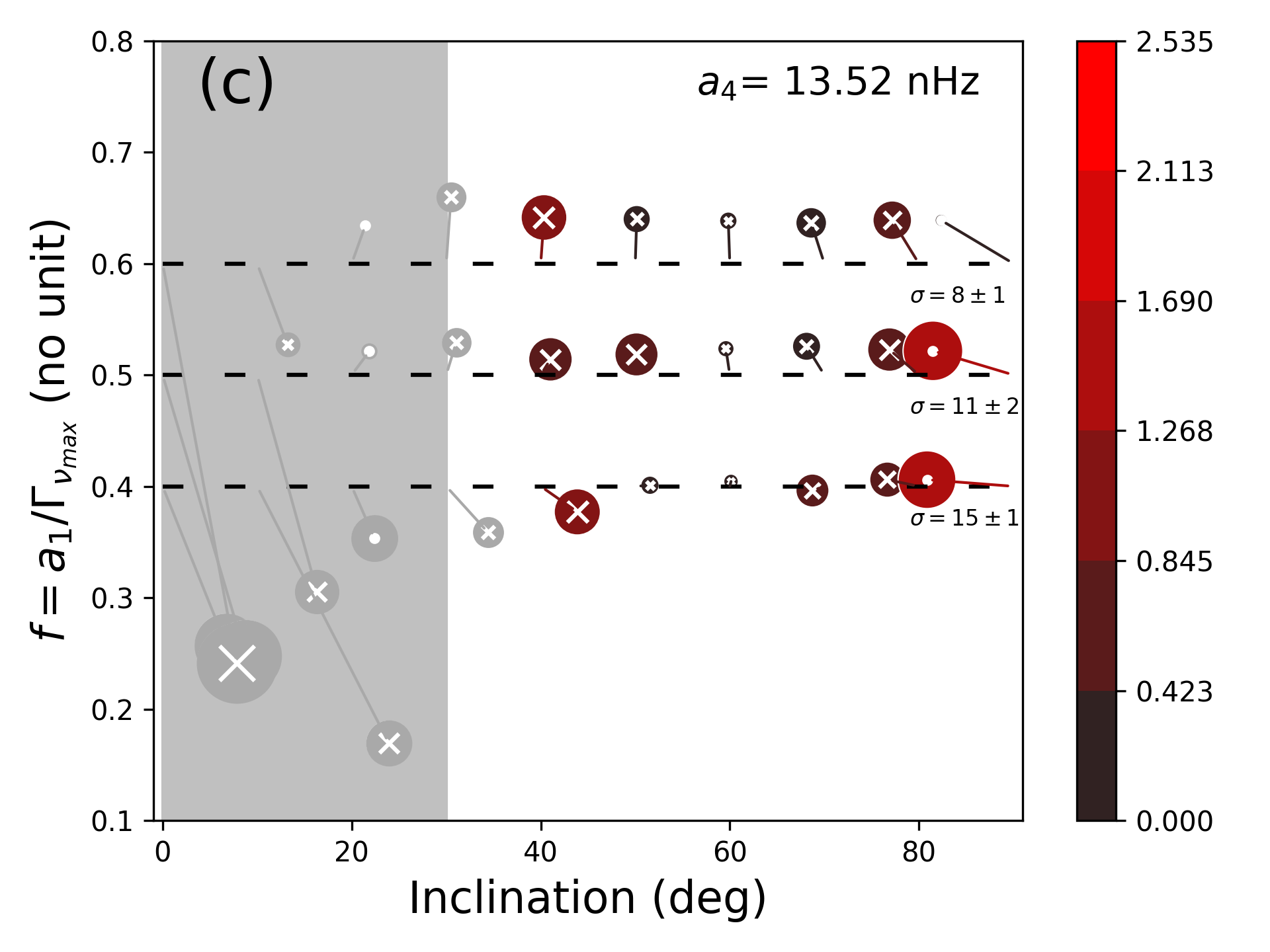} \\
\includegraphics[angle=0, scale=0.37]{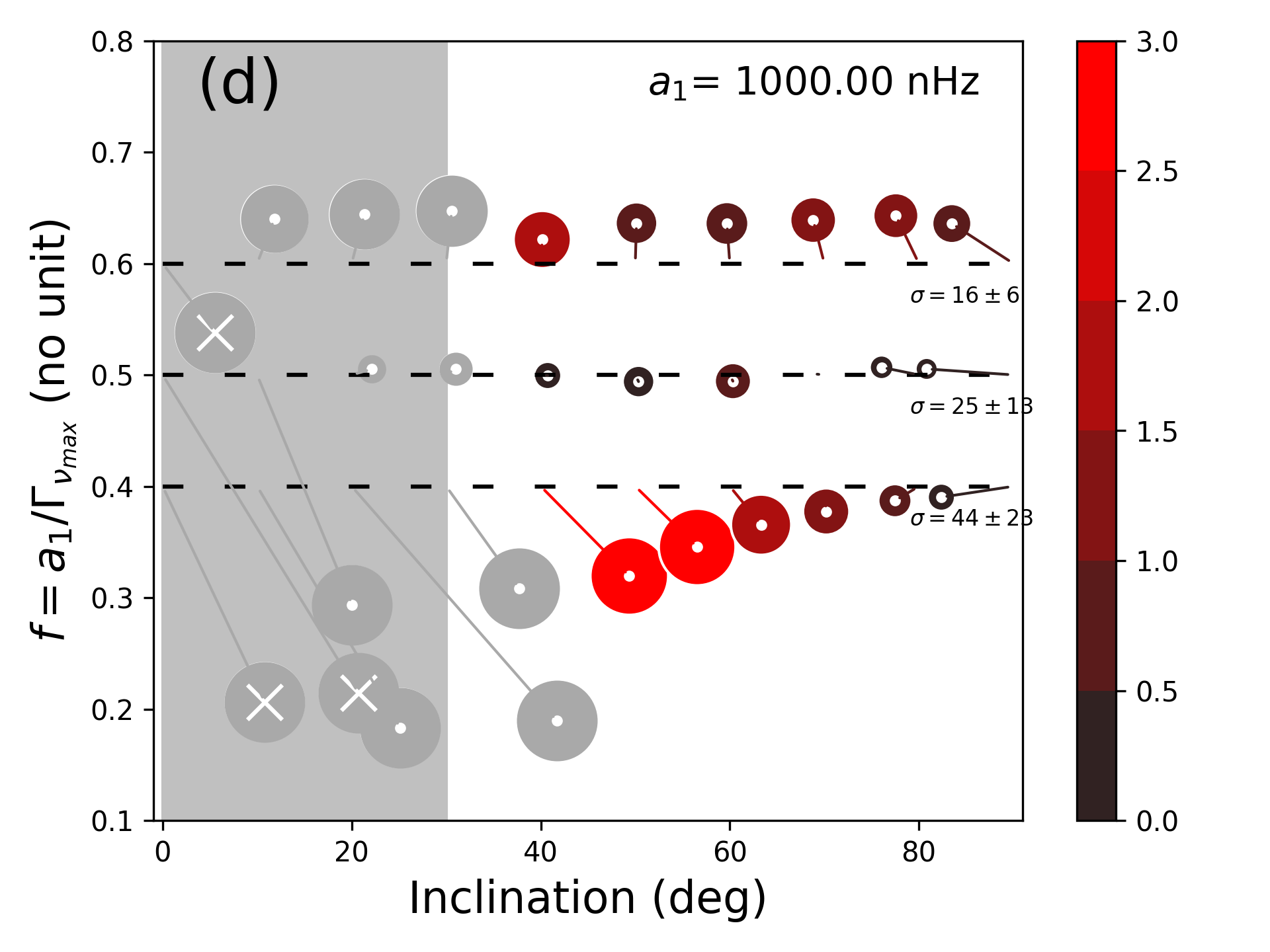}
\includegraphics[angle=0, scale=0.37]{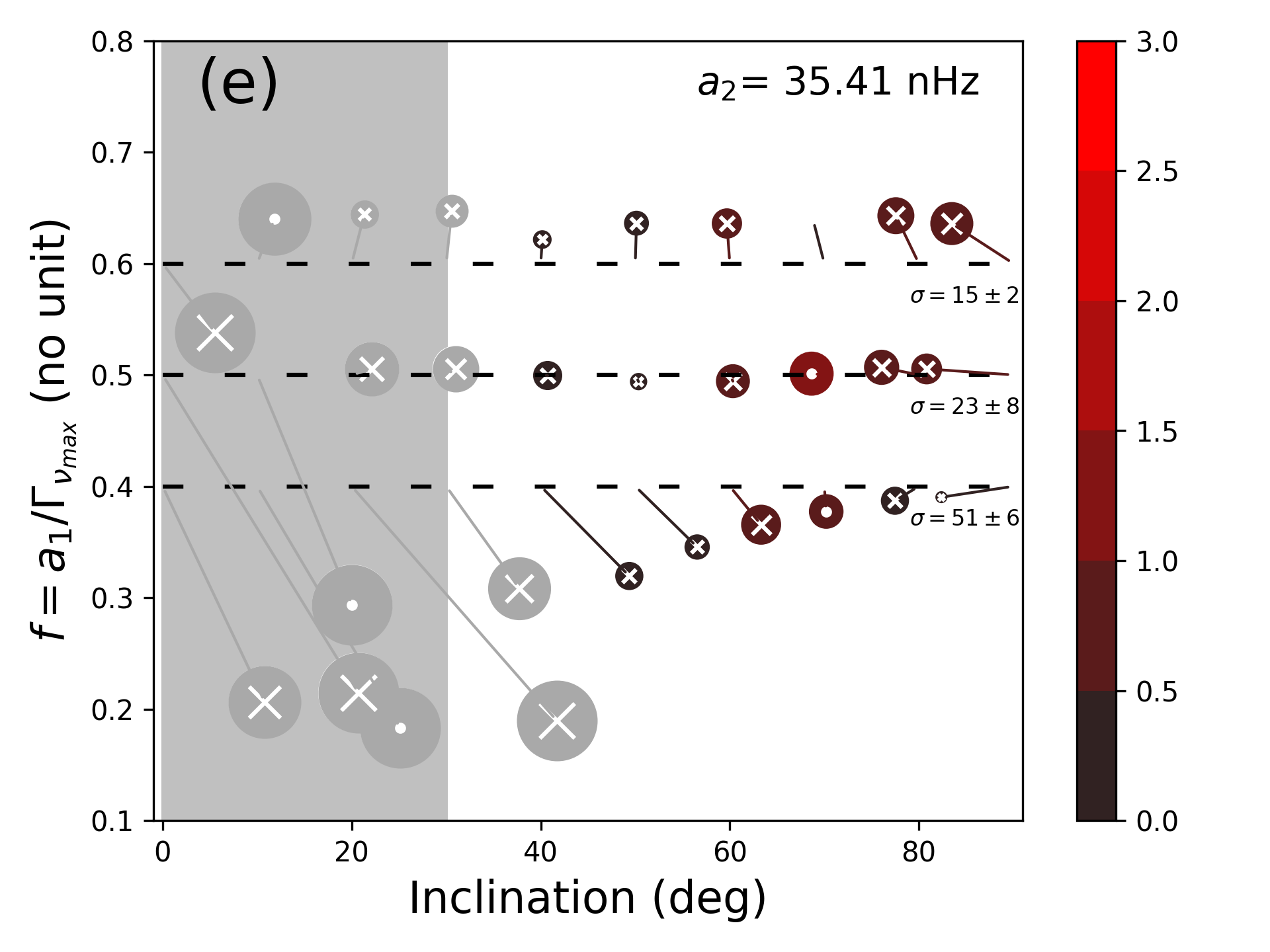}
\includegraphics[angle=0, scale=0.37]{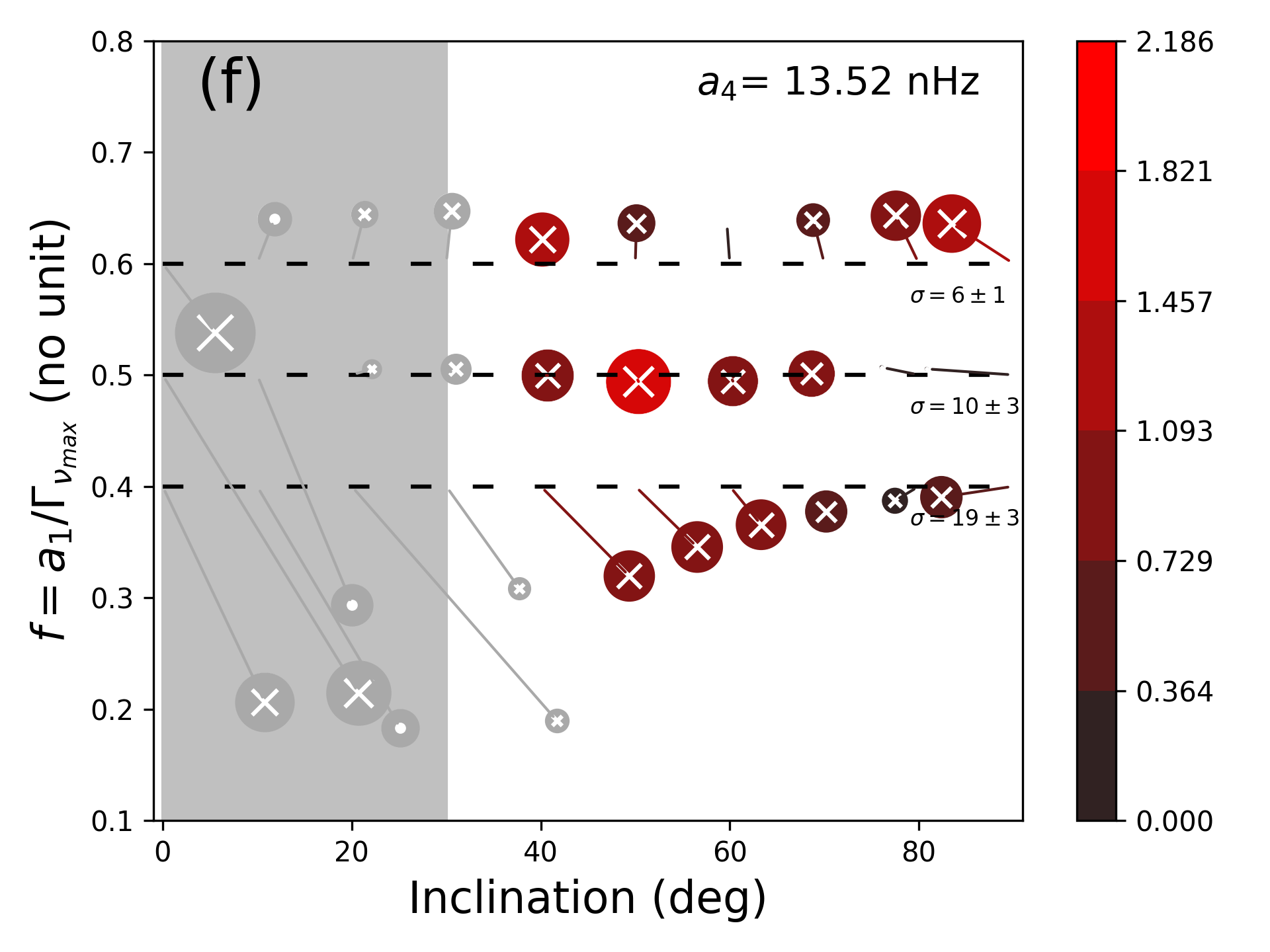}
\caption{Bias analysis for $\widehat{HNR}=20$ for an equatorial activity band ($\theta_0=85$,$\delta=10$) of similar intensity to the Sun ($\epsilon_{nl} = 5.10^{-4}$), for $T_{obs} = 2$ years (top) and $T_{obs} = 4$ years (bottom).}
\label{fig:bias:hnr20:Eq}
\end{center}
\end{figure*}

\begin{figure*}[ht]
\begin{center}
\includegraphics[angle=0, scale=0.37]{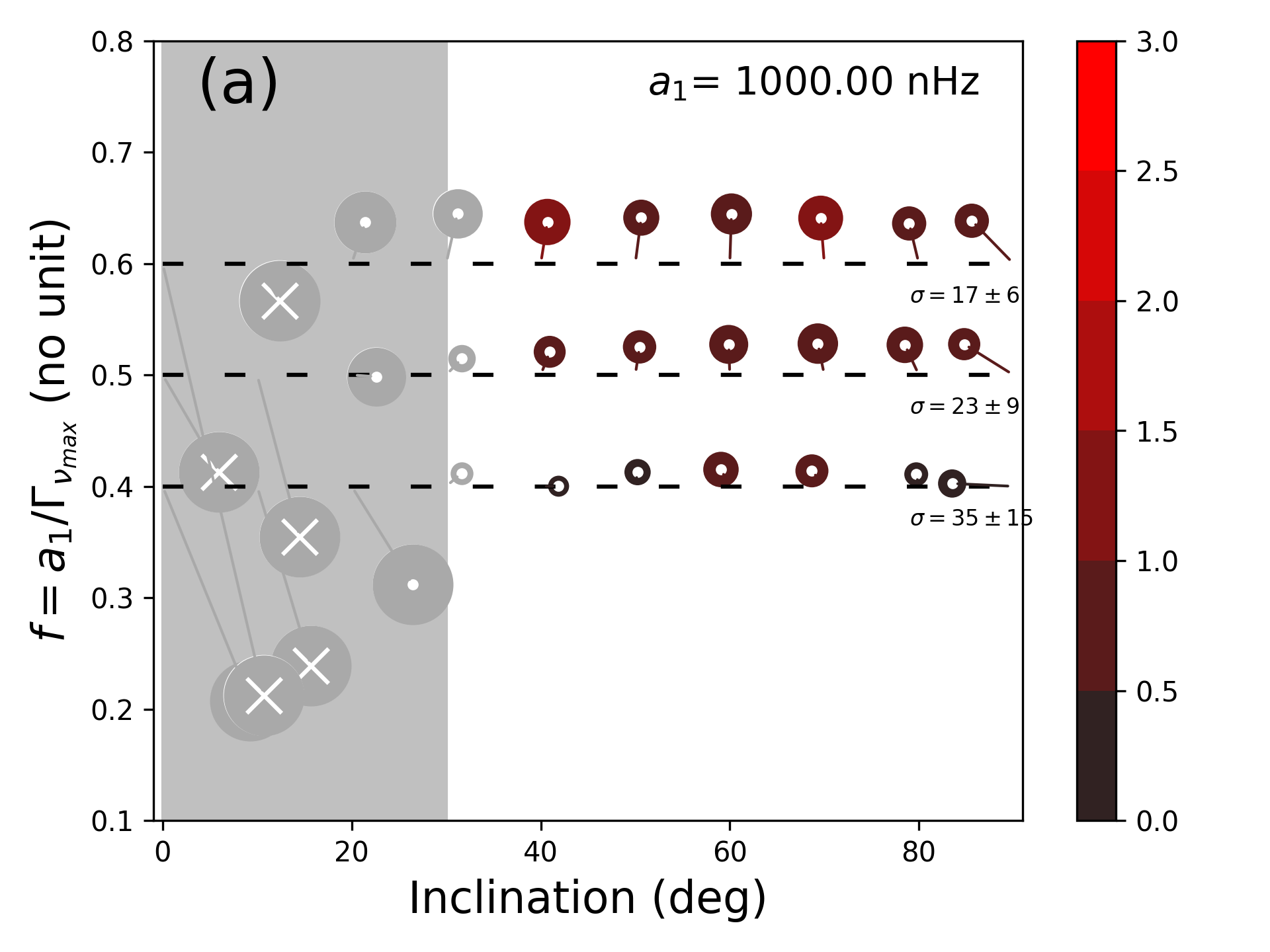}
\includegraphics[angle=0, scale=0.37]{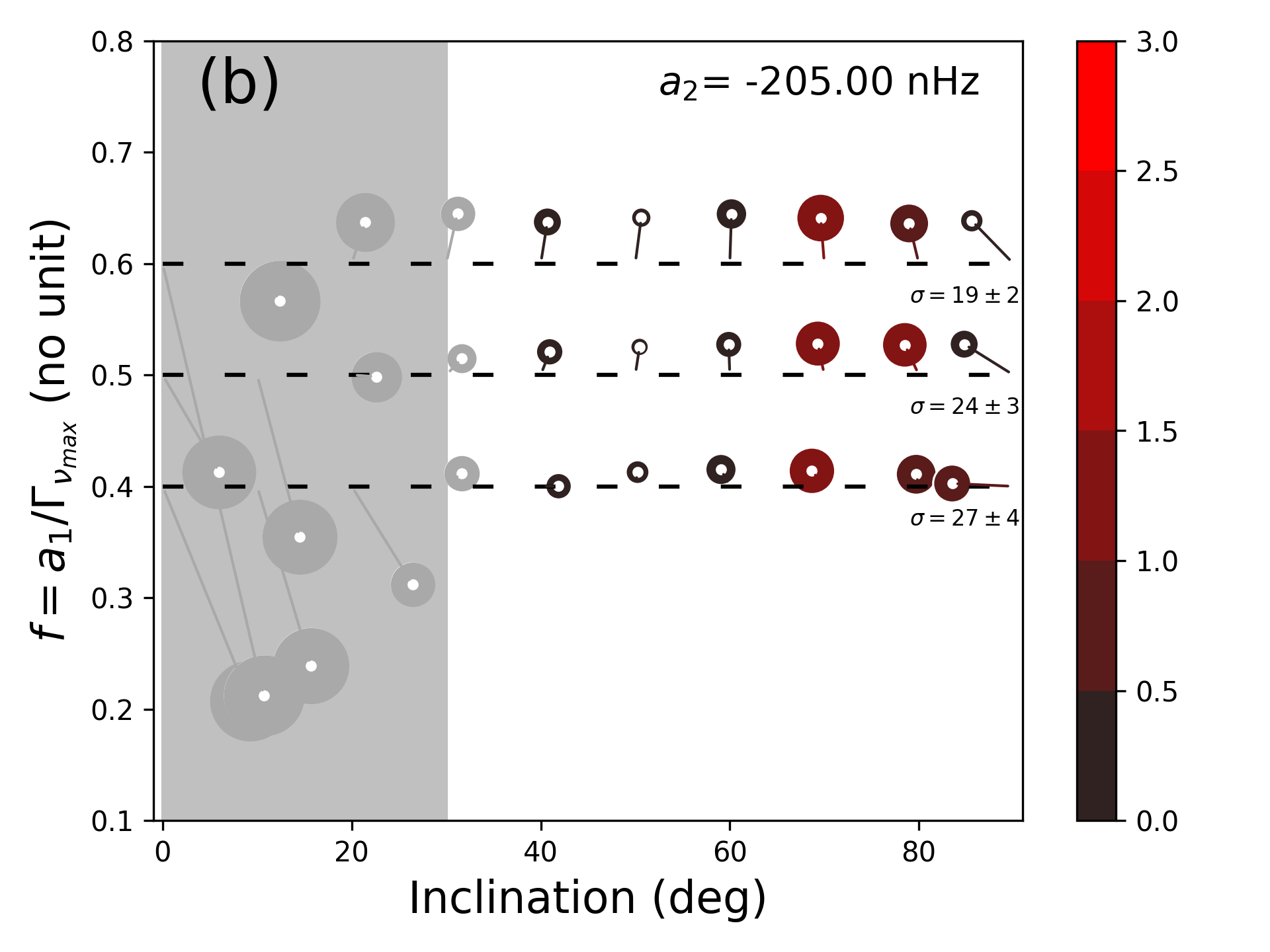}
\includegraphics[angle=0, scale=0.37]{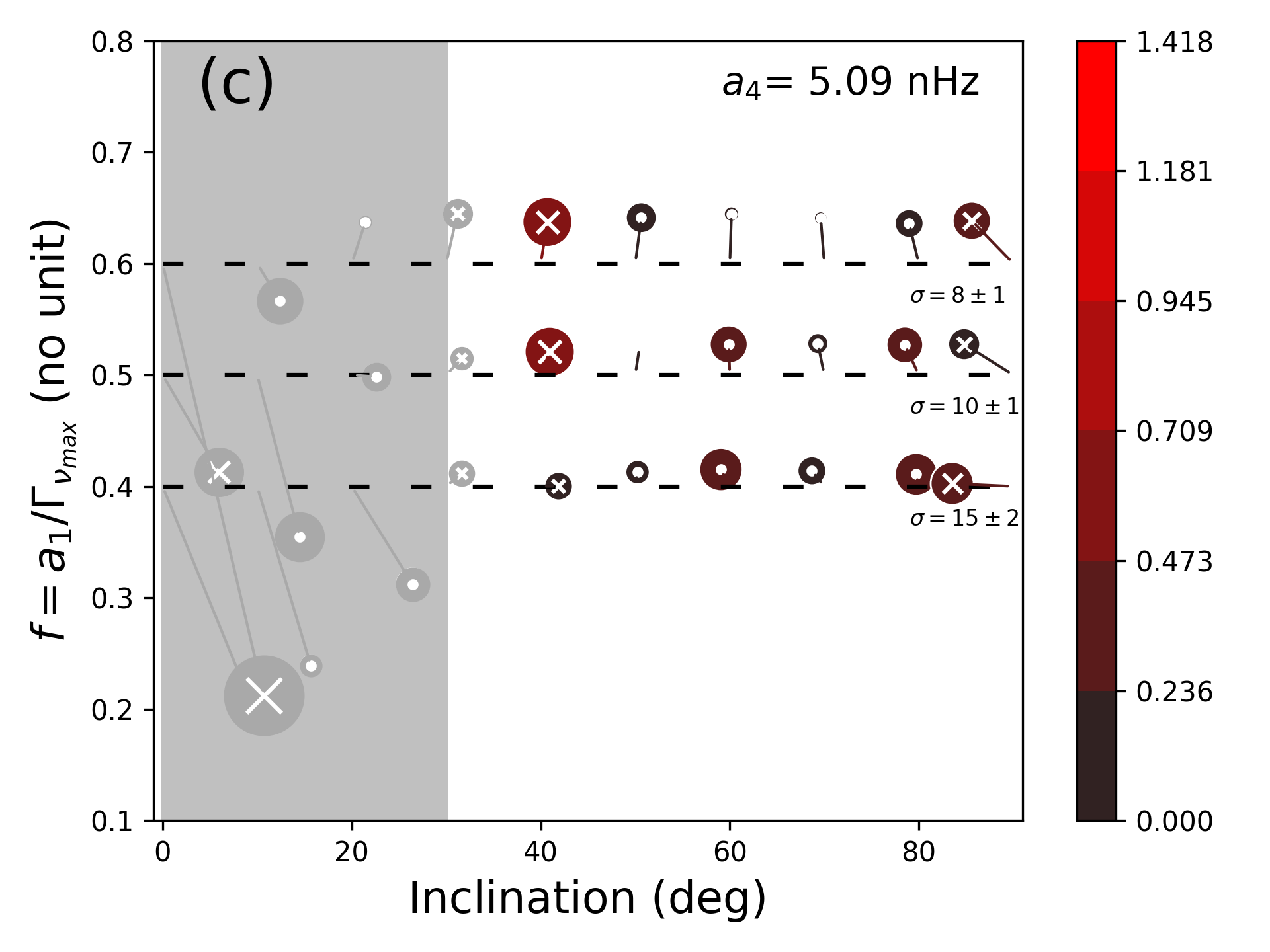} \\
\includegraphics[angle=0, scale=0.37]{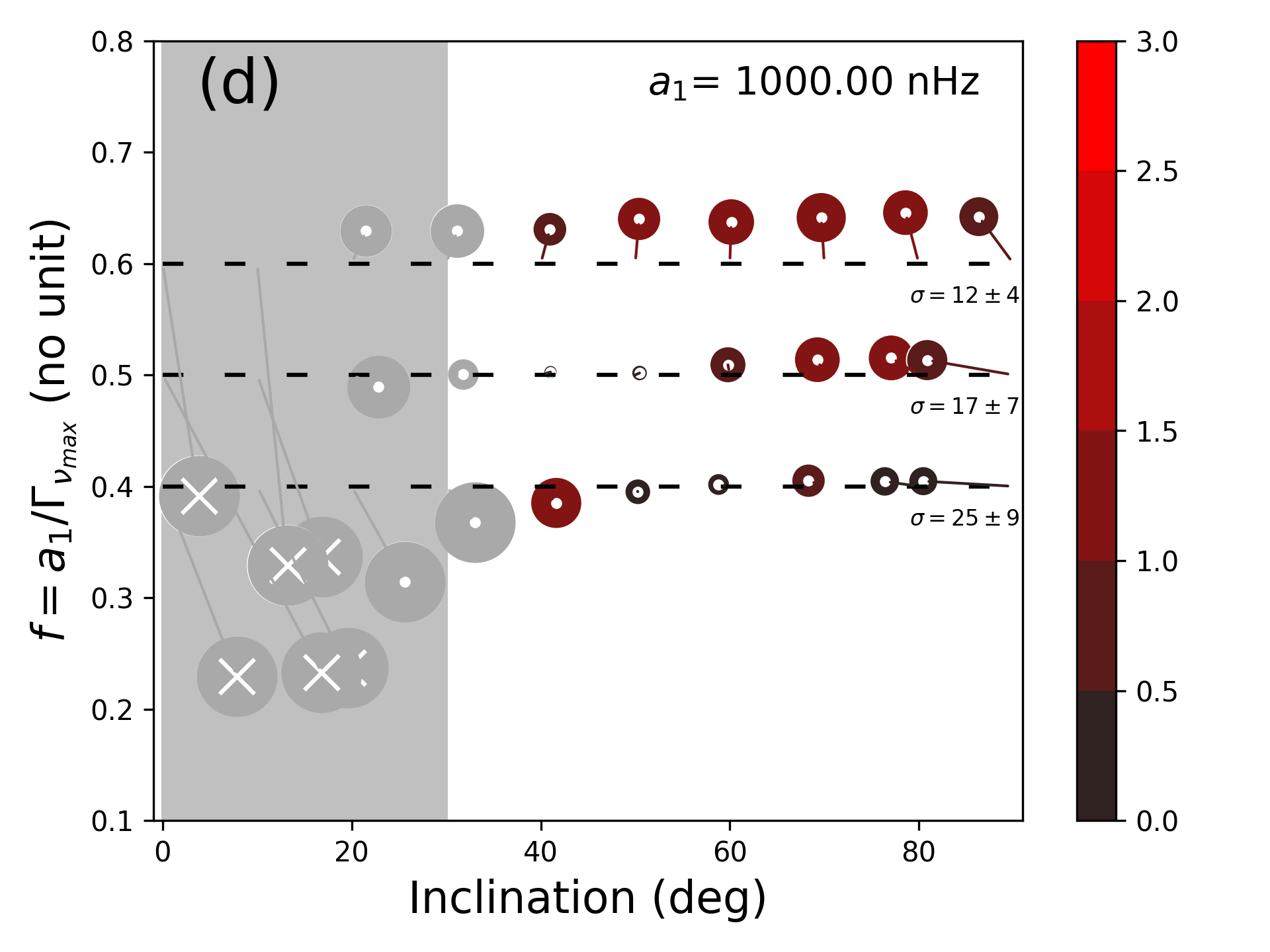}
\includegraphics[angle=0, scale=0.37]{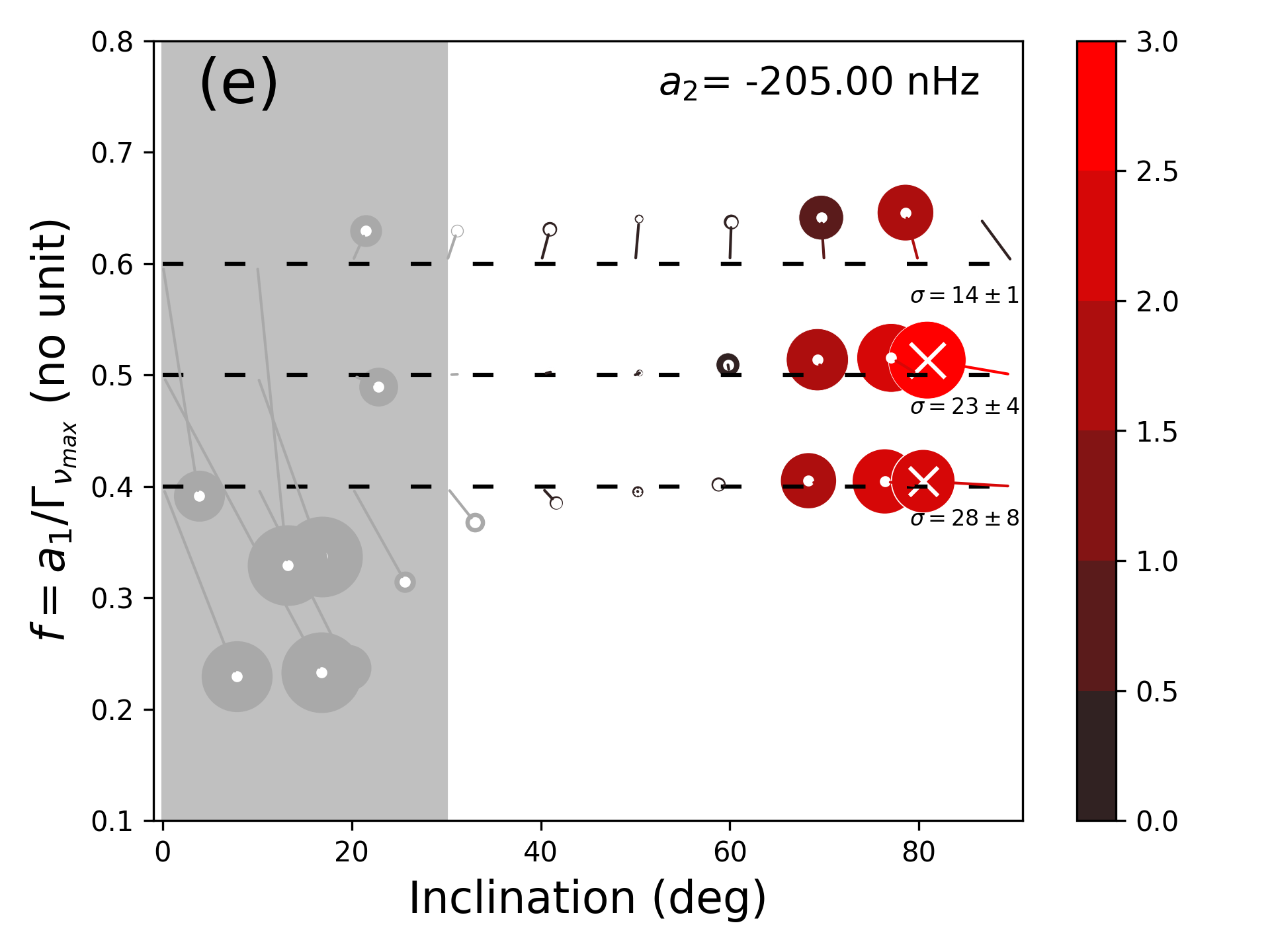}
\includegraphics[angle=0, scale=0.37]{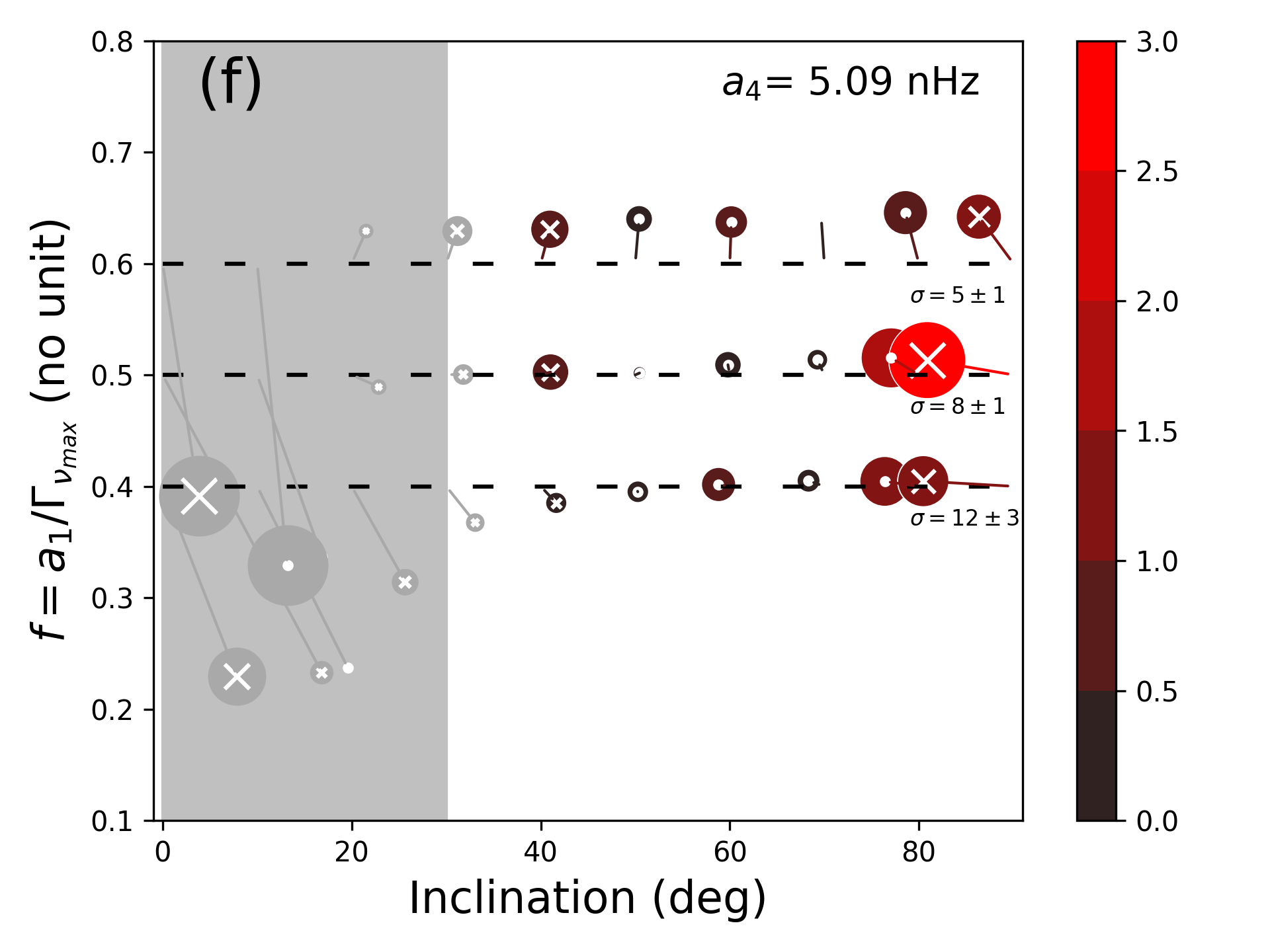}
\caption{Bias analysis for $\widehat{HNR}=20$ for a large polar activity cap ($\theta_0=22.5$,$\delta=45$) of similar intensity to the Sun ($\epsilon_{nl} = 5.10^{-4}$), for $T_{obs} = 2$ years (top) and $T_{obs} = 4$ years (bottom). }
\label{fig:bias:hnr20:Pol}
\end{center}
\end{figure*}

 This section shows the bias on the $a_1$, $a_2$ and $a_4$ coefficients in the case of  $\widehat{HNR}=10$ or $\widehat{HNR}=20$. The figure \ref{fig:bias:hnr10:Eq} and \ref{fig:bias:hnr20:Eq}  are for the case of an equatorial band of activity and figure \ref{fig:bias:hnr10:Pol} and \ref{fig:bias:hnr20:Pol}, for a polar cap. It is noticeable that the inaccuracy of the fit remains smaller than standard deviation in the majority of the parameter space and for $i>30^\circ$, $a_1/\Gamma_{\nu_{max}}>0.4$. Interestingly, the relative-to-error bias on $a_1$ is more pronounced in the case of an equatorial band of activity and for $\{a_1/\Gamma_{\nu_{max}} \le 0.4, i=[30,60]\}$ (Figure \ref{fig:bias:hnr10:Eq}d and \ref{fig:bias:hnr20:Eq}d) and can exceed 3 times the uncertainty. The accuracy of the inference of activity based on the a-coefficients is expected to weakly depend on the $\widehat{HNR}$ (although potentially with large uncertainty) for most stars, except within this regime (in addition to the gray area), where the centrifugal effect will be overestimated.
 
\section{{Figures of the best fits}} \label{appendix:best_fits}

\begin{figure*}[ht]
\begin{center}
\includegraphics[angle=0]{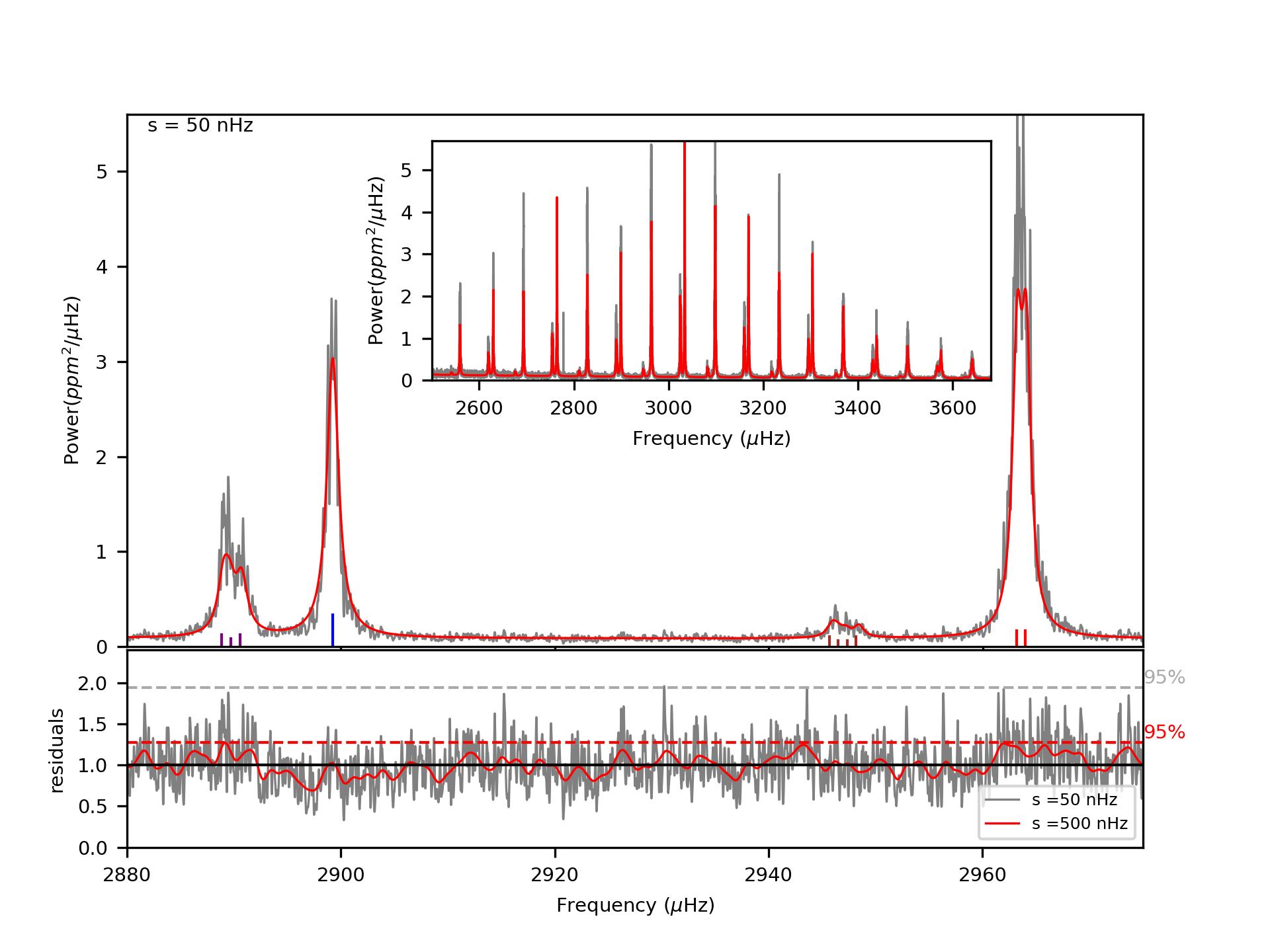}
\caption{Solar power spectrum for observation between 1999 and 2002 after Gaussian smoothing over a kernel width s=50nHz. Superimposed is the best MCMC fit. {Top.} Highlight of a $l=2,0,3,1$ mode group (left to right) and their fit (red). The power asymmetry due to non-symmetric $m$ components is visible in the data. {Inset.} Overall view of the modes. {Bottom.} Residual of the fit with two level s of Gaussian smoothing.}
\label{fig:bestfit:Sun1999}
\end{center}
\end{figure*}

\begin{figure*}[ht]
\begin{center}
\includegraphics[angle=0]{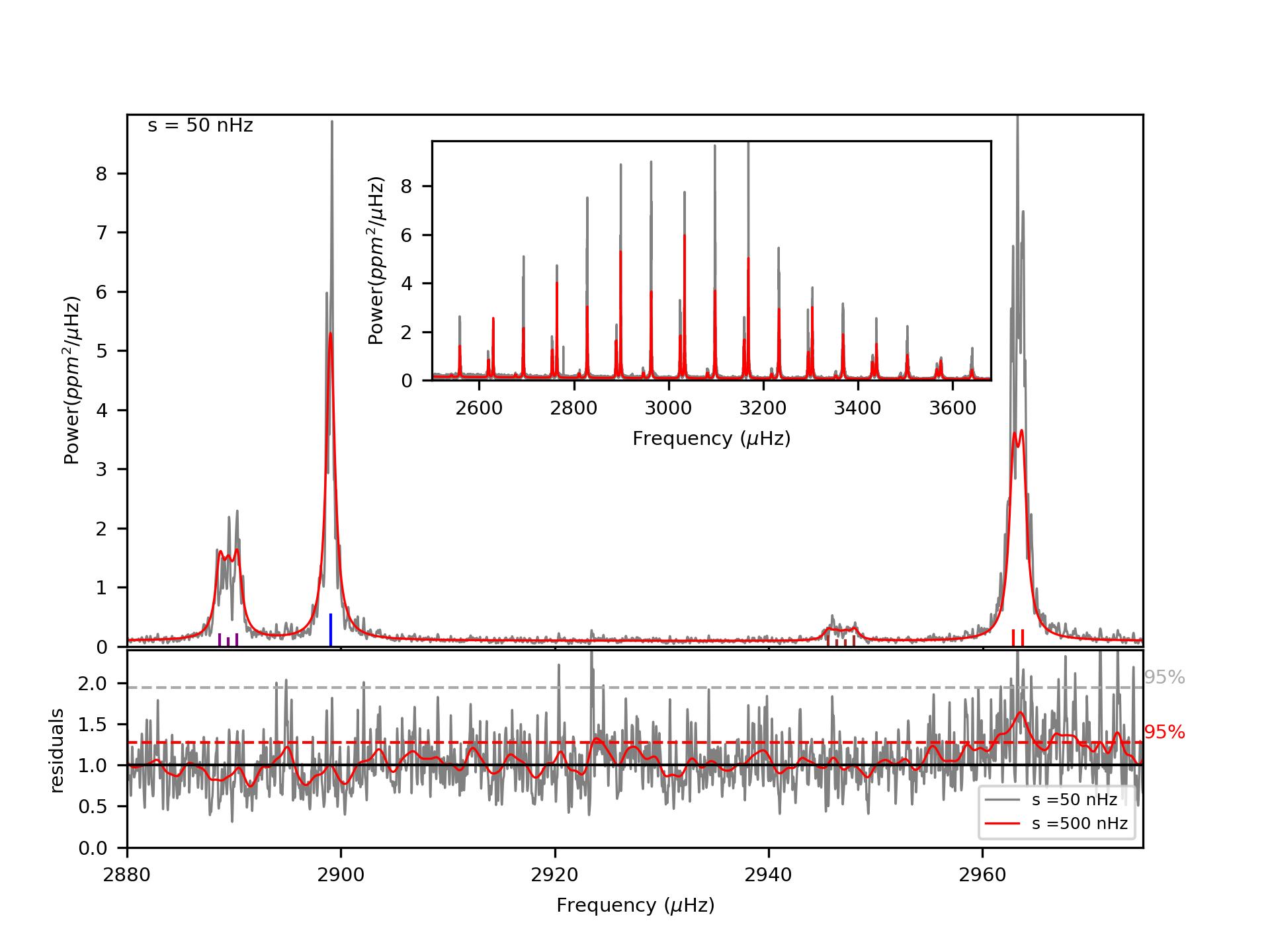}
\caption{Same as Figure \ref{fig:bestfit:Sun1999} but between 2006 and 2009. {Top.} Highlight of a $l=2,0,3,1$ mode group (left to right) and their fit (red). The power asymmetry seen for 1999-2002 is not apparent in the data. {Inset.} Overall view of the power spectrum. {Bottom.} Residual of the fit with two level s of Gaussian smoothing. The residual on $l=1$ is a bit high here due to the fix visibility for all modes.}
\label{fig:bestfit:Sun2006}
\end{center}
\end{figure*}

\begin{figure*}[ht]
\begin{center}
\includegraphics[angle=0]{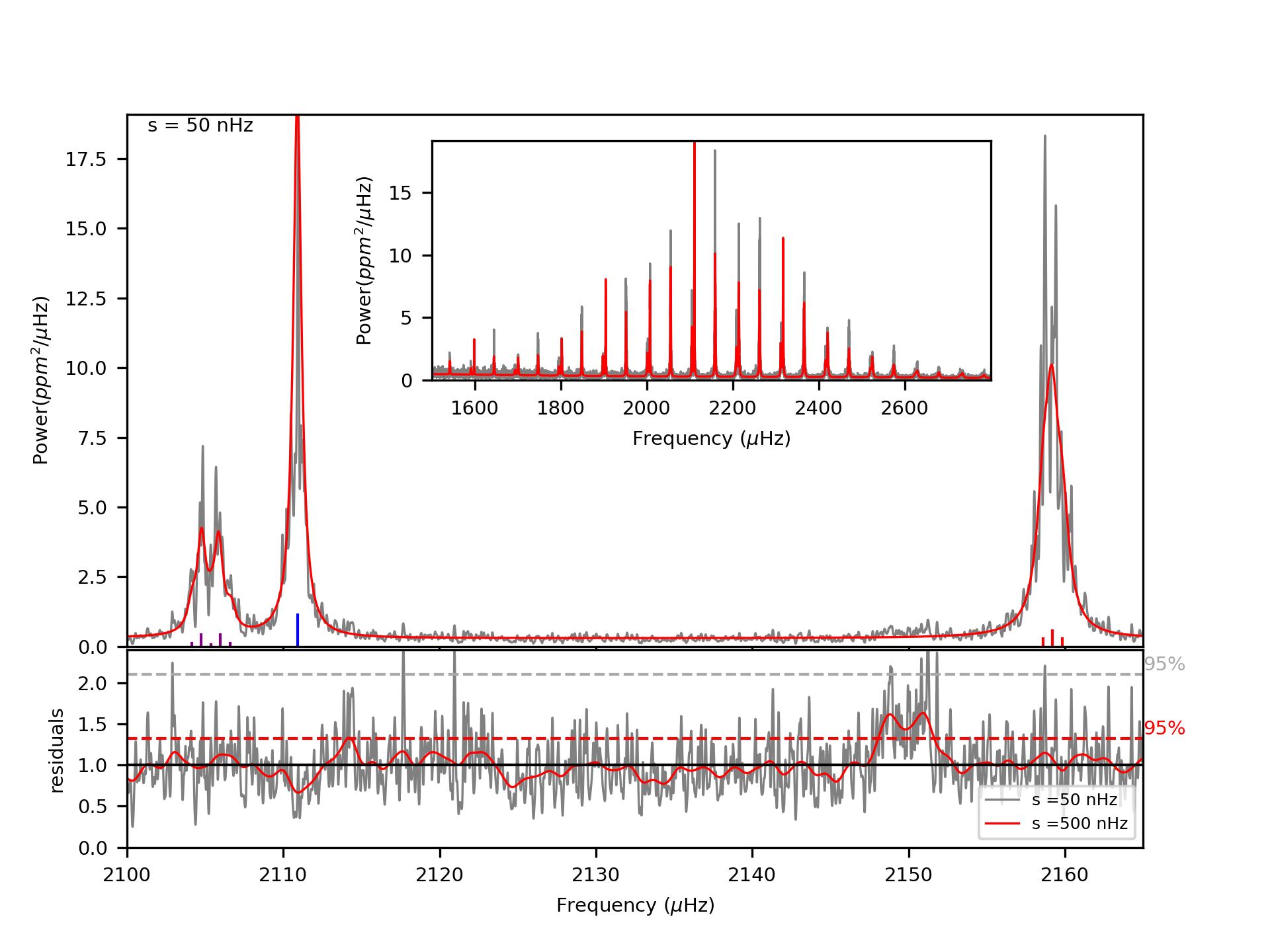}
\caption{Same as Figure \ref{fig:bestfit:Sun1999} but for 16 Cyg A. {Top.} Highlight of a $l=2,0,1$ mode group (left to right) and their fit (red). A very mild power asymmetry is seen on the $l=2$ data. {Inset.} Overall view of the power spectrum. {Bottom.} Residual of the fit with two level s of Gaussian smoothing. The residual show an excess of power due to the low HNR$\simeq 1.7$ $l=3$ (not fitted here). }
\label{fig:bestfit:16CygA}
\end{center}
\end{figure*}

\begin{figure*}[ht]
\begin{center}
\includegraphics[angle=0]{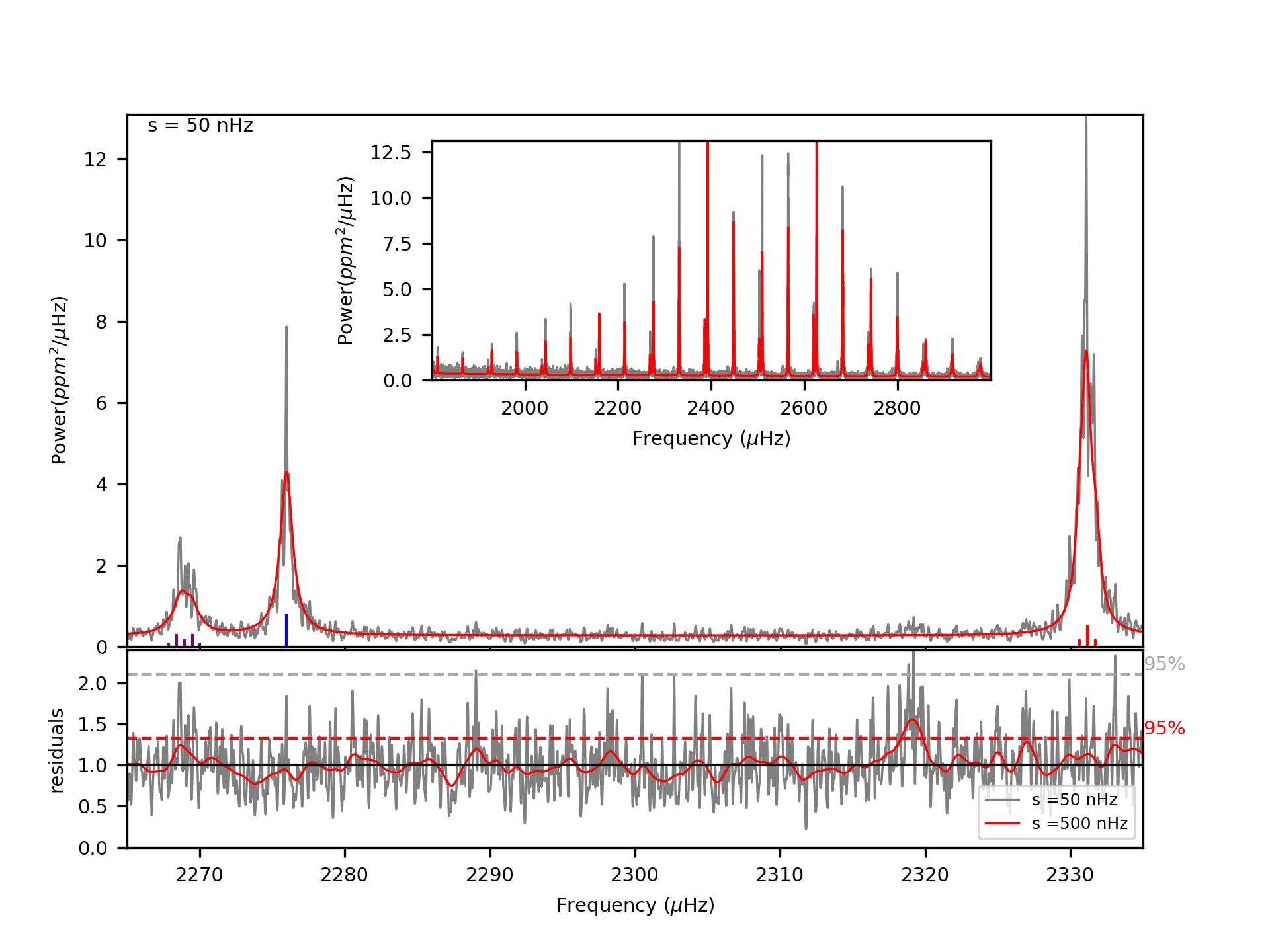}
\caption{Same as Figure \ref{fig:bestfit:Sun1999} but for 16 Cyg B, $a_4<-21$ nHz. {Top.} Highlight of a $l=2,0,1$ mode group (left to right) and their fit (red). A mild power asymmetry is seen on the $l=2$ data. {Inset.} Overall view of the power spectrum. {Bottom.} Residual of the fit with two level s of Gaussian smoothing. The residual show an excess of power due to the low HNR$\simeq 1.6$ $l=3$ (not fitted here).}
\label{fig:bestfit:16CygB_L}
\end{center}
\end{figure*}

\begin{figure*}[ht]
\begin{center}
\includegraphics[angle=0]{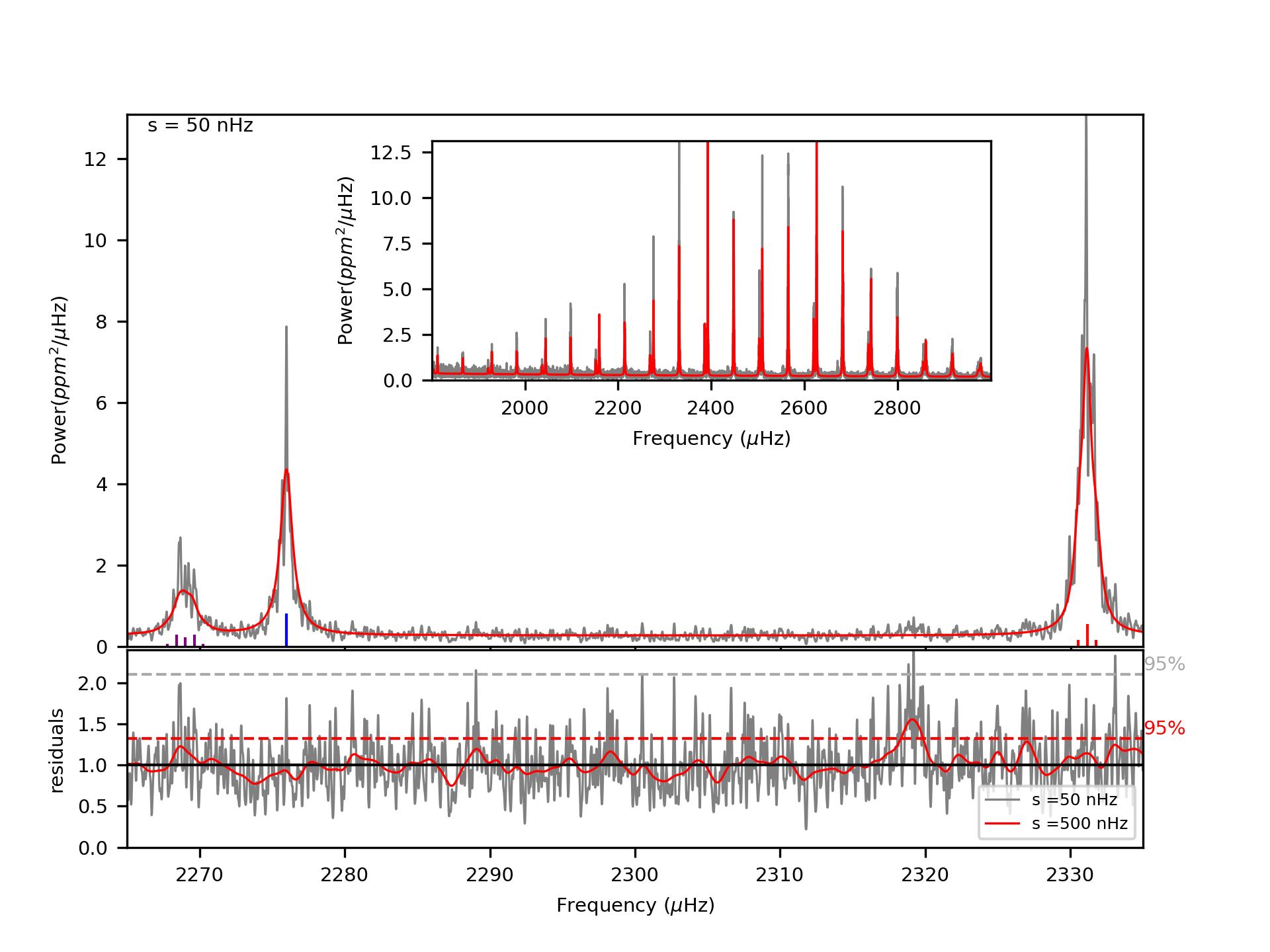}
\caption{Same as Figure \ref{fig:bestfit:Sun1999} but for 16 Cyg B, $a_4>-21$ nHz. {Top.} Highlight of a $l=2,0,1$ mode group (left to right) and their fit (red). {Inset.} Overall view of the power spectrum. {Bottom.} Residual of the fit with two level s of Gaussian smoothing. The residual show an excess of power due to the low HNR$\simeq 1.6$ $l=3$ (not fitted here).}
\label{fig:bestfit:16CygB_U}
\end{center}
\end{figure*}

{This section presents visuals on the best MCMC fits for the power spectrum of the active Sun (Figure \ref{fig:bestfit:Sun1999}), quiet Sun (Figure \ref{fig:bestfit:Sun2006}), 16 Cyg A (Figure \ref{fig:bestfit:16CygA}) and for the two $a_4$ solutions of 16 Cyg B (Figures \ref{fig:bestfit:16CygB_L} and \ref{fig:bestfit:16CygB_U}). Fog 16 Cyg B, the solutions are here separated by selecting the median for the samples only below or above $a_4 = 21 nHz$.
In the case of the Sun, the impact of the activity on the profile of $l=2$ is very clear, as the shifts of the m-components introduces an asymmetry in power for the overall $l=2$ mode profile during the active phase. During the quiet phase such an asymmetry is not visible.}

{Similarly and although it is less pronounced than in the Sun, in the case of 16 Cyg A and B, a weak asymmetry is visible on $l=2$ modes. For 16 Cyg B, the two solutions of $a_4$ provide sensibly similar mode profiles as the difference is weakly apparent to the eye only for $l=2$. The residuals do not show striking differences. This visually supports the fact that goodness of the fit is similar and explain that the two solutions co-exists in statistical terms.
}

\end{appendix}

\clearpage
\bibliography{mybib}
\bibliographystyle{aa}

\end{document}